\documentclass[onecolumn,sort&compress,numbers]{els-mrw} % For numbered references Style 

\usepackage{amsmath,amssymb,amsfonts,amsthm,makeidx,graphicx}
\usepackage{txfonts}
\usepackage{helvet}
\usepackage{braket}
\usepackage{hyperref}

\hypersetup{
    colorlinks=true,
    linkcolor=blue,
    filecolor=magenta,      
    urlcolor=cyan,
}

\begin{document}

%%%%%%%%%%%%%%%%%%%%%%%%%%%%%%%%%%%%%%%%%%%%%%%%%%%%%%%%%%%%%%%%
%% the following items are mandatory: 
%% - title
%% - author names
%% - affiliation details
%% - abstract
%% - keywords

%% Precise, concise, and informative description of the focus of this work. Avoid abbreviations and formulae in the title
\chapter{The Constituent Quark Model}\label{chap1}

%% All author names and affiliations, and email address for corresponding author
\author[1,2]{D.R. Entem}%
\author[2]{F. Fern\'andez}%
\author[1]{P.G. Ortega}%
\author[3]{J. Segovia}%

%\author[1,2]{Third Author}%

\address[1]{\orgname{Universidad de Salamanca}, \orgdiv{Departamento de F\'\i sica Fundamental}, \orgaddress{E-37008 Salamanca, Spain}}
\address[2]{\orgname{Universidad de Salamanca}, \orgdiv{Instituto de F\'\i sica Fundamental y Matem\'aticas}, \orgaddress{E-37008 Salamanca, Spain}}
\address[3]{\orgname{Universidad Pablo de Olavide}, \orgdiv{Departamento de Sistemas F\'\i sicos, Qu\'\i micos y Naturales}, \orgaddress{E-41013 Sevilla, Spain}}

\articletag{The Constituent Quark Model}

\maketitle

%%%%%%%%%%%%%%%%%%%%%%%%%%%%%%%%%%%%%%%%%%%%%%%%%%%%%%%%%%%%%%%%
%% the following item is mandatory: 
%% 100-150 word summary of the chapter
\begin{abstract}[Abstract]
	In this chapter we give a pedagogical introduction to the constituent quark model. The explanation of magnetic moments of the nucleons was crucial to introduce an effective quark
    mass for light quarks that nowadays are understood as an effect of Spontaneous Chiral Symmetry Breaking in QCD. We give an overview of the first applications of the model and an introduction to the
    most modern developments studying states beyond the naive quark model as tetraquarks and pentaquarks.
\end{abstract}

%% 5-10 words that embody the key topics in the chapter. What terms would someone put into a search engine if they were looking for a chapter like this?
\begin{keywords}
 	Constituent quark model, hadron spectra, hadron-hadron interactions, tetraquarks, pentaquarks
\end{keywords}

%%%%%%%%%%%%%%%%%%%%%%%%%%%%%%%%%%%%%%%%%%%%%%%%%%%%%%%%%%%%%%%%
%% the following item is optonal: 
%% - Single figure visually illustrating the key topic/method/outcome described in the chapter
%\begin{figure}[h]
%	\centering
%	\includegraphics[width=7cm,height=4cm]{blankfig}
%	\caption{Optional: Single figure visually illustrating the key topic/method/outcome described in the chapter. 
%		     Please add here some text explaining the pic...}
%	\label{fig:titlepage}
%\end{figure}

%%%%%%%%%%%%%%%%%%%%%%%%%%%%%%%%%%%%%%%%%%%%%%%%%%%%%%%%%%%%%%%%
%% the following item is optional: 
%% - System of abbreviations/terms/symbols used in the specific field of study/community. List and define
\begin{glossary}[Nomenclature]
	\begin{tabular}{@{}lp{34pc}@{}}
		LHC & Large Hadron Collider\\
        OGE & One-gluon-exchange \\
        OPE & One-pion-exchange \\
        QCD & Quantum Chromodynamics \\
        QED & Quantum Electrodynamics \\
        QFT & Quantum Field Theory \\
        RGM & Resonating Group Method \\
        SSB & Spontaneous Symmetry Breaking \\
	\end{tabular}
\end{glossary}

%%%%%%%%%%%%%%%%%%%%%%%%%%%%%%%%%%%%%%%%%%%%%%%%%%%%%%%%%%%%%%%%
%% the following item is mandatory: 
%% List of the key points and topics a reader can expect to learn from this chapter 
\section*{Objectives}
\begin{itemize}
	\item Section 2: Explain the classification of mesons and baryons within the Constituent Quark Model, highlighting the relevance of the SU(2) and SU(3) flavor and spin symmetry groups for the description of their properties and structure.
	\item Section 3: Analyze the contribution of quarks to the magnetic moment of light baryons and justify the need for an effective mass for light quarks within the framework of the Constituent Quark Model.
	\item Section 4: Describe the interaction between quarks and antiquarks mediated by gluons using the theory of Quantum Chromodynamics (QCD), with emphasis on the exchange of a single gluon and its instantaneous nonrelativistic reduction.
    \item Section 5: Introduce the historically most successful CQM for mesons and baryons.
    \item Section 6: Analyze the consequences of the spontaneous Chiral Symmetry Breaking of QCD at quark level.
    \item Section 7: Describe the models that include SSB effects and their applications to the hadron spectra.
    \item Section 8: Explore the extension of the naive quark model including higher Fock states, with emphasis on the meson-meson and meson-baryon channel influence on the meson and baryon spectrum, respectively.
    \item Section 9: Analyze compact multiquark states, focusing on tetraquark and pentaquark compact states.
    \item Section 10: Understand the basis of the RGM in one of the first applications at quark level, the $NN$ system.
\end{itemize}

%%%%%%%%%%%%%%%%%%%%%%%%%%%%%%%%%%%%%%%%%%%%%%%%%%%%%%%%%%%%%%%%
%% the following items are mandatory: 
%% - Section: Introduction 
%% - further sections
%% - Section: Conclusion
\section{Introduction}\label{intro}

The constituent quark model is a phenomenological model that aims to incorporate the most important features of quarks and QCD into a simple framework that allows the calculation of meson and baryon observables and baryon-baryon interactions. 
Most of these properties of QCD has been explained in the former chapters.  Let summarize the most relevant for the model.

Nowadays it is accepted that hadrons are composite particles of quarks although these ultimate blocks of matter have never been directly observed or found isolated. Hadrons, mesons and baryons, are organized in multiplets following  irreducible representations of the SU(3) group. All the hadrons of each multiplet have the same spin and parity quantum numbers and are distinguished by the isospin and the hypercharge. However, since quarks are fermions, the wave function describing quark systems should be totally antisymmetric in all variables and an  artificial, in principle,  variable called color is needed for this picture to be compatible with the Fermi statistic. The wave function corresponding to this degree of freedom should be totally antisymmetric, i.e., hadrons are color singlets.
Having in mind that quarks carry color and hadrons do not, one can argue that forces between quarks depend on color. As we need three different colors it was assumed the color group symmetry to be the SU$_{\rm c}$(3). However as we cannot do experiments with isolated quarks as, for example, with electrons, it is impossible to study this force between colors in a traditional way. We know that the requirement of a local U(1) invariance led to the local gauge theory of QED. Then, if the color SU$_{\rm c}$(3) group is an exact symmetry of the hamiltonian governing quark dynamics, we can go an step further assuming that the quark wave equation is invariant under local SU$_{\rm c}(3)$ tranformations, which leads to the lagrangian of QCD. Contrarily to the case of QED, as the SU(3) is a non abelian group with eighth generators, we have in the theory eight massless vector (spin 1) gluons that interact with themselves. This last property is related to the confinement of quarks.

The Constituent Quark Model is largely based on the concept of constituent quark mass. The different ways of introduce this mass gives place to different variants of the model. In the early times it was realized that to reproduce the baryon magnetic moments it was necessary a mass larger than the current mass of the QCD lagrangian which is generated by the Higgs mechanism. Such effective mass was attributed to a cloud of low-momentum gluons attaching themselves to the current-quark.

Once one admit the existence of this mass it is straightforward to write a Hamiltonian including one gluon exchange and a confinement interaction between quarks. This correspond with the hamiltonian of the naive constituent quark model.
The model has been able to make a first description of the baryon and meson spectrum and provide an explanation of the hard core of the nucleon-nucleon interaction.

A second step forward occurs when the concept of dynamical quark mass is introduced. Chiral symmetry is a symmetry of the Lagrangian of QCD. However, it is spontaneously broken in nature, as demonstrated by the nonexistence of hadronic parity doublets. There are several candidates to explain this dynamical effect, but regardless of their origin, they have two immediate physical consequences. First, the appearance of a dynamical quark mass which, at high momentum, coincide with the current quark of perturbative QCD and evolve into a constituent quark mass as their momentum becomes smaller. Second, the appearance of Goldstone bosons which mediate an interaction between quarks. This second version of the constituent quark model is some times called Chiral Constituent Quark Model. This version become important to describe possible  molecular structures of the recent discovered exotic states.

In the following sections we will develop these two versions of the model and explain its more relevant applications.

%\begin{itemize}
%\item Mencionar modelo quark y simetrias SU(2) y SU(3)
%\item Introducir modelo quark constituyente a traves de momentos magneticos
%\item Introducir interacciones entre quarks constituyentes: confinamiento y OGE
%\item Introducir modelos quirales e interacciones entre quarks con Bosones de Goldstone (JPG 1993 y otros)
%\item Introducir modelos quark en el sector pesado (JPG 2005 Vijande y otros como Cornell)
%\item Interaccion entre hadrones, hard core NN
%\end{itemize}

%%%%%%%%%%%%%%%%%%%%%%%%%%%%%%%%%%%%%%%%%%%%%%%%%%%%%%%%%%%%%%%%
%% in the following we showcase possible style elements
%%
%%
%%%%%%%%%%%%%%%%%%%%%%%%%%%%%%%%%%%%%%%%%%%%%%%%%%%%%%%%%%%%%%%%%%%%%%%%%%%%%%%%
%%%%%%%%%%%%%%%%%%%%%%%%%%%%%%%%%%%%%%%%%%%%%%%%%%%%%%%%%%%%%%%%%%%%%%%%%%%%%%%%

\section{Mesons and Baryons in the Constituent Quark Model}

One century of fundamental research in atomic physics has demonstrated that ordinary matter is corpuscular, with the atoms themselves containing a dense nuclear core composed of protons and neutrons, collectively named as nucleons, which are members of a broader class of femtometre-scale particles, called hadrons. These, at the same time, are complicated bound states of quarks and gluons whose strong nuclear interactions are described by Quantum Chromodynamics (QCD), which is a quantum field theory with a SU($N_c$) local \emph{color} gauge symmetry with $N_c=3$ colors. Since the $N_{c}^2-1=8$ non-Abelian gauge fields, the gluons, do not carry any intrinsic quantum number beyond color charge and spin, and because color is believed to be permanently confined, a very successful classification scheme for hadrons in terms of their constituent (valence) quarks and antiquarks was independently proposed by Murray Gell-Mann~\cite{GellMann:1964nj} and George Zweig~\cite{Zweig:1964CERN} in 1964. It basically separates hadrons in mesons and baryons which are, respectively, quark-antiquark and three-quark bound states, located at the multiplets of the flavor symmetry and whose quantum numbers are given by those of their constituent quarks and antiquarks.

\begin{table}[!t]
\TBL{\caption{\label{tab:QuarkQuantumNumbers} Quark quantum numbers.}}{
\begin{tabular}{llrrrrrr}
& & $u$ & $d$ & $c$ & $s$ & $t$ & $b$ \\
\hline
$Q$ & - electric charge & $+\frac{2}{3}$ & $-\frac{1}{3}$ & $+\frac{2}{3}$ & $-\frac{1}{3}$ & $+\frac{2}{3}$ & $-\frac{1}{3}$ \\
$I_z$ & - isospin z-component & $+\frac12$ & $-\frac12$ & $0$ & $0$ & $0$ & $0$ \\
${\cal B}$ & - baryon number & $+\frac{1}{3}$ & $+\frac{1}{3}$ & $+\frac{1}{3}$ & $+\frac{1}{3}$ & $+\frac{1}{3}$ & $+\frac{1}{3}$ \\
$S$ & - strangeness & $0$ & $0$ & $0$ & $-1$ & $0$ & $0$ \\
$C$ & - charmness & $0$ & $0$ & $+1$ & $0$ & $0$ & $0$ \\
$B$ & - bottomness & $0$ & $0$ & $0$ & $0$ & $0$ & $-1$ \\
$T$ & - topness & $0$ & $0$ & $0$ & $0$ & $+1$ & $0$ \\
\end{tabular}}{}
\end{table}

Therefore, if one wants to specify the quantum numbers of a hadron following the constituent quark model formalism one must first specify the quantum numbers of its valence quarks. Quarks are spin-$1/2$ fermions and have by convention positive parity, while antiquarks have negative parity. There are three families, or generations, of quarks, made of the up, down, charm, strange, top and bottom flavors,
\begin{equation}
\left(\begin{matrix} u \\ d \end{matrix}\right), \quad \left(\begin{matrix} c \\ s \end{matrix}\right) ,\quad \left(\begin{matrix} t \\ b \end{matrix}\right),
\end{equation}
and the three families (generations) of corresponding antiquarks. The baryon number ${\cal B}$ is defined as ${\cal B} = 1/3$ for quarks and ${\cal B} = -1/3$ for antiquarks in such a way that baryons (antibaryons) have ${\cal B}=1 (-1)$ and mesons ${\cal B}=0$. Table~\ref{tab:QuarkQuantumNumbers} gives the other additive quantum numbers (flavors) for the three generations of quarks. They are related to the charge $Q$ (in units of the elementary charge, $e$, which is the absolute value of the electron's charge) through the generalized Gell-Mann--Nishijima formula
\begin{equation}
Q = I_z + \frac{{\cal B}+S+C+B+T}{2} \,,
\end{equation}
being the hypercharge, defined as
\begin{equation}
Y \equiv {\cal B}+S-\frac{1}{3}(C-B+T) \,,
\end{equation}
equal to $\frac{1}{3}$ for the $u$ and $d$ quarks, $-\frac{2}{3}$ for the $s$ quark and $0$ for the heavier quarks. By convention the flavor $S$, $C$, $B$, $T$ of a quark has the same sign as its charge $Q$. Hence the flavor carried by a charged meson has the same sign as its charge, \emph{e.g.} the strangeness of the $K^+$, $(u\bar s)$-meson, is $+1$ and the charm and strangeness of the $D_s^-$, $(s\bar c)$-meson, are each $-1$. Note that hadrons containing the $t$-quark do not bind due to its very short lifetime of about $10^{-25}\,\text{s}$ when decaying into $bW^+$.

Within the constituent quark model framework, \emph{viz.} ignoring gluon degrees of freedom, the total wave function of a meson, as a hadron with baryonic number equal zero and integer total angular momentum, can be expanded as follows
\begin{equation}
\label{eq:FockMeson}
|{\cal B}=0\rangle = \sum_{n=0}^{\infty} {\cal C}_{n} |(q\bar q)(q\bar q)^n \rangle = {\cal C}_0 |q\bar q\rangle + {\cal C}_1 |qq\bar q\bar q\rangle + \ldots \,,
\end{equation}
whereas the wave function of a baryon, which has baryonic number equal one and non-intenger total angular momentum, can be written as 
\begin{equation}
\label{eq:FockBaryon}
|{\cal B}=1\rangle = \sum_{n=0}^{\infty} {\cal D}_{n} |(qqq)(q\bar q)^{n}\rangle = {\cal D}_{0} |qqq\rangle + {\cal D}_{1} |qqqq\bar q\rangle + \ldots
\end{equation}
In the so-called naive constituent quark model, states with $n=0$ in Eqs.~\eqref{eq:FockMeson} and~\eqref{eq:FockBaryon} are named ordinary mesons and baryons, and those with $n\geq1$ exotic multiquark systems. The particular case $n=1$ within both meson and baryon sectors are known as tetraquarks and pentaquarks, respectively, and they will be discussed later. It is worth noting herein that exotic states can have quantum numbers that are not allowed by the naive constituent quark model picture.

%%%%%%%%%%%%%%%%%%%%%%%%%%%%%%%%%%%%%%%%%%%%%%%%%%%%%%%%%%%%%%%%%%%%%%%%%%%%%%%%

\begin{figure}[!t]
\includegraphics[width=0.60\textwidth]{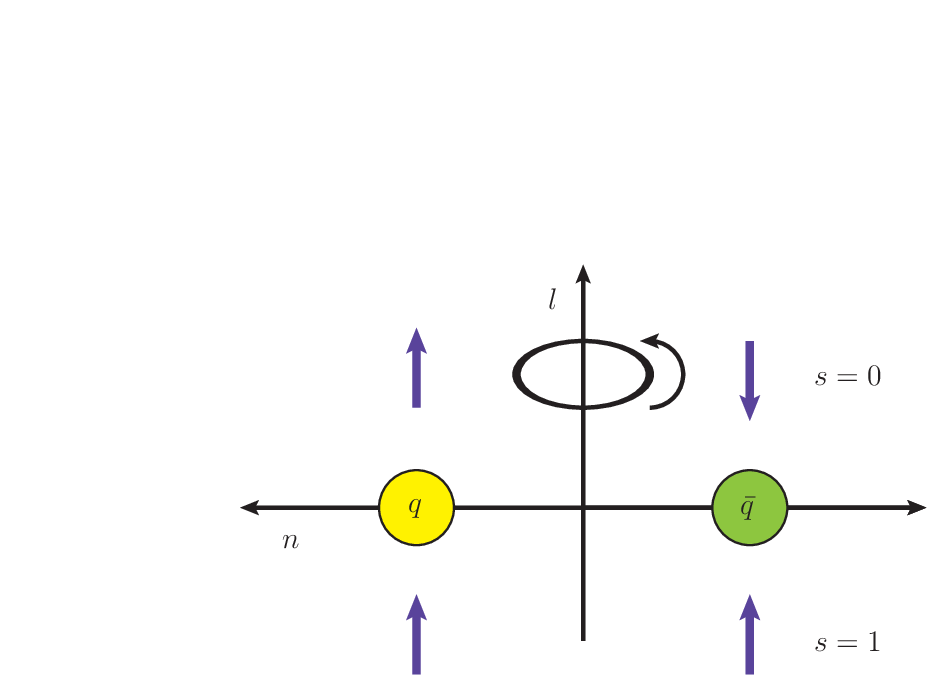}
\caption{\label{fig:MesonQuantumNumbers} An ordinary meson can be in a spin-singlet state ($s=0$, antiparallel quark spins) or in a spin-triplet
state ($s=1$, parallel quark spins). Excited states are obtained by switching angular momentum $\ell>0$ within the pair (orbital excitation) or by inducing vibrations $n>1$ (radial excitations).}
\end{figure}

\subsection{Mesons}

For a meson made by a quark, $q$, and an antiquark, $\bar q$, the total angular momentum, $\vec{j}$, is constructed by adding the spins of the constituent quark-antiquark pair, $\vec{s}=\vec{s}_q+\vec{s}_{\bar q}$, and their relative angular momentum, $\vec{\ell}$, in such a way that $\vec{j}=\vec{s}+\vec{\ell}$. As one can pictorially see in Fig.~\ref{fig:MesonQuantumNumbers}, the corresponding quantum numbers are $s=0$ (spin-singlet) or $1$ (spin-triplet) and $\ell=0$, $1$, $2$, $\ldots$ The meson's total angular momentum, $j$ is an integer number with $|\ell-s| \leq j \leq \ell+s$. Radial excitations, or vibrations, are usually denoted by the quantum number $n\geq1$\footnote{This is not the $n$ introduced in Eqs.(\ref{eq:FockMeson}-\ref{eq:FockBaryon})}. Then, the quantum numbers of an ordinary meson are represented using the following notation
\begin{equation}
n^{2s+1}\ell_j \quad\text{or}\quad i^G(j^{PC}) \,.
\end{equation}
Here, $P$ represents parity, $C$ denotes charge conjugation (or $C$-parity), $i$ is the isospin, and $G$ stands for $G$-parity. The two notations are not fully equivalent, as the former omits the isospin $i$, while the latter does not specify $n$. Hadron's total spin and isospin are often written in capital letters, leading to the common notation $I^G(J^{PC})$~\cite{ParticleDataGroup:2024cfk}.

A brief explanation of the $P$, $C$ and $G$ parities is now due. $P$-parity essentially tests whether a system looks the same or changes under mirror reflection, in such a way that $P=+1$ means that the system remains unchanged when all spatial coordinates are reversed and $P=-1$ indicates that the system changes sign when spatial coordinates are reversed. Charge conjugation ($C$-parity) transforms a particle into its corresponding antiparticle, whereby $i_3$, the electric charge $Q$, the baryon number ${\cal B}$ and the flavour quantum numbers ($S$, $C$, $B$ and $T$) reverse sign. Only neutral particles (those with no net electric charge) can have a well-defined $C$-parity, being $C=+1$ the case in which the particle is unchanged under charge conjugation and $C=-1$ otherwise. Finally, $G$-parity is defined as the product of $C$-parity and an isospin rotation of $180^{\circ}$ around the second isospin axis, $G=Ce^{i\pi i_2}$. It is specifically defined for particles in isospin multiplets (sets of particles with similar mass and properties but different charges, like the negative, neutral and positive pions). Knowing that the parity of a quark can be chosen arbitrarily, $+1$ or $-1$, and that the relative parity between a fermion and its antifermion is predicted by the Dirac equation to be negative, the convention is to assign a positive parity to quarks and a negative parity to antiquarks, whereby the parity of the meson is defined as
\begin{equation}
P (q\bar q) = (-1)^{\ell+1} \,,
\end{equation}
and the remaining $C$ and $G$ parities as
\begin{align}
C(q\bar q) &= (-1)^{\ell+s} \,, \\
G(q\bar q) &= (-1)^{\ell+s+i} \,.
\end{align}

\begin{table}[!t]
\caption{\label{tab:QuarkComposition} Quark composition of the quark-antiquark ground state mesons where $S$ denotes strangeness.}
%\begin{ruledtabular}
\begin{tabular}{crcrr}
$J^P=0^{-}$ & & $J^P=1^{-}$ & & S \\
\hline
$\pi^+$ & $|u\bar d\rangle$ & $\rho^+$ & $|u\bar d\rangle$ & 0 \\
$\pi^0$ & $\frac{1}{\sqrt{2}}|d\bar d-u\bar u\rangle$ & $\rho^0$ & $\frac{1}{\sqrt{2}}|d\bar d-u\bar u\rangle$ & 0 \\
$\pi^-$ & $-|d\bar u\rangle$ & $\rho^-$ & $-|d\bar u\rangle$ & 0 \\
$K^+$ & $|u\bar s\rangle$ & $K^{\ast+}$ & $|u\bar s\rangle$ & +1 \\
$K^0$ & $|d\bar s\rangle$ & $K^{\ast0}$ & $|d\bar s\rangle$ & +1 \\
$\overline{K}^0$ & $-|s\bar d\rangle$ & $\overline{K}^{\ast0}$ & $-|s\bar d\rangle$ & -1 \\
$K^-$ & $-|s\bar u\rangle$ & $K^{\ast-}$ & $-|s\bar u\rangle$ & -1 \\
\end{tabular}
%\end{ruledtabular}
\end{table}

Once the quantum numbers that characterize a meson has been established, one is in the position to build the meson's wave function. Let us first deal with mesons made of the three lightest quarks ($u$, $d$, $s$) and their antiquarks ($\bar u$, $\bar d$, $\bar s$). One can construct the following quark-antiquark combinations:
\begin{equation}
|u\bar u \rangle,\,
|u\bar d \rangle,\,
|u\bar s \rangle,\,
|d\bar u \rangle,\,
|d\bar d \rangle,\,
|d\bar s \rangle,\,
|s\bar u \rangle,\,
|s\bar d \rangle,\,
|s\bar s \rangle,\,
\end{equation}
which are orthogonal and unit-normalized:
\begin{align}
\langle u\bar u|u\bar d\rangle &= \langle u\bar u|u\bar s\rangle = \ldots = 0 \,, \\
\langle u\bar u|u\bar u\rangle &= \langle u\bar d|u\bar d\rangle = \ldots = 1 \,,
\end{align}
and form a nonet of the flavor SU(3) symmetry for ground, and excited, states of each meson's spin-parity quantum numbers. Table~\ref{tab:QuarkComposition} shows the quark content assignments of the ground state pseudoscalar and vector mesons following the flavor nonet classification. The $\pi^0$ and $\rho^0$ are linear combinations of $u\bar u$ and $d\bar d$ pairs, without containing any $s\bar s$. The minus signs shall be clarified later. Note also the abuse of notation $|d\bar d\rangle - |u\bar u\rangle \equiv |d\bar d - u \bar u\rangle$. The two missing isoscalar mesons (for each spin-parity quantum numbers) are expected to mix since all their quantum numbers are the same and thus their wave functions should be linear superpositions of the octet and singlet wave functions
\begin{align}
\text{SU(3)-singlet:} \quad & |1\rangle \equiv \frac{1}{\sqrt{3}} |u\bar u + d\bar d + s\bar s \rangle \,, \\
\text{SU(3)-octet:} \quad & |8\rangle \equiv \frac{1}{\sqrt{6}} |u\bar u + d\bar d - 2s\bar s \rangle \,,
\end{align}
which are normalized and orthogonal to the ones listed in Table~\ref{tab:QuarkComposition}. The linear superposition involves an angle that can be measured, the mixing angle $\theta$. That is to say, the physically observed states should be actually
\begin{align}
\label{eq:mixing}
|\psi \rangle &\equiv \hspace*{0.30cm}  \cos\theta \, |1\rangle + \sin\theta \, |8\rangle  \,, \\
|\psi'\rangle &\equiv - \sin\theta \, |1\rangle + \cos\theta \, |8\rangle  \,.
\end{align}

The condition of ideal mixing implies that the $s\bar s$ component decouples from the $u\bar u$ and $d\bar d$ ones. This happens for
\begin{equation}
\tan\theta = \frac{1}{\sqrt{2}} \quad\Rightarrow\quad \cos\theta = \sqrt{\frac{2}{3}} \,,\quad \sin\theta=\sqrt{\frac{1}{3}} \,,
\end{equation}
where
\begin{align}
|\psi\rangle &= \frac{\sqrt{2}}{3} | u\bar u + d\bar d + s\bar s \rangle + \frac{1}{3\sqrt{2}} | u\bar u + d\bar d - 2 s\bar s \rangle = \frac{1}{\sqrt{2}} |d\bar d + u\bar u \rangle \,, \\
|\psi'\rangle &= - \frac{1}{3} | u\bar u + d\bar d + s\bar s \rangle + \frac{1}{3} | u\bar u + d\bar d - 2 s\bar s \rangle = -|s\bar s\rangle \,. \\
\end{align}
But also for the case in which 
\begin{equation}
\tan\theta = -\sqrt{2} \quad\Rightarrow\quad \cos\theta = \sqrt{\frac{1}{3}} \,,\quad \sin\theta=-\sqrt{\frac{2}{3}} \,,
\end{equation}
where the quark content of $\psi$ and $\psi'$ are swapped
\begin{align}
|\psi\rangle &= \frac{1}{3} | u\bar u + d\bar d + s\bar s \rangle - \frac{1}{3} | u\bar u + d\bar d - 2 s\bar s \rangle = |s\bar s\rangle \,, \\
|\psi'\rangle &= \frac{\sqrt{2}}{3} | u\bar u + d\bar d + s\bar s \rangle + \frac{1}{3\sqrt{2}} | u\bar u + d\bar d - 2 s\bar s \rangle = \frac{1}{\sqrt{2}} |d\bar d + u\bar u \rangle \,.
\end{align}

There is then an ambiguity in the mixing of Eq.~\eqref{eq:mixing}, with $\theta$ equals to either $35.4^\circ$ or $-54.7^\circ$, depending on which observed state is ascribed to the $\psi$ and which to the $\psi'$. Note that $180^\circ$ can always be added or subtracted (which flips the signs of the wavefunctions in Eq.~\eqref{eq:mixing}), but the convention that $\theta$ lies between $-90^\circ$ and $+90^\circ$ is adopted. For vector mesons the mixing is usually written as
\begin{align}
|\omega \rangle &\equiv \hspace*{0.30cm}  \cos\theta_V \, |1\rangle + \sin\theta_V \, |8\rangle  \,, \\
|\phi\rangle &\equiv - \sin\theta_V \, |1\rangle + \cos\theta_V \, |8\rangle  \,,
\end{align}
where $\theta_V$ is close to ideal, $\theta_V\approx35.4^\circ$; hence
\begin{align}
|\omega \rangle \approx \frac{1}{\sqrt{2}} |u\bar u + d\bar d \rangle \,, \quad
|\phi\rangle \approx -|s\bar s\rangle \,.
\end{align}
There are exceptions, the most paradigmatic one is the case of pseudoscalars, where the convention is
\begin{align}
|\eta' \rangle &\equiv \hspace*{0.30cm}  \cos\theta_P \, |1\rangle + \sin\theta_P \, |8\rangle  \,, \\
|\eta\rangle &\equiv - \sin\theta_P \, |1\rangle + \cos\theta_P \, |8\rangle  \,,
\end{align}
with $\theta_P$ in the range $-10^\circ$ to $-20^\circ$, \emph{viz.} it is far from the ideal mixing and indicates that $\eta$ and $\eta'$ are almost pure octet and pure singlet, respectively.

\subsection{The groups SU(2), SU(3) and their representations}

In the lowest dimension nontrivial representation of the rotation group $(j=1/2)$, the generators may be written as
\begin{equation}
\label{eq:genSU2}
J_i = \frac{1}{2} \sigma_i \,, \quad \text{with } i=1,\,2,\,3,
\end{equation} 
with $J_{\pm}=J_{1}\pm iJ_2$ and $\sigma_i$ are the Pauli matrices
\begin{equation}
\sigma_1 = \left(\begin{matrix} 0 & 1 \\ 1 & 0 \end{matrix}\right) \,, \quad
\sigma_2 = \left(\begin{matrix} 0 & -i \\ i & 0 \end{matrix}\right) \,, \quad
\sigma_3 = \left(\begin{matrix} 1 & 0 \\ 0 & -1 \end{matrix}\right) \,.
\end{equation}
The basis (or set of base states) for this representation is conventionally chosen to be the eigenvectors of $\sigma_3$, that is, the column vectors
\begin{equation}
\left(\begin{matrix} 1 \\ 0 \end{matrix}\right) \,, \quad\text{and}\quad
\left(\begin{matrix} 0 \\ 1 \end{matrix}\right) \,,
\end{equation}
describing a spin-1/2 particle of spin projection up ($m=+1/2$ or $\uparrow$) and spin projection down ($m=-1/2$ or $\downarrow$) along the 3-axis, respectively.

The Pauli matrices $\sigma_i$ are hermitian, and the transformation matrices
\begin{equation}
U(\theta_i) = e^{-i\theta_i\sigma_i/2} \,,
\end{equation}
are then unitary, and $\det(U)=\pm 1$. The set of all unitary $2\times 2$ matrices is known as the group U(2). However, U(2) is larger than the group of matrices $U(\theta_i)$, since the generators $\sigma_i$ all have zero trace. Now, for any hermitian traceless matrix $\sigma$, we can show that
\begin{equation}
\det(e^{i\sigma}) = e^{i{\rm Tr}(\sigma)} = 1 \,.
\end{equation} 
The set of unitary $2\times 2$ matrices with determinant 1, form the SU(2) subgroup of U(2), known as the special unitary group in two dimensions. The SU(2) algebra is just the algebra of the generators $J_i$ in Eq.~\eqref{eq:genSU2}. There are thus $1$, $2$, $3$, $4$, $\ldots$, dimensional representations of SU(2) corresponding to $j=0$, $1/2$, $1$, $3/2$, $\ldots$, respectively. The two-dimensional representation is, of course, just the $\sigma$-matrices themselves. It is called the fundamental representation of SU(2), the first non-trivial representation from which all others can be built.

Looking at the masses of hadrons one can realized that there are copies of the same particle whose only difference seems to be their electric charge. We are talking about the proton and neutron or the three pionic states. Nuclear physicists took this as a hint that they are in fact manifestations of the same particle, assigning to them an additional internal degree of freedom which the nuclear strong interaction does not distinguish. Indeed the mathematical structure used to discuss the similarity between neutron and proton is almost a carbon copy of spin, and is called isospin. The isospin generators satisfy
\begin{equation}
[I_i,I_j] = i \epsilon_{ijk} I_k \,,
\end{equation}
with $\epsilon_{ijk}$ the Levi-Civita symbol. The proton and neutron form an example of the SU(2) isospin fundamental representation 
\begin{equation}
p = \left(\begin{matrix} 1 \\ 0 \end{matrix}\right) \,, \quad\text{and}\quad
n = \left(\begin{matrix} 0 \\ 1 \end{matrix}\right) \,,
\end{equation}
in such a way that the generators are $I_i = \frac{1}{2}\tau_i$, where
\begin{equation}
\tau_1 = \left(\begin{matrix} 0 & 1 \\ 1 & 0 \end{matrix}\right) \,, \quad
\tau_2 = \left(\begin{matrix} 0 & -i \\ i & 0 \end{matrix}\right) \,, \quad
\tau_3 = \left(\begin{matrix} 1 & 0 \\ 0 & -1 \end{matrix}\right) \,,
\end{equation}
are the isospin version of the Pauli matrices. Concerning the three pionic states: $\pi^+$, $\pi^0$ and $\pi^-$; they are located in a weight diagram (multiplet) of the $3$-dimensional representation of SU(2) isospin and then they can be represented as the following eigenvectors
\begin{equation}
\pi^+ = \left(\begin{matrix} 1 \\ 0 \\ 0 \end{matrix}\right) \,, \quad
\pi^0 = \left(\begin{matrix} 0 \\ 1 \\ 0 \end{matrix}\right) \,, \quad 
\pi^- = \left(\begin{matrix} 0 \\ 0 \\ 1 \end{matrix}\right) \,,
\end{equation}
that correspond to the eigenvalues $i_3=+1,0,-1$ of the operator $I_3$ represented by the matrix
\begin{equation}
I_3 = \left( \begin{matrix} 1 & 0 & 0 \\ 0 & 0 & 0 \\ 0 & 0 & -1 \end{matrix} \right) \,.
\end{equation}

\begin{figure}[!t]
\includegraphics[width=0.80\textwidth]{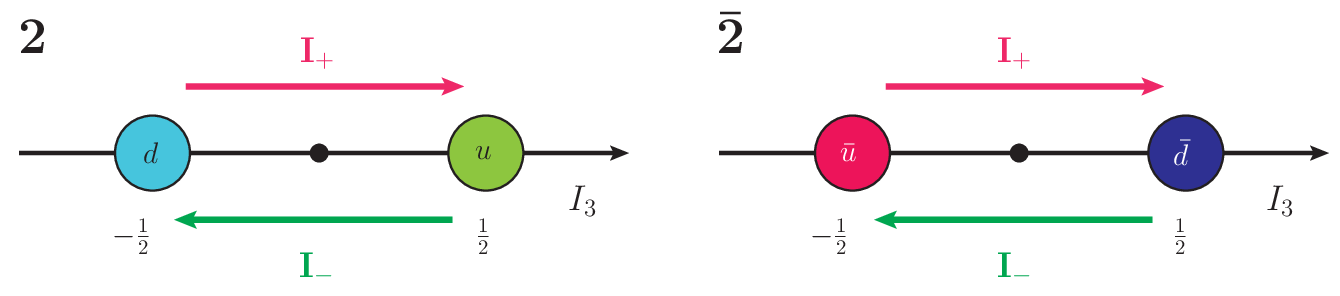}
\caption{\label{fig:WeightDiagramsSU2} Weight diagrams for SU(2).}
\end{figure}

Since mesons and baryons are made up of quarks, and we aim to generalize the discussion of isospin to encompass any hadron, let's derive the isospinor expressions for the $u$ and $d$ quarks (and their antiquark counterparts). Our first goal is to symmetrize the wave functions of the light mesons listed in Table~\ref{tab:QuarkComposition}, and to ensure that the $i=0$ and $i=1$ wave functions are eigenstates of $G$-parity. Figure~\ref{fig:WeightDiagramsSU2} displays the weight diagrams for both the fundamental representation of SU(2) and its conjugate representation. Now, let's apply a passive rotation about the $y$-axis to the quark isospinor with $u$ and $d$ components:
\begin{equation}
\label{eq:Isoquarks}
\left(\begin{matrix} u' \\ d' \end{matrix}\right) = \left(\begin{matrix} \cos\frac{\theta}{2} & \sin\frac{\theta}{2} \\ -\sin\frac{\theta}{2} & \cos\frac{\theta}{2} \end{matrix}\right)  \left(\begin{matrix} u \\ d \end{matrix}\right)
\end{equation}
Charge conjugation $Cu=\bar u$ and $Cd=\bar d$ flips the sign of $i_3$ as shown in Fig.~\ref{fig:WeightDiagramsSU2}. Reversing the order of Eq.~\eqref{eq:Isoquarks} and applying $C$ gives
\begin{equation}
\left(\begin{matrix} \bar{d}' \\ -\bar{u}' \end{matrix}\right) = \left(\begin{matrix} \cos\frac{\theta}{2} & \sin\frac{\theta}{2} \\ -\sin\frac{\theta}{2} & \cos\frac{\theta}{2} \end{matrix}\right)  \left(\begin{matrix} \bar{d} \\ -\bar{u} \end{matrix}\right) \,.
\end{equation}
Therefore, the isospinor of the antiquark with components $(\bar d, -\bar u)$ has the same transformation properties as those of the quark with components $(u, d)$. Therefore, we shall label the isospinor of the  $u$ antiquark as $|-\bar u\rangle$ (or $-|\bar u\rangle$).

The $G$-parity operation, $G=Ce^{i\pi i_2}$, on the $u$ and $d$ quarks gives
\begin{equation}
C \begin{pmatrix} 0 & 1 \\ -1 & 0 \end{pmatrix} \begin{pmatrix} u \\ d \end{pmatrix} = \begin{pmatrix} \bar d \\ -\bar u \end{pmatrix},
\quad\quad \text{and} \quad\quad
C \begin{pmatrix} 0 & 1 \\ -1 & 0 \end{pmatrix} \begin{pmatrix} \bar d \\ -\bar u \end{pmatrix} = \begin{pmatrix} -u \\ -d \end{pmatrix} \,,
\end{equation}
leading to the transformations
\begin{equation}
G u = \bar d, \quad G d = -\bar u, \quad G \bar{u} = d, \quad G \bar{d} = -u \,.
\end{equation}
and so $G$-parity operator reverses the sign when applied to a $d$ or $\bar{d}$ quark. Furthermore, we find that $G^2 u = -u$ and $G^2 d = -d$. For example, the following combinations are eigenstates of the $G$-parity:
\begin{align}
G (|u\bar d\rangle - |\bar d u\rangle) &= +1 \cdot (|u\bar d\rangle - |\bar d u\rangle)\,, \\
G (|u\bar d\rangle + |\bar d u\rangle) &= -1 \cdot (|u\bar d\rangle + |\bar d u\rangle)\,,
\end{align}
with positive and negative $G$-parities, respectively. We have seen that the pion has negative $G$-parity. Therefore, the symmetrized SU(2) wavefunctions are obtained by adding the permuted pairs. With proper normalization, the case of the pion, with negative G-parity, is as follows
\begin{align}
|\pi^+\rangle &= \frac{1}{\sqrt{2}} \big( |u\bar d\rangle + |\bar d u\rangle \big) \,, \\
|\pi^0\rangle &= \frac{1}{2} \big( |d\bar d - u\bar u\rangle + |\bar d d - \bar u u\rangle \big) \,, \\
|\pi^-\rangle &= -\frac{1}{\sqrt{2}} \big( |d\bar u\rangle + |\bar u d\rangle \big) \,.
\end{align}
Note now that, within the quark model picture, a pseudoscalar meson has a toal wave function given by
\begin{equation}
|\pi^+\rangle = \Phi(\ell=0) \cdot \frac{1}{\sqrt{2}} \big( |\uparrow\downarrow \rangle - |\downarrow\uparrow \rangle \big) \cdot \frac{1}{\sqrt{2}} \big( |u\bar d\rangle + |\bar d u\rangle \big) \,,
\end{equation}
\emph{viz.} a total wave function made up of a symmetric orbital wave function, an antisymmetric spin wave function and a flavor one which is symmetric.

\begin{table}[!t]
\caption{\label{tab:QuarkComposition2} Symmetrized flavor wave functions of the light-quark ground state $1S_0$ and $3S_1$ mesons.}
%\begin{ruledtabular}
\begin{tabular}{lrlr}
$1^1S_0(0^{-+})$ & & $1^3S_1(1^{--})$ & \\
\hline
$\ket{\pi^+}$ & $\frac{1}{\sqrt{2}}\ket{u\bar d + \bar d u}$ & $\ket{\rho^+}$ & $\frac{1}{\sqrt{2}}\ket{u\bar d - \bar d u}$ \\
$\ket{\pi^0}$ & $\frac{1}{2}(\ket{d\bar d - u\bar u} + \ket{\bar d d-\bar u u })$ & $\ket{\rho^0}$ & $\frac{1}{2}(\ket{d\bar d - u\bar u} - \ket{\bar d d-\bar u u })$ \\
$\ket{\pi^-}$ & $-\frac{1}{\sqrt{2}} \ket{d\bar u + \bar u d}$ & $\ket{\rho^-}$ & $-\frac{1}{\sqrt{2}} \ket{d\bar u - \bar u d}$ \\
$\ket{8}$ & $\frac{1}{2\sqrt{3}}(\ket{u\bar u+d\bar d-2s\bar s}+\ket{\bar u u+\bar d d - 2\bar s s})$ & $\ket{8}$ & $\frac{1}{2\sqrt{3}}(\ket{u\bar u+d\bar d-2s\bar s} - \ket{\bar u u+\bar d d - 2\bar s s})$ \\
$\ket{1}$ & $\frac{1}{\sqrt{6}}(\ket{u\bar u+d\bar d+s\bar s}+\ket{\bar u u+ \bar d d + \bar s s})$ & $\ket{1}$ & $\frac{1}{\sqrt{6}}(\ket{u\bar u+d\bar d+s\bar s} - \ket{\bar u u+ \bar d d + \bar s s})$ \\
$\ket{K^+}$ & $\frac{1}{\sqrt{2}}\ket{u\bar s + \bar s u}$ &  $\ket{K^{\ast+}}$ & $\frac{1}{\sqrt{2}}\ket{u\bar s - \bar s u}$ \\
$\ket{K^0}$ & $\frac{1}{\sqrt{2}}\ket{d\bar s + \bar s d}$ & $\ket{K^{\ast0}}$ & $\frac{1}{\sqrt{2}}\ket{d\bar s - \bar s d}$ \\
$\ket{\overline{K}^0}$ & $-\frac{1}{\sqrt{2}}\ket{s \bar d + \bar d s}$ & $\ket{\overline{K}^{\ast0}}$ & $-\frac{1}{\sqrt{2}}\ket{s \bar d - \bar d s}$ \\
$\ket{K^-}$ & $-\frac{1}{\sqrt{2}}\ket{s \bar u + \bar u s}$ & $\ket{K^{\ast-}}$ & $-\frac{1}{\sqrt{2}}\ket{s \bar u - \bar u s}$ \\
\end{tabular}
%\end{ruledtabular}
\end{table}

Table~\ref{tab:QuarkComposition2} presents the symmetrized flavor wave functions for both pseudoscalar and vector mesons for $i=1$. For completeness, the table also lists the singlet and octet isoscalar flavor wave functions as well as those of the two kaon isodoublets. Note that the two kaon doublets, $(K^+,K^0)$ and $(\overline{K}^0,K^-)$, constructed as in Table~\ref{tab:QuarkComposition2} have the following $G$ and $C$ parity transformations 
\begin{align}
&
G\ket{K^+} = - \ket{\overline{K}^0}, \quad
G\ket{K^0} = + \ket{K^-}, \quad
G\ket{\overline{K}^0} = + \ket{K^+}, \quad
G\ket{K^-} = - \ket{K^0}, \\
&
C\ket{K^0} = - \ket{\overline{K}^0}, \quad
C\ket{\overline{K}^0} = - \ket{K^0}, \quad
C\ket{K^\pm} = -\ket{K^\mp},
\end{align}
for the pseudoscalar kaons, and
\begin{align}
&
G\ket{K^{\ast+}} = + \ket{\overline{K}^{\ast0}}, \quad
G\ket{K^{\ast0}} = - \ket{K^{\ast-}}, \quad
G\ket{\overline{K}^{\ast0}} = - \ket{K^{\ast+}}, \quad
G\ket{K^{\ast-}} = + \ket{K^{\ast0}}, \\
&
C\ket{K^{\ast0}} = + \ket{\overline{K}^{\ast0}}, \quad
C\ket{\overline{K}^{\ast0}} = + \ket{K^{\ast0}}, \quad
C\ket{K^{\ast\pm}} = +\ket{K^{\ast\mp}},
\end{align}
for the vector ones.

\begin{table}
\caption{\label{tab:SU3fijk} SU(3) structure constants.}
%\begin{ruledtabular}
\begin{tabular}{cccccccccc}
$ijk$ & $123$ & $147$ & $156$ & $246$ & $257$ & $345$ & $367$ & $458$ & $678$ \\ 
\hline
$f_{ijk}$ & $1$ & $1/2$ & $-1/2$ & $1/2$ & $1/2$ & $1/2$ & $-1/2$ & $\sqrt{3}/2$ & $\sqrt{3}/2$ \\
\end{tabular}
%\end{ruledtabular}
\end{table}

\begin{figure}[!t]
\includegraphics[width=0.80\textwidth]{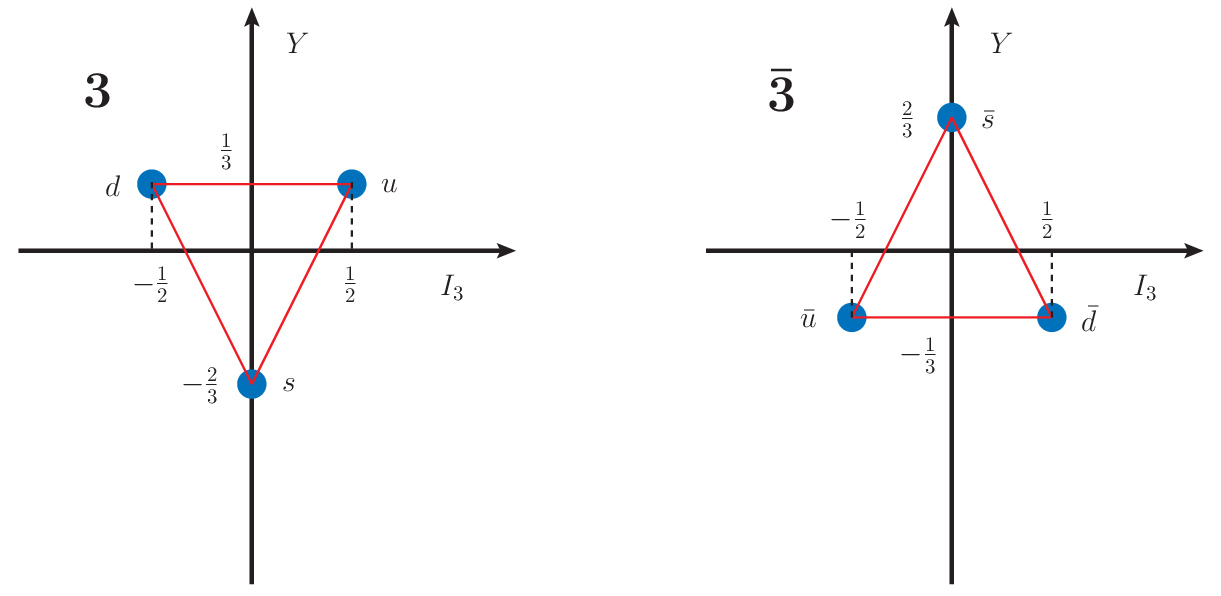}
\caption{\label{fig:SU3quarks} \emph{Left panel:} Weight diagram of the fundamental representation of SU(3). \emph{Right panel:} weight diagram of the conjugate representation.}
\end{figure}

\begin{figure}[!t]
\includegraphics[width=0.40\textwidth]{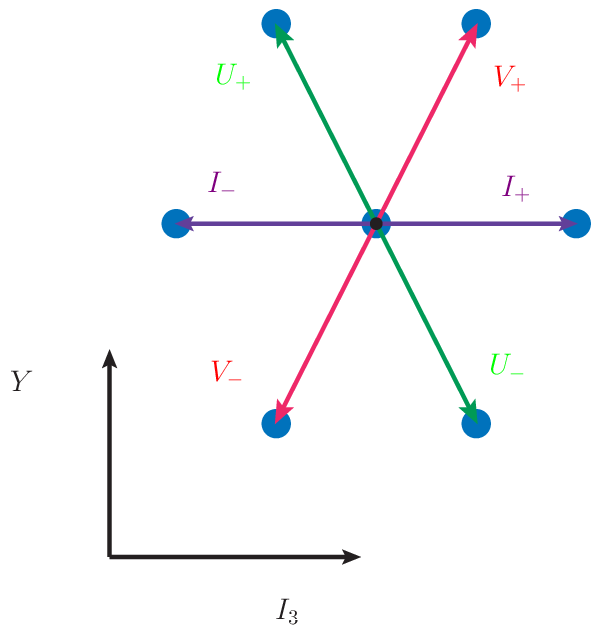}
\caption{\label{fig:IVUspin} The raising and lowering operators $I_\pm$ increase or decrease $i_3$ by one unit, while $V_\pm$ and $U_\pm$ increase or decrease $y$ by one unit and $i_3$ by $\frac{1}{2}$.}
\end{figure}

As suggested by the previous discussions, the isospin symmetry group for the $u$ and $d$ quarks can be expanded to include the $s$-quark. In addition to isospin's third component, $i_3$, strangeness $S$ is also conserved in strong interactions. However, it is often more practical to use the hypercharge $Y$ as the conserved quantity, whose expression is
\begin{equation}
Y = B + S = \frac{1}{3} + S \,.
\end{equation}
We are therefore seeking a group of unitary operators which commute with the
Hamiltonian, but for which two of the generators are simultaneously diagonal:
\begin{equation}
[I_3,H]=0,\quad [Y,H]=0, \quad \text{and}\quad [I_3,Y]=0.
\end{equation}
The symmetry group is SU(3) with the unitary operators,
\begin{equation}
\label{eq:U-SU3}
U = e^{i \sum_{n=1}^{8} \alpha_n G_n } \,,
\end{equation}
where the eight generators $G_n$ obey the commutation relations,
\begin{equation}
[G_i,G_j] = i f_{ijk} G_k \,.
\label{su3struct}
\end{equation}
The structure constants $f_{ijk}$ are given in Table~\ref{tab:SU3fijk} and the fundamental representation of SU(3) is described by $G_i\equiv \frac{1}{2}\lambda_i$ with
\begin{align}
&
\lambda_1 = \begin{pmatrix}
0 & 1 & 0 \\
1 & 0 & 0 \\
0 & 0 & 0
\end{pmatrix}, \quad
\lambda_2 = \begin{pmatrix}
0 & -i & 0 \\
i & 0 & 0 \\
0 & 0 & 0
\end{pmatrix}, \quad
\lambda_3 = \begin{pmatrix}
1 & 0 & 0 \\
0 & -1 & 0 \\
0 & 0 & 0
\end{pmatrix}, \nonumber \\
&
\lambda_4 = \begin{pmatrix}
0 & 0 & 1 \\
0 & 0 & 0 \\
1 & 0 & 0
\end{pmatrix}, \quad
\lambda_5 = \begin{pmatrix}
0 & 0 & -i \\
0 & 0 & 0 \\
i & 0 & 0
\end{pmatrix}, \quad
\lambda_6 = \begin{pmatrix}
0 & 0 & 0 \\
0 & 0 & 1 \\
0 & 1 & 0
\end{pmatrix}, \nonumber \\
&
\lambda_7 = \begin{pmatrix}
0 & 0 & 0 \\
0 & 0 & -i \\
0 & i & 0
\end{pmatrix}, \quad
\lambda_8 = \frac{1}{\sqrt{3}} \begin{pmatrix}
1 & 0 & 0 \\
0 & 1 & 0 \\
0 & 0 & -2
\end{pmatrix}.
\end{align}
the Gell-Mann matrices that replace the Pauli matrices of SU(2). Notice that 
\begin{align}
    \tilde G_i = -G_i^* = - G_i^T
\end{align}
also fulfills Eq.(\ref{su3struct}) which means that is also a representation of SU(3) and it is called the dual or conjugate representation.
The diagonal operators are $I_3$ and $Y$ with
\begin{equation}
I_3 \equiv G_3, \quad\quad Y \equiv \frac{2}{\sqrt{3}} G_8 \,.
\end{equation}
The quantum numbers $i_3$ and $y$ for the three light quarks can be read off the diagonal elements of the corresponding matrices $I_3$ and $Y$. The weight diagram of the fundamental representation is shown in left-panel of Fig.~\ref{fig:SU3quarks}. SU(3) includes three SU(2) subgroups, those of the $I$-spin, and the so-called $V$-spin and $U$-spin, respectively. The corresponding ladder operators are
\begin{equation}
I_\pm = G_1 \pm i G_2 ,\quad 
V_\pm = G_4 \pm i G_5 ,\quad
U_\pm = G_6 \pm i G_7 ,
\end{equation}
which, acting on the wave functions, the ladder operators increment or decrement the quantum numbers $i_3$ and $y$ as illustrated in Fig.~\ref{fig:IVUspin}. The conjugate representation of SU(3) is not equal to the fundamental one as shown in the right panel of Fig.~\ref{fig:SU3quarks}. That is to say, the group elements of the conjugate representation are obtained by complex conjugating the unitary operator $U$ in Eq.~\eqref{eq:U-SU3}, \emph{i.e.} by replacing the generators $G_i$ by $G_i^\prime=-G_i^\ast$

\begin{figure}[!t]
\includegraphics[width=0.40\textwidth]{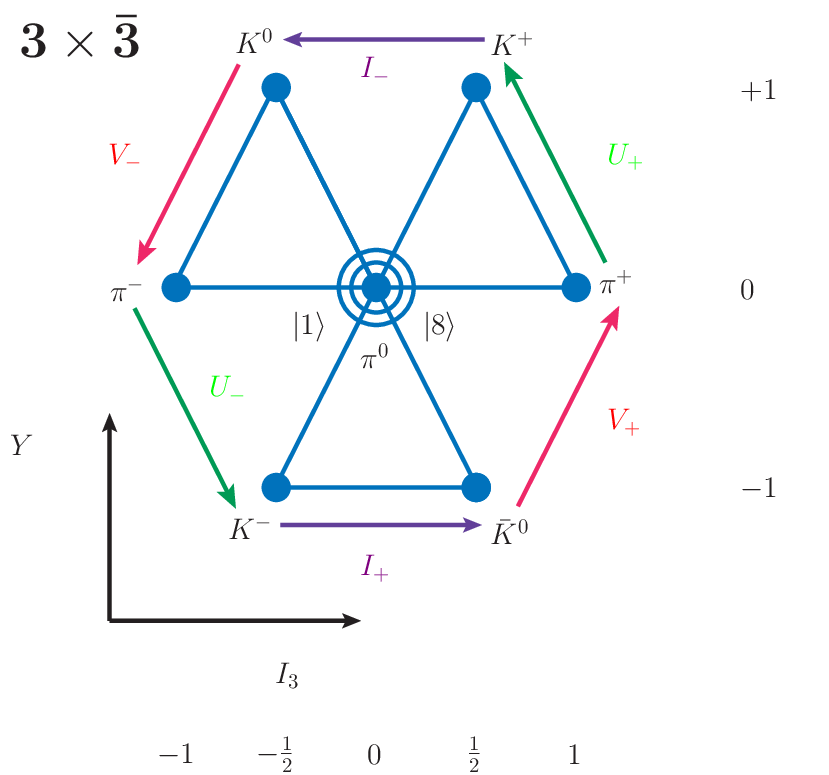}
\caption{\label{fig:PseudoscalarNonet} The pseudoscalar nonet $\bf 3\times \bar 3$ representation decomposed into an octet and a singlet.}
\end{figure}

We have then explained all the tools needed to construct the SU(3) flavor wave functions of quark-antiquark mesons shown in Table~\ref{tab:QuarkComposition}. That is to say, coupling the fundamental and conjugate representations gives an octet plus a singlet ones. The corresponding weight diagram is easily obtained by superimposing the three triangles of the conjugate representation to the corners of the fundamental one (see Fig.~\ref{fig:PseudoscalarNonet} for pseudoscalar mesons). The center of the hexagon $(i_3=y=0)$ is occupied by three states, the $i=1$ and the two $i=0$ states.

\subsection{Baryons}

Baryons are composed of three quarks, while their charge-conjugated counterparts, the antibaryons, consist of the corresponding three antiquarks. Baryons carry a baryon number $B=1$, whereas antibaryons have $B=-1$. Baryons made only of light quarks ($u$ and $d$) are classified as $N$ for isospin $i = \frac{1}{2}$ and $\Delta$ for $i = \frac{3}{2}$. Hyperons are baryons that include at least one strange ($s$) quark. Baryons with two light quarks are designated $\Lambda$ for $i=0$ and $\Sigma$ for $i=1$. Baryons with only one light quark are called $\Xi$ and form isospin doublets, while those without any light quarks are referred to as $\Omega$ and are isospin singlets.

\begin{figure}[!t]
\includegraphics[width=0.30\textwidth]{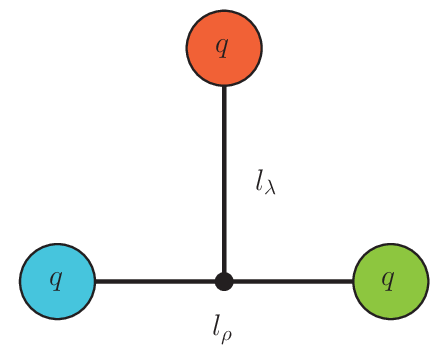}
\caption{\label{fig:BaryonQuantumNumbers} Angular momenta $\ell_\rho$ and $\ell_\lambda$ in a $qqq$ baryon.}
\end{figure}

Following parity conventions, the baryon's parity is given by
\begin{equation}
P(qqq) = (-1)^{\ell_\rho} (-1)^{\ell_\lambda} \,,
\end{equation}
whereas the one for its antibaryon by
\begin{equation}
P(\bar q\bar q\bar q) = - (-1)^{\ell_\rho} (-1)^{\ell_\lambda} \,.
\end{equation}
In this context, $\ell_\rho$ refers to the angular momentum of a quark-quark (diquark) pair, while $\ell_\lambda$ is the angular momentum between the diquark and the third quark (see Fig.~\ref{fig:BaryonQuantumNumbers}). For ground state baryons, both $\ell_\rho$ and $\ell_\lambda$ are zero, leading to quantum numbers $J^P = \frac{1}{2}^+$ or $\frac{3}{2}^+$. The corresponding antibaryons exhibit opposite parities. When additional angular momentum is included, higher spin states such as $\frac{5}{2}$, $\frac{7}{2}$, and so on, become possible, along with changes in parity. The $C$ and $G$ parities are not well-defined.

In order to derive the flavor wave functions of $3$-quark ground states, we first need to study their spin structure. The simplest configuration is that of the $\Delta$ resonance with $j=\frac{3}{2}$ and parallel spins:
\begin{equation}
\ket{\chi_{jm}} = |\chi_{\frac{3}{2}+\frac{3}{2}}\rangle = \ket{\uparrow\uparrow\uparrow} \,.
\end{equation}
From now on, starting from the $\Delta$ wave function, we apply the ladder operators on the quarks and then normalize the result:
\begin{align}
|\chi_{\frac{3}{2}+\frac{1}{2}}\rangle &= \frac{1}{\sqrt{3}} \big( \ket{\downarrow\uparrow\uparrow} + \ket{\uparrow\downarrow\uparrow} + \ket{\uparrow\uparrow\downarrow}\big) \,, \label{eq:32+12} \\
|\chi_{\frac{3}{2}-\frac{1}{2}}\rangle &= \frac{1}{\sqrt{3}} \big( \ket{\downarrow\downarrow\uparrow} + \ket{\downarrow\uparrow\downarrow} + \ket{\uparrow\downarrow\downarrow}\big) \,, \\
|\chi_{\frac{3}{2}-\frac{3}{2}}\rangle &= \ket{\downarrow\downarrow\downarrow} \,.
\end{align}
These four spin-$\frac{3}{2}$ wave functions build the symmetric quadruplet predicted by the decomposition of the spin coupling
\begin{equation}
\bf 2 \times 2 \times 2 = 2 + 2 + 4 \,.
\end{equation}
and they are given in notation of angular coupling as
\begin{eqnarray}
\ket{\chi_{\frac 3 2 m}} &=& |(\frac{1}{2}\frac{1}{2})1\frac{1}{2};\frac 3 2 m\rangle = 
\sum_{m_{s_1}m_{s_2}m_{s_{12}}m_{s_3}} (\frac 1 2 m_{s_1} \frac 1 2 m_{s_2}|1 m_{s_{12}})(1 m_{s_{12}} \frac 1 2 m_{s_3}|\frac 3 2 m) 
\ket{\frac 1 2 m_{s_1}} \ket{\frac 1 2 m_{s_2}}\ket{\frac 1 2 m_{s_3}}
\label{j32}
\end{eqnarray}
where $(j_1 m_1 j_2 m_2|jm)$ are Clebsch-Gordan coeficients and should be read as angular momentum $\frac 1 2$ of quarks 1 and 2 coupled to 1 and the result coupled with spin $\frac 1 2$ of quark 3 to the total spin $\frac 3 2$ with component $m$.

On the other hand, we also expect two mixed-symmetric doublets with spin $\frac{1}{2}$ that are orthogonal to the four previous states. Consider the configurations with $m=+\frac{1}{2}$ in which the spin projection of a diquark vanishes, while the third quark has spin up. The diquark can be either in the antisymmetric $s=0$,
\begin{equation}
\ket{00} = \frac{1}{\sqrt{2}} \big( \ket{\uparrow\downarrow} - \ket{\downarrow\uparrow} \big) \,,
\end{equation}
state or in the symmetric $s=1$ state,
\begin{equation}
\ket{10} = \frac{1}{\sqrt{2}} \big( \ket{\uparrow\downarrow} + \ket{\downarrow\uparrow} \big) \,.
\end{equation}
The wave function of the so-called mixed symmetric case (in which the permutation of the first two quarks is symmetric) reads as
\begin{equation}
|\chi_{\frac{1}{2}+\frac{1}{2}}\rangle = \frac{1}{\sqrt{2}} \big(\ket{\uparrow\downarrow\uparrow} + \ket{\downarrow\uparrow\uparrow}\big) + \alpha \ket{\uparrow\uparrow\downarrow} \,.
\end{equation}
The inclusion of the third term with $\alpha \neq 0$ is necessary; otherwise, the wave function cannot be orthogonal to Eq.~\eqref{eq:32+12}. By multiplying from the left with the bra of Eq.~\eqref{eq:32+12} and imposing the orthogonality condition we get
\begin{equation}
2\frac{1}{\sqrt{2}}\frac{1}{\sqrt{3}} + \frac{\alpha}{\sqrt{3}} = 0 \,,
\end{equation}
one is lead to $\alpha=-2$. The normalized wave function is then equal to
\begin{equation}
|\chi_{\frac{1}{2}+\frac{1}{2}}\rangle = \frac{1}{\sqrt{6}} \big( \ket{\uparrow\downarrow\uparrow} + \ket{\downarrow\uparrow\uparrow} - 2 \ket{\uparrow\uparrow\downarrow} \big) \,.
\end{equation}
Operating with $J_-$ on each quark gives
\begin{equation}
|\chi_{\frac{1}{2}-\frac{1}{2}}\rangle = -\frac{1}{\sqrt{6}} \big( \ket{\uparrow\downarrow\downarrow} + \ket{\downarrow\uparrow\downarrow} - 2 \ket{\downarrow\downarrow\uparrow} \big) \,.
\end{equation}
For the mixed antisymmetric case, the spin wavefunction is
\begin{equation}
|\chi_{\frac{1}{2}+\frac{1}{2}}\rangle^\prime = \frac{1}{\sqrt{2}} \big( \ket{\uparrow\downarrow\uparrow} - \ket{\downarrow\uparrow\uparrow} \big) \,,
\end{equation}
and, again, operating with $J_-$ on each quark gives
\begin{equation}
|\chi_{\frac{1}{2}-\frac{1}{2}}\rangle^\prime = \frac{1}{\sqrt{2}} \big( \ket{\uparrow\downarrow\downarrow} - \ket{\downarrow\uparrow\downarrow} \big) \,,
\end{equation}

The isospin wave functions for baryons with $u$ and $d$ quarks are immediately
found by formally replacing $\uparrow$ by $u$ and $\downarrow$ by $d$. That is to say, one has for the fully symmetric case, $\ket{\phi_S}$,
\begin{align}
\Delta^{++}:\quad & |\phi_{\frac{3}{2}+\frac{3}{2}}\rangle = \ket{uuu}  \,, \nonumber \\
\Delta^+:\quad & |\phi_{\frac{3}{2}+\frac{1}{2}}\rangle = \frac{1}{\sqrt{3}}\big(\ket{duu}+\ket{udu}+\ket{uud}\big)  \,, \nonumber \\
\Delta^0:\quad & |\phi_{\frac{3}{2}-\frac{1}{2}}\rangle = \frac{1}{\sqrt{3}}\big(\ket{ddu}+\ket{dud}+\ket{udd}\big)  \,, \nonumber \\
\Delta^-:\quad & |\phi_{\frac{3}{2}-\frac{3}{2}}\rangle = \ket{ddd} \,;
\label{DelatF}
\end{align}
for the mixed-symmetric one, $\ket{\phi_{MS}}$:
\begin{align}
p:\quad & |\phi_{\frac{1}{2}+\frac{1}{2}}\rangle = +\frac{1}{\sqrt{6}}\big(\ket{udu}+\ket{duu}-2\ket{uud}\big) \,, \\
n:\quad & |\phi_{\frac{1}{2}-\frac{1}{2}}\rangle =-\frac{1}{\sqrt{6}}\big(\ket{udd}+\ket{dud}-2\ket{ddu}\big) \,;
\end{align}
and for the mixed-antisymmetric case, $\ket{\phi_{MA}}$:
\begin{align}
p:\quad & |\phi_{\frac{1}{2}+\frac{1}{2}}\rangle^\prime = \frac{1}{\sqrt{2}}\big(\ket{udu}-\ket{duu}\big) \,, \\
n:\quad & |\phi_{\frac{1}{2}-\frac{1}{2}}\rangle^\prime = \frac{1}{\sqrt{2}}\big(\ket{udd}-\ket{dud}\big) \,.
\end{align}
The quadruplets are assigned to the isospin-$\frac{3}{2}$ resonance and the doublets to the isospin-$\frac{1}{2}$ nucleon. As we shall see later, both doublets of spin and isospin are needed to describe the nucleon.

\begin{figure}[!t]
\includegraphics[width=0.48\textwidth]{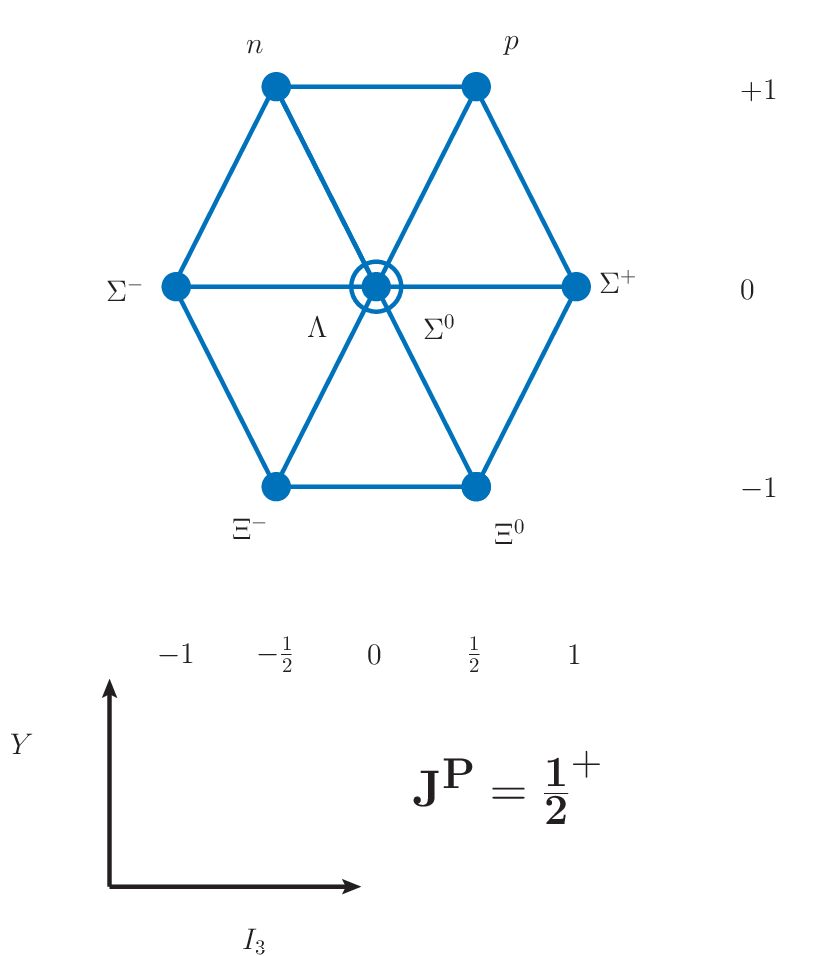}
\includegraphics[width=0.48\textwidth]{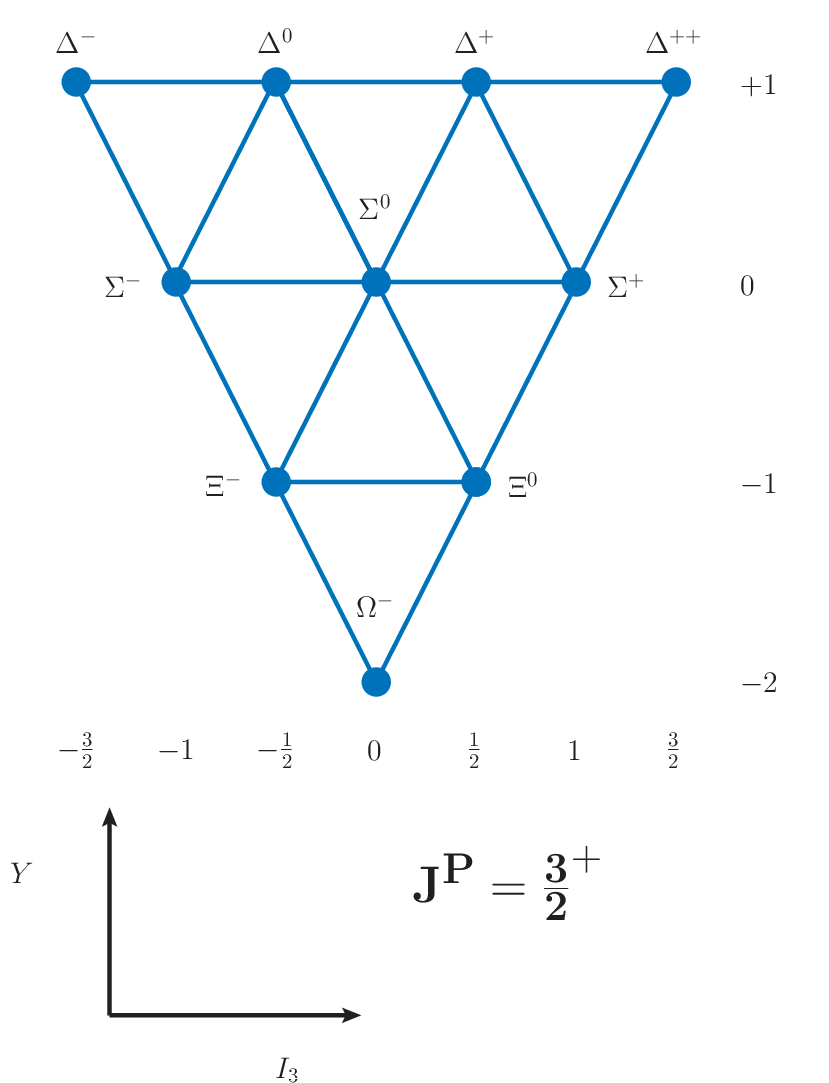}
\caption{\label{fig:OctetDecupletBaryons} Weight diagrams of the octet and decuplet ground state baryons. The arrows refer to the electric charges}
\end{figure}

In order to construct the SU(3) flavor wave functions of the three light quarks, one needs to start from the flavor wave functions above and derive the octet and decuplet cases by using the ladder operators $I_\pm$, $U_\pm$ and $V_{\pm}$ as shown in Fig.~\ref{fig:OctetDecupletBaryons}. For example, the flavor wave function of the $\Sigma^+$ baryons is
\begin{equation}
\ket{\Sigma^+} = U_- \Big[\frac{1}{\sqrt{6}}\big(\ket{udu}+\ket{duu}-2\ket{uud}\big)\Big] = \frac{1}{\sqrt{6}} \big(\ket{usu}+\ket{suu}-2\ket{uus}\big) \,,
\end{equation}
for the mixed-symmetric (MS) case and
\begin{equation}
\ket{\Sigma^+} = U_- \Big[\frac{1}{\sqrt{2}}\big(\ket{udu}-\ket{duu}\big)\Big] = \frac{1}{\sqrt{2}} \big(\ket{usu}-\ket{suu}\big) \,,
\end{equation}
for the mixed-antisymmetric (MA) one. Table~\ref{tab:SU3-12Baryons} summarizes the SU$_f$(3) wave functions of the spin-$\frac{1}{2}$ ground state baryons made of $u$, $d$ and $s$ quarks. On the other hand, Table~\ref{tab:SU3-32Baryons} summarizes the SU$_f$(3) wave functions of the spin-$\frac{3}{2}$ ground state baryons made of $u$, $d$ and $s$ quarks. 

\begin{table}
\caption{\label{tab:SU3-12Baryons} The SU(3)$_f$ wave functions of the spin-$\frac{1}{2}$ ground state baryons made of $u$, $d$ and $s$ quarks.}
%\begin{ruledtabular}
\begin{tabular}{lll}
Baryons & $\phi_{MS}$ & $\phi_{MA}$ \\
\hline
$p$ & $\frac{1}{\sqrt{6}}\ket{(ud+du)u-2uud}$ & $\frac{1}{\sqrt{2}}\ket{(ud-du)u}$ \\ 
$n$ & $-\frac{1}{\sqrt{6}}\ket{(ud+du)d-2ddu}$ & $\frac{1}{\sqrt{2}}\ket{(ud-du)d}$ \\
$\Sigma^+$ & $\frac{1}{\sqrt{6}}\ket{(us+su)u-2uus}$ & $\frac{1}{\sqrt{2}}\ket{(us-su)u}$ \\
$\Sigma^0$ & $\frac{1}{2\sqrt{3}}\ket{(sd+ds)u+(su+us)d-2(du+ud)s}$ & $\frac{1}{2}\ket{(ds-sd)u+(us-su)d}$ \\
$\Sigma^-$ & $\frac{1}{\sqrt{6}}\ket{(ds+sd)d-2dds}$ & $\frac{1}{\sqrt{2}}\ket{(ds-sd)d´}$ \\
$\Lambda^0$ & $\frac{1}{2}\ket{(sd+ds)u-(su+us)d}$ & $\frac{1}{2\sqrt{3}}\ket{(sd-ds)u+(us-su)d+2(ud-du)s}$ \\
$\Xi^-$ & $-\frac{1}{\sqrt{6}}\ket{(ds+sd)s-2ssd}$ & $\frac{1}{\sqrt{2}}(ds-sd)s$ \\
$\Xi^0$ & $-\frac{1}{\sqrt{6}}\ket{(us+su)s-2ssu}$ & $\frac{1}{\sqrt{2}}(us-su)s$ \\
\end{tabular}
%\end{ruledtabular}
\end{table}

\begin{table}
\caption{\label{tab:SU3-32Baryons} The SU(3)$_f$ wave functions of the spin-$\frac{3}{2}$ ground state baryons made of $u$, $d$ and $s$ quarks.}
%\begin{ruledtabular}
\begin{tabular}{ll}
Baryons & $\phi_{S}$ \\
\hline
$\Delta^{++}$ & $\ket{uuu}$ \\
$\Delta^{+}$ & $\frac{1}{\sqrt{3}}\ket{uud+udu+duu}$ \\
$\Delta^{0}$ & $\frac{1}{\sqrt{3}}\ket{udd+dud+ddu}$ \\
$\Delta^{-}$ & $\ket{ddd}$ \\
$\Sigma^+$ & $\frac{1}{\sqrt{3}}\ket{uus+usu+suu}$ \\
$\Sigma^0$ & $\frac{1}{\sqrt{3}}\ket{uds+usd+sdu}$\\
$\Sigma^-$ & $\frac{1}{\sqrt{3}}\ket{dds+dsd+sdd}$\\
$\Xi^0$ & $\frac{1}{\sqrt{3}}\ket{uss+sus+ssu}$ \\
$\Xi^-$ & $\frac{1}{\sqrt{3}}\ket{dss+sds+ssd}$ \\
$\Omega^-$ & $\ket{sss}$ \\
\end{tabular}
%\end{ruledtabular}
\end{table}

In the quark model the total wavefunction of a baryon should be symmetric
under the permutation of any pair of quarks. Colour is then added to make the
wavefunction antisymmetric, as required for fermions. A fully symmetric wavefunction for spin-$\frac{1}{2}$ baryons is constructed by combining the spin states $\chi_{MS}$ and $\chi_{MA}$ with the flavor states $\phi_{MS}$ and $\phi_{MA}$:
\begin{equation}
|\frac{1}{2}^+\rangle = \frac{1}{\sqrt{2}} \big( |\chi_{MS}\phi_{MS}\rangle + |\chi_{MA}\phi_{MA}\rangle \big) \,.
\end{equation}
The total wavefunction for spin-$\frac{3}{2}$ baryons must be also fully symmetric and thus
\begin{equation}
|\frac{3}{2}^+\rangle = |\chi_S\phi_S\rangle \,.
\end{equation}

The last SU$_f$(3) orthogonal state that remains to be discussed is the singlet case, whose totally antisymmetric wave function is given by:
\begin{equation}
|\phi_A\rangle = \frac{1}{\sqrt{6}} |s(du-ud)+(usd-dsu)+(du-ud)s\rangle \,.
\end{equation}
As we have seen, there is no fully antisymmetric spin wave function for a baryon and thus $\phi_A$ cannot be combined with any spin state to generate a totally symmetric wave function for ground state baryons. Therefore, $\phi_A$ is not realized for ground state baryons.

\section{Magnetic moments of light baryons: quark effective mass}

The magnetic moment of a particle with electric charge $e$ and mass $m$ is equal to its $g$-factor multiplied by $(e/2m)\vec{s}$. The $g$-factor is known as the gyromagnetic factor which match the value $2$ for a spin-1/2 fundamental particle as predicted by the Dirac equation. 

Within the quark model, the magnetic dipole moment of a baryon is obtained by computing the expectation value of the operator $\vec{\mu}_i=g(q_i e/2m_i)\vec{s}$ on quark $i$ with charge $q_i e$, constituent mass $m_i$, g=2 for point-like fermions, and by adding the contributions of the three quarks. We assume that the constituent masses of the $u$ and $d$ quarks are equal, hence that $m_u = m_d \equiv m$, while the $s$ quark has the mass $m_s$. The magnetic moments of the quarks are then given by
\begin{equation}
\mu_u = +\frac{2}{3} \left(\frac{e}{2m}\right) ,\quad
\mu_d = -\frac{1}{3} \left(\frac{e}{2m}\right) ,\quad
\mu_s = -\frac{1}{3} \left(\frac{e}{2m_s}\right) ,\quad
\end{equation}
The magnetic dipole moment of the proton, obtained by adding the quark contributions, is
\begin{equation}
\mu_p = \sum_{i=1}^3 \langle p\uparrow | 2\mu_is_{zi} | p\uparrow \rangle = \sum_{i=1}^3 \langle p\uparrow | \mu_i \sigma_{zi} | p\uparrow \rangle
\end{equation}
where $|p\uparrow\rangle$ is the wave function for a proton with spin along the $z$-axis and $\vec{s}\equiv\frac{1}{2}\vec{\sigma}$. According to the accepted convention, the magnetic dipole operator is always evaluated between states with $m_S=+S$. Therewith, the proton wave function is
\begin{align}
| p \uparrow \rangle &= \frac{1}{\sqrt{2}} \frac{1}{\sqrt{6}} \frac{1}{\sqrt{6}} | u d u + d u u - 2 u u d \rangle | \uparrow \downarrow \uparrow + \downarrow \uparrow \uparrow - 2 \uparrow \uparrow \downarrow \rangle + \frac{1}{\sqrt{2}} \frac{1}{\sqrt{2}} \frac{1}{\sqrt{2}} | u d u - d u u \rangle | \uparrow \downarrow \uparrow - \downarrow \uparrow \uparrow \rangle \nonumber \\
&
= \frac{1}{6} \frac{1}{\sqrt{2}} \Big[ | u d u + d u u - 2 u u d \rangle | \uparrow \downarrow \uparrow + \downarrow \uparrow \uparrow - 2 \uparrow \uparrow \downarrow \rangle 
+ | 3 u d u - 3 d u u \rangle | \uparrow \downarrow \uparrow - \downarrow \uparrow \uparrow \rangle \Big].
\end{align}
Collecting the $udu$, $duu$ and $uud$ terms, equation above gives
\begin{align}
\label{eq:dipole1}
| p \uparrow \rangle = \frac{1}{6} \frac{1}{\sqrt{2}} \Big[ &+ 4 | u d u \uparrow \downarrow \uparrow \rangle - 2 | u d u \downarrow \uparrow \uparrow \rangle - 2 | u d u \uparrow \uparrow \downarrow \rangle \nonumber \\
& 
- 2 | d u u \uparrow \downarrow \uparrow \rangle + 4 | d u u \downarrow \uparrow \uparrow \rangle - 2 | d u u \uparrow \uparrow \downarrow \rangle \nonumber \\
&
- 2 | u u d \uparrow \downarrow \uparrow \rangle - 2 | u u d \downarrow \uparrow \uparrow \rangle + 4 | u u d \uparrow \uparrow \downarrow \rangle \Big].
\end{align}
The expectation values for the dipole magnetic moment of each term in Eq.~\eqref{eq:dipole1} are
\begin{align}
&
\langle u d u \uparrow \downarrow \uparrow | \sum_i \mu_i\sigma_{zi} | u d u \uparrow \downarrow \uparrow \rangle = \mu_u - \mu_d + \mu_u = 2\mu_u - \mu_d \,, \nonumber \\
&
\langle u d u \downarrow \uparrow \uparrow | \sum_i \mu_i\sigma_{zi} | u d u \downarrow \uparrow \uparrow \rangle = - \mu_u + \mu_d + \mu_u = \mu_d \,, \nonumber \\
&
\langle u d u \uparrow \uparrow \downarrow | \sum_i \mu_i\sigma_{zi} | u d u \uparrow \uparrow \downarrow \rangle = + \mu_u + \mu_d - \mu_u = \mu_d \,, \nonumber \\
&
\langle d u u \uparrow \downarrow \uparrow | \sum_i \mu_i\sigma_{zi} | d u u \uparrow \downarrow \uparrow \rangle = \mu_d - \mu_u + \mu_u = \mu_d \,, \nonumber \\
&
\langle d u u \downarrow \uparrow \uparrow | \sum_i \mu_i\sigma_{zi} | d u u \downarrow \uparrow \uparrow \rangle = - \mu_d + \mu_u + \mu_u = 2\mu_u - \mu_d \,, \nonumber \\
&
\langle d u u \uparrow \uparrow \downarrow | \sum_i \mu_i\sigma_{zi} | d u u \uparrow \uparrow \downarrow \rangle = + \mu_d + \mu_u - \mu_u = \mu_d \,, \nonumber \\
&
\langle u u d \uparrow \downarrow \uparrow | \sum_i \mu_i\sigma_{zi} | u u d \uparrow \downarrow \uparrow \rangle = \mu_u - \mu_u + \mu_d = \mu_d \,, \nonumber \\
&
\langle u u d \downarrow \uparrow \uparrow | \sum_i \mu_i\sigma_{zi} | u u d \downarrow \uparrow \uparrow \rangle = - \mu_u + \mu_u + \mu_d = \mu_d \,, \nonumber \\
&
\langle u u d \uparrow \uparrow \downarrow | \sum_i \mu_i\sigma_{zi} | u u d \uparrow \uparrow \downarrow \rangle = + \mu_u + \mu_u - \mu_d = 2\mu_u - \mu_d \,,
\end{align}
in such a way that
\begin{align}
\label{eq:muProton}
\mu_p &= \left( \frac{1}{6} \frac{1}{\sqrt{2}} \right)^2 \big[ 3 \cdot 16\cdot (2\mu_u - \mu_d) + 6 \cdot 4 \cdot \mu_d \big] \nonumber \\
&
= \left( \frac{1}{6} \frac{1}{\sqrt{2}} \right)^2 (96\mu_u-24\mu_d) \nonumber \\
&
= \frac{4}{3} \mu_u - \frac{1}{3} \mu_d \nonumber \\
&
= \frac{e}{2m} \,.
\end{align}

The magnetic dipole moment of the neutron can be calculated similarly to that of the proton. The wave function of the neutron is:
\begin{align}
| n \uparrow \rangle &= - \frac{1}{\sqrt{2}} \frac{1}{\sqrt{6}} \frac{1}{\sqrt{6}} | u d d + d u d - 2 d d u \rangle | \uparrow \downarrow \uparrow + \downarrow \uparrow \uparrow - 2 \uparrow \uparrow \downarrow \rangle + \frac{1}{\sqrt{2}} \frac{1}{\sqrt{2}} \frac{1}{\sqrt{2}} | u d d - d u d \rangle | \uparrow \downarrow \uparrow - \downarrow \uparrow \uparrow \rangle \nonumber \\
&
= \frac{1}{6} \frac{1}{\sqrt{2}} \Big[ - | u d d + d u d - 2 d d u \rangle | \uparrow \downarrow \uparrow + \downarrow \uparrow \uparrow - 2 \uparrow \uparrow \downarrow \rangle + | 3 u d d - 3 d u d \rangle | \uparrow \downarrow \uparrow - \downarrow \uparrow \uparrow \rangle \Big] \,.
\end{align}
Again, collecting terms involving the same quark ordering, one arrives at
\begin{align}
| n \uparrow \rangle = \frac{1}{6} \frac{1}{\sqrt{2}} \Big[ &+2 | u d d \uparrow \downarrow \uparrow \rangle - 4 | u d d \downarrow \uparrow \uparrow \rangle + 2 | u d d \uparrow \uparrow \downarrow \rangle \nonumber \\
&
- 4 | d u d \uparrow \downarrow \uparrow \rangle + 2 | d u d \downarrow \uparrow \uparrow \rangle + 2 | d u d \uparrow \uparrow \downarrow \rangle \nonumber \\ 
&
+ 2 | d d u \uparrow \downarrow \uparrow \rangle + 2 | d d u \downarrow \uparrow \uparrow \rangle - 4 | d d u \uparrow \uparrow \downarrow \rangle \Big] \,.
\end{align}
The expectation values of $\sum \mu_i \sigma_{z_i}$ are obtained in a manner similar to the proton in such a way that the neutron magnetic moment is given by
\begin{align}
\mu_n &= \left( \frac{1}{6} \frac{1}{\sqrt{2}} \right)^2 \big[ 6 \cdot 4 \cdot \mu_u + 3 \cdot 16 \cdot (2\mu_d-\mu_u) \big] \nonumber \\
&
= \left( \frac{1}{6} \frac{1}{\sqrt{2}} \right)^2 (96 \mu_d - 24 \mu_u) \nonumber \\
&
= \frac{4}{3} \mu_d - \frac{1}{3} \mu_u \nonumber \\
&
= -\frac{2}{3} \left(\frac{e}{2m}\right) \,.
\end{align}

Hence, within the quark model, one arrives at the prediction for the ratio of the magnetic dipole moments:
\begin{equation}
\frac{\mu_n}{\mu_p} = \frac{g_n}{g_p} = -\frac{2}{3}
\end{equation}
which compares astonishingly well with the experimental value for the ratio $g_n/g_p$, $-0.68497934 \pm  0.00000016$~\cite{ParticleDataGroup:2024cfk}. Besides, the magnetic moment of a baryon $B$ is conventionally written relative to that of the proton and thus expressed as
\begin{equation}
\label{eq:muBaryon}
\vec{\mu}_B = g_B \mu_N \vec{s} \,,
\end{equation}
where
\begin{equation}
\mu_N = \frac{e}{2m_p} 
\end{equation}
is the nuclear magneton. Using Eqs.~\eqref{eq:muProton} and~\eqref{eq:muBaryon}, the mass $m$ of the $u$ and $d$ quarks can be estimated to be
\begin{equation}
\label{eq:Mu}
\mu_p = \frac{e}{2m} = \frac{g_p}{2} \mu_N = 2.79 \left(\frac{e}{2m_p}\right)  \quad\Rightarrow\quad m = \frac{m_p}{2.79} = 336\,\text{MeV} \,.
\end{equation}

The magnetic moment of the lightest hyperon $\Lambda^0$ is easy to predict. The isospin-$\frac{1}{2}$ $u$ and $d$ quarks couple to $i=0$, since $i(s)=0$ and $i(\Lambda)=0$, and thus they are in an antisymmetric state. Hence, their spins also couple to zero and the magnetic moment of the $\Lambda$ stems from the $s$ quark only, expecting that
\begin{equation}
\mu_{\Lambda} = \mu_s = - \frac{1}{3} \left( \frac{e}{2m_s} \right) \,.
\end{equation}
This equation, together with Eq.~\eqref{eq:muProton}, can be used to estimate the mass of the $s$ quark:
\begin{equation}
\label{eq:Ms}
\mu_\Lambda = - \frac{1}{3} \left( \frac{e}{2m_s} \right) = -0.22\mu_p \quad\Rightarrow\quad m_s = -\frac{1}{0.22} \left(\frac{2m}{e}\right) \left(-\frac{e}{6}\right) = 1.51m = 507\,\text{MeV} \,.
\end{equation}

As one can see in Eqs.~\eqref{eq:Mu} and~\eqref{eq:Ms}, the masses of the quarks calculated within the quark model picture differ considerably from the current (bare) quark masses that appear in the QCD Lagrangian:
\begin{align}
&
\bar{m}_u \approx 2.2\,\text{MeV}, \quad\quad
\bar{m}_d \approx 4.7\,\text{MeV}, \quad\quad
\bar{m}_s \approx 100\,\text{MeV}, \nonumber \\[1ex]
&
\bar{m}_c \approx 1280\,\text{MeV}, \quad\quad
\bar{m}_b \approx 4180\,\text{MeV}, \quad\quad
\bar{m}_t \approx 173\,\text{GeV}\,. \nonumber 
\end{align}
This is because bare quark masses are provided by the so-called Higgs mechanism but most of the mass of the nucleon, and generally of any hadron, is not due to the Higgs boson but to the strong interaction dynamics carried by gluons. As it shall be shown later, the interaction between quarks through gluons produces a rich plethora of low-energy (nonperturbative) emergent phenomena such as the spontaneous breaking of chiral symmetry which provides a natural explanation for the constituent masses of dynamically-dressed quarks:
\begin{align}
&
m_u \approx 350\,\text{MeV}, \quad\quad
m_d \approx 350\,\text{MeV}, \quad\quad
m_s \approx 500\,\text{MeV}, \nonumber \\[1ex]
&
m_c \approx 1500\,\text{MeV}, \quad\quad
m_b \approx 4700\,\text{MeV}, \quad\quad
m_t \approx 173\,\text{GeV}\,, \nonumber 
\end{align}
which roughly correspond to the ones derived above from the baryon magnetic moments using the quark model framework.

\section{The quark--(anti-)quark interaction mediated by gluons}
\label{sec4}

The short-range (high-energy) quark--(anti-)quark interaction can be calculated via perturbative QCD. Therefore, we explicitly calculate one-gluon exchange
potentials and discuss some basic problems in the derivation.

\begin{figure}[!t]
\begin{center}
\includegraphics[trim = 0mm 0mm 0mm 0mm, clip, height=0.25\textheight, width=0.60\textwidth]{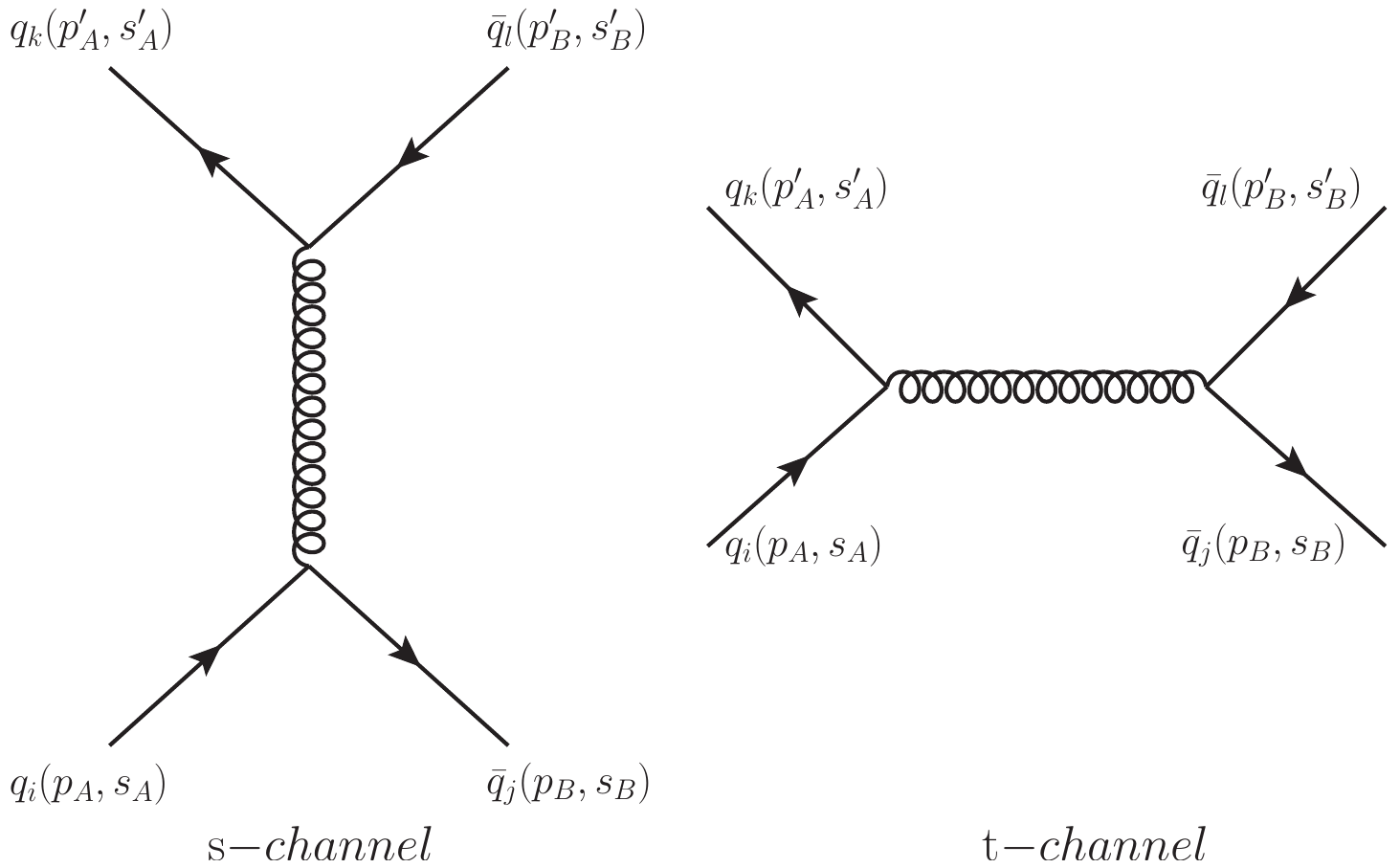}
\caption{\label{fig:OGEdiagrams} One-gluon exchange diagrams for
quark-antiquark. The $s$-channel ($t$-channel) is depicted in the left (right)
panel.}
\end{center}
\end{figure}

The so-called Breit interaction was firstly deduced by Gregory Breit~\cite{Breit:1929zz, Breit:1930zza} for the electron-electron scattering process. The resulting potential included, apart from a leading Coulomb term, relativistic corrections which had their origin in the one-photon exchange process and an expansion of the propagator, leading to retardation. Likewise, we shall re-derive the Breit interaction for the quark-antiquark scattering process
\begin{equation}
q_{i}(p_{A},s_{A}) + \bar{q}_{j}(p_{B},s_{B}) \to q_{k}(p'_{A},s'_{A}) +
\bar{q}_{l}(p'_{B},s'_{B}),
\label{eq:scattering}
\end{equation}
with four-momenta $p_{X}$, $p'_{X}$ and spins $s_{X}$, $s'_{X}$ as indicated,
and $i,\,j,\,k,\,l=1,\,2,\,3$ being color indices. In this case the Breit
interaction is equivalent to the potential arising from one-gluon exchange,
including retardation corrections. The scattering process
Eq.~(\ref{eq:scattering}) is described, at lowest order perturbative QCD, by
two Feynman diagrams, an $s$-channel diagram (left panel of
Fig.~\ref{fig:OGEdiagrams}) and a $t$-channel diagram (right panel of Fig.~\ref{fig:OGEdiagrams}). 

We are going to derive, as an example, the potential for quarkonium (mesons with equal flavor quark-antiquark content). Mesons are color neutral objects and so we posses already some knowledge about the color wave function
\begin{equation}
\left|{\rm Meson}\right\rangle_{\text{color}} \sim \frac{1}{\sqrt{3}} \sum_{i=1}^{3} \left|
\bar{q}_{i}q_{i} \right\rangle,
\end{equation}
resulting for Eq.~(\ref{eq:scattering}) in the conditions:
\begin{equation}
\frac{1}{\sqrt{3}} \delta_{ij} \quad \mbox{ and } \quad \frac{1}{\sqrt{3}} \delta_{kl} \,.
\label{eq:colorwave}
\end{equation}
Taking this condition into account the $s$-channel diagram depicted in the left panel of Fig.~\ref{fig:OGEdiagrams} does not contribute since its invariant amplitude in Feynman gauge reads as (Feynman rules are used from~\cite{Peskin:1995ev}):
\begin{equation}
\begin{split}
{\cal M}_{fi} &= \left[ \bar{u}(p'_{A},s'_{A}) \left( ig\gamma^{\mu}\frac{\lambda_{kl}^{a}}{2} \right) v(p'_{B},s'_{B}) \right] \frac{-g_{\mu\nu}}{(p_{A}+p_{B})^{2}+i\epsilon} \left[ \bar{v}(p_{B},s_{B}) \left(ig\gamma^{\nu}\frac{\lambda_{ij}^{a}}{2} \right) u(p_{A},s_{A}) \right] \\
&
= \frac{g^{2}}{(p_{A}+p_{B})^{2}+i\epsilon} \frac{\lambda_{kl}^{a}}{2}\frac{\lambda_{ij}^{a}}{2}  \left[ \bar{u}(p'_{A},s'_{A}) \gamma_{\mu} v(p'_{B},s'_{B}) \right] \left[ \bar{v}(p_{B},s_{B}) \gamma^{\mu} u(p_{A},s_{A}) \right] \,,
\end{split}
\end{equation}
that, using Eq.~(\ref{eq:colorwave}), vanishes as
\begin{equation}
\frac{1}{\sqrt{3}}\delta_{ij}\frac{\lambda_{ij}^{a}}{2} = \frac{1}{2\sqrt{3}}
{\rm Tr}(\lambda^{a}) = 0 \,,
\end{equation}
or, in other words, the gluon as color octet does not couple to a color-singlet. This leaves the $t$-channel diagram as the only one to deal with. Employing Feynman rules, the $t$-channel invariant amplitude in Feynman gauge is given by
\begin{equation}
\begin{split}
{\cal M}_{fi} &= \left[ \bar{u}(p'_{A},s'_{A}) \left(ig\gamma^{\mu}\frac{\lambda_{ki}^{a}}{2} \right) u(p_{A},s_{A}) \right] \frac{g_{\mu\nu}}{q^{2}+i\epsilon} \left[ \bar{v}(p_{B},s_{B}) \left(ig\gamma^{\nu}\frac{\lambda_{jl}^{a}}{2} \right) v(p'_{B},s'_{B}) \right] \\
&
= - \frac{g^{2}}{q^{2}+i\epsilon} \frac{\lambda_{ki}^{a}}{2} \frac{\lambda_{jl}^{a}}{2} \left[ \bar{u}(p'_{A},s'_{A}) \gamma_{\mu} u(p_{A},s_{A})\right] \left[ \bar{v}(p_{B},s_{B}) \gamma^{\mu} v(p'_{B},s'_{B}) \right]
\end{split}
\end{equation}
with $q^{2}=(p_{A}-p'_{A})^{2}=(p'_{B}-p_{B})^{2}$ and
\begin{equation}
\frac{1}{\sqrt{3}} \sum_{i,j=1}^{3} \delta_{ij} \frac{1}{\sqrt{3}} \sum_{k,l=1}^{3} \delta_{kl} \sum_{a=1}^{8} \frac{\lambda_{ki}^{a}}{2} \frac{\lambda_{jl}^{a}}{2} = \frac{1}{12} \sum_{i,k=1}^{3} \sum_{a=1}^{8} \lambda_{ki}^{a} \lambda_{ik}^{a} = \frac{1}{12} \sum_{a=1}^{8} {\rm Tr} [(\lambda^{a})^{2}] = \frac{4}{3}.
\end{equation}
Therefore, for color neutral initial and final meson states the process described in Eq.~(\ref{eq:scattering}) leads to the amplitude
\begin{equation}
{\cal M}_{fi} = -\frac{4}{3} \frac{g^{2}}{q^{2}+i\epsilon} \left[ \bar{u}(p'_{A},s'_{A}) \gamma_{\mu} u(p_{A},s_{A})\right] \left[ \bar{v}(p_{B},s_{B}) \gamma^{\mu} v(p'_{B},s'_{B}) \right].
\label{eq:amplitude}
\end{equation}
 
In order to compute the amplitude in Eq.~(\ref{eq:amplitude}), we evaluate the Dirac currents $\left[\bar{u}(p'_{A},s'_{A}) \gamma_{\mu} u(p_{A},s_{A})\right]$ and $\left[ \bar{v}(p_{B},s_{B}) \gamma^{\mu} v(p'_{B},s'_{B}) \right]$. After applying the normalization conventions for Dirac spinors:
\begin{align}
& 
u(p_{A},s_{A})=\sqrt{\frac{E_{p_{A}}+m_{A}}{2m_{A}}} \left(\begin{matrix}\chi_{s_{A}} \\ \frac{\vec{\sigma}\cdot\vec{p}_{A}}{E_{A}+m_{A}}\chi_{s_{A}}\end{matrix}\right) , \\
&
\bar{u}(p'_{A},s'_{A})=\sqrt{\frac{E_{p'_{A}}+m_{A}}{2m_{A}}} \left(\begin{matrix}\chi^{\dagger}_{s'_{A}} & ,-\chi^{\dagger}_{s'_{A}}\frac{\vec{\sigma}\cdot \vec{p\,}'_{A}}{E'_{A}+m_{A}}\end{matrix} \right), \\
&
v(p'_{B},s'_{B})=\sqrt{\frac{E_{p'_{B}}+m_{B}}{2m_{B}}} \left(\begin{matrix}\frac{\vec{\sigma}\cdot\vec{p\,}'_{B}}{E'_{B}+m_{B}} \chi^{c}_{s'_{B}} \\ \chi^{c}_{s'_{B}} \end{matrix}\right), \\
&
\bar{v}(p_{B},s_{B})=\sqrt{\frac{E_{p_{B}}+m_{B}}{2m_{B}}} \left(\begin{matrix} \chi^{c\dagger}_{s_{B}} \frac{\vec{\sigma}\cdot\vec{p}_{B}}{E_{B}+m_{B}} & ,-\chi^{c\dagger}_{s_{B}}\end{matrix}\right),
\end{align}
with energies $E_{p_{X}}=\sqrt{\vec{p\,}_{X}^{2}+m_{X}^{2}}$ and quark masses
$m_{X}$; and using the auxiliary calculations
\begin{align}
&
\left(\begin{matrix}\chi^{\dagger}_{s'_{A}} & ,-\chi^{\dagger}_{s'_{A}}\frac{\vec{\sigma}\cdot \vec{p\,}'_{A}}{E_{p'_{A}}+m_{A}}\end{matrix} \right) \left(\begin{matrix} 0 & \sigma_{i} \\ -\sigma_{i} & 0 \end{matrix}\right) \left(\begin{matrix}\chi_{s_{A}} \\ \frac{\vec{\sigma}\cdot\vec{p}_{A}}{E_{p_{A}}+m_{A}}\chi_{s_{A}}\end{matrix} \right) = \chi^{\dagger}_{s'_{A}} \left[ \frac{\vec{\sigma}\cdot \vec{p\,}'_{A}}{E_{p'_{A}}+m_{A}}\sigma_{i} + \sigma_{i} \frac{\vec{\sigma}\cdot\vec{p}_{A}}{E_{p_{A}}+m_{A}} \right] \chi_{s_{A}} \,, \\
&
\left(\begin{matrix} \chi^{c\dagger}_{s_{B}} \frac{\vec{\sigma}\cdot\vec{p}_{B}}{E_{p_{B}}+m_{B}} & ,-\chi^{c\dagger}_{s_{B}} \end{matrix}\right) \left(\begin{matrix} 0 & \sigma_{i} \\ -\sigma_{i} & 0 \end{matrix}\right) \left(\begin{matrix}\frac{\vec{\sigma}\cdot\vec{p\,}'_{B}}{E_{p'_{B}}+m_{B}} \chi^{c}_{s'_{B}} \\ \chi^{c}_{s'_{B}} \end{matrix}\right)
=\chi^{c\dagger}_{s_{B}} \left[  \sigma_{i}\frac{\vec{\sigma}\cdot\vec{p\,}'_{B}}{E_{p'_{B}}+m_{B}} + \frac{\vec{\sigma}\cdot\vec{p}_{B}}{E_{p_{B}}+m_{B}} \sigma_{i} \right] \chi^{c}_{s'_{B}} \,,
\end{align}
and
\begin{align}
(\vec{\sigma}\cdot\vec{\epsilon})(\vec{\sigma}\cdot\vec{x}) + (\vec{\sigma}\cdot\vec{y})(\vec{\sigma}\cdot\vec{\epsilon}) &= \vec{\epsilon}\cdot\vec{x} + i \vec{\sigma} \cdot (\vec{\epsilon}\times\vec{x}) + \vec{y}\cdot\vec{\epsilon} + i \vec{\sigma} \cdot (\vec{y}\times\vec{\epsilon}) = \vec{\epsilon}\cdot(\vec{y}+\vec{x}) + i\vec{\epsilon}\cdot\vec{\sigma} \times (\vec{y}-\vec{x}) \,,
\end{align}
one can easily verify for the Dirac currents
\begin{align}
&
\bar{u}(p'_{A},s'_{A})\gamma^{\mu} u(p_{A},s_{A}) = \left(\begin{matrix} u^{\dagger}(p'_{A},s'_{A})u(p_{A},s_{A}) \\ u^{\dagger}(p'_{A},s'_{A})\gamma_{0}\vec{\gamma}u(p_{A},s_{A}) \end{matrix}\right) \nonumber \\
&
= \sqrt{\frac{E_{p'_{A}}+m_{A}}{2m_{A}}} \sqrt{\frac{E_{p_{A}}+m_{A}}{2m_{A}}} \left(\begin{matrix} \chi^{\dagger}_{s'_{A}} \left[ 1 + \frac{(\vec{\sigma}\cdot \vec{p\,}'_{A})(\vec{\sigma}\cdot\vec{p}_{A})}{(E_{p'_{A}}+m_{A})(E_{p_{A}}+m_{A}) } \right] \chi_{s_{A}} \\ \chi^{\dagger}_{s'_{A}} \left[ \left( \frac{\vec{p}_{A}}{E_{p_{A}}+m_{A}} + \frac{\vec{p\,}'_{A}}{E_{p'_{A}}+m_{A}} \right) + i\vec{\sigma} \times \left( \frac{\vec{p\,}'_{A}}{E_{p'_{A}}+m_{A}}-\frac{\vec{p}_{A}}{E_{p_{A}}+m_{A}}\right) \right] \chi_{s_{A}} \end{matrix} \right),
\label{eq:currentsA}
\end{align}
and
\begin{align}
&
\bar{v}(p_{B},s_{B}) \gamma^{\mu} v(p'_{B},s'_{B}) = \left(\begin{matrix} v^{\dagger}(p_{B},s_{B})v(p'_{B},s'_{B}) \\ v^{\dagger}(p_{B},s_{B})\gamma_{0}\gamma_{\mu} v(p'_{B},s'_{B}) \end{matrix}\right) \nonumber \\
&
= \sqrt{\frac{E_{p'_{B}}+m_{B}}{2m_{B}}} \sqrt{\frac{E_{p_{B}}+m_{B}}{2m_{B}}} \left(\begin{matrix} \chi^{c\dagger}_{s_{B}} \left[ 1 + \frac{(\vec{\sigma}\cdot \vec{p}_{B})(\vec{\sigma}\cdot\vec{p\,}'_{B})}{(E_{p_{B}}+m_{B})(E_{p'_{B}}+m_{B}) } \right] \chi^{c}_{s'_{B}} \\ \chi^{c\dagger}_{s_{B}} \left[  \left( \frac{\vec{p\,}'_{B}}{E_{p'_{B}}+m_{B}} + \frac{\vec{p}_{B}}{E_{p_{B}}+m_{B}} \right) +i \vec{\sigma} \times \left( \frac{\vec{p}_{B}}{E_{p_{B}}+m_{B}} - \frac{\vec{p\,}'_{B}}{E_{p'_{B}}+m_{B}} \right) \right] \chi^{c}_{s'_{B}} \end{matrix} \right).
\label{eq:currentsB}
\end{align}
where the first row gives the time component and the second the spatial components.

For further calculation, we substitute
\begin{align}
\label{eq:KinematicChoice}
&
\vec{p}_{A} = \vec{p}_{1} + \frac{1}{2}\vec{q}, \quad
\vec{p\,}'_{A} = \vec{p}_{1} - \frac{1}{2}\vec{q}, \quad
\vec{s}_{A} = s_{1}, \quad m_{A} = m_{1}, \\
&
\vec{p}_{B} = \vec{p}_{2} - \frac{1}{2}\vec{q}, \quad
\vec{p\,}'_{B} = \vec{p}_{2} + \frac{1}{2}\vec{q}, \quad
\vec{s}_{B} = s_{2}, \quad m_{B} = m_{2} .
\end{align}
where $\vec q$ is the three momentum transfered. This choice of kinematics is crucial because, although the amplitude remains time reversal invariant when expressed in Lorentz covariants, its nonrelativistic expansion and subsequently the potential are not. By using the new variables in Eq.~\eqref{eq:KinematicChoice}, the resulting potential becomes invariant under time reversal transformation. The time-like and space-like components of Eqs.~(\ref{eq:currentsA}) and~(\ref{eq:currentsB}) read now as
\begin{itemize}
\item The $\bar{u}(p'_{A},s'_{A}) \gamma^{0} u(p_{A},s_{A})$ term:
\begin{equation}
u^{\dagger}(p'_{A},s'_{A}) u(p_{A},s_{A}) = \sqrt{\frac{E_{p'_{A}}+m_{1}}{2m_{1}}} \sqrt{\frac{E_{p_{A}}+m_{1}}{2m_{1}}} \chi^{\dagger}_{s'_{1}} \left[ 1 + \frac{\vec{p\,}_{1}^{2} - \frac{1}{4}\vec{q\,}^{2} - i\vec{\sigma}\cdot (\vec{q}\times\vec{p}_{1})}{(E_{p'_{A}}+m_{1})(E_{p_{A}}+m_{1})} \right] \chi_{s_{1}} \,.
\end{equation}
\item The $\bar{v}(p_{B},s_{B}) \gamma^{0} v(p'_{B},s'_{B})$ term:
\begin{equation}
v^{\dagger}(p_{B},s_{B}) v(p'_{B},s'_{B}) = \sqrt{\frac{E_{p'_{B}}+m_{2}}{2m_{2}}} \sqrt{\frac{E_{p_{B}}+m_{2}}{2m_{2}}} \chi^{c\dagger}_{s_{2}} \left[ 1 + \frac{\vec{p\,}_{2}^{2} - \frac{1}{4}\vec{q\,}^{2} - i\vec{\sigma}\cdot (\vec{q}\times\vec{p}_{2})}{(E_{p'_{B}}+m_{2})(E_{p_{B}}+m_{2})} \right] \chi^{c}_{s'_{2}} \,.
\end{equation}
\item The $\bar{u}(p'_{A},s'_{A}) \vec{\gamma} u(p_{A},s_{A})$ term:
\begin{equation}
\bar{u}(p'_{A},s'_{A}) \vec{\gamma} u(p_{A},s_{A}) = \frac{1}{2m_{1}} \chi^{\dagger}_{s'_{1}} \left[2\vec{p}_{1} - i\vec{\sigma}\times\vec{q}\,\right] \chi_{s_{1}} + {\cal O}\left(\frac{1}{m_{1}^{2}}\right) \,.
\end{equation}
\item The $\bar{v}(p_{B},s_{B}) \vec{\gamma} v(p'_{B},s'_{B})$ term:
\begin{equation}
\bar{v}(p_{B},s_{B}) \vec{\gamma} v(p'_{B},s'_{B}) = \frac{1}{2m_{2}}
\chi^{c\dagger}_{s_{2}} \left[2\vec{p}_{2} -
i\vec{\sigma}\times\vec{q}\,\right] \chi^{c}_{s'_{2}} + {\cal
O}\left(\frac{1}{m_{2}^{2}}\right) \,.
\end{equation}
\end{itemize}
For the space-like part of the currents we have neglected corrections of
order ${\cal O}(m_{i}^{-2})$ because they provide ${\cal O}(m_{i}^{-2}m_{j}^{-1})$ and ${\cal O}(m_{i}^{-2}m_{j}^{-1})$ contributions in the amplitude and thus they are negligible as we expand the amplitude up to order ${\cal O}(m_{i}^{-1}m_{j}^{-1})$. We denote the spectation values of the spin operators as
\begin{align}
&
\vec{s}_{1}  \equiv \frac 1 2 \chi^{\dagger}_{s'_{1}}\vec{\sigma}\chi_{s_{1}} \,, \\
&
\vec{s}_{2} \equiv
\frac 1 2 \chi^{c\dagger}_{s_{2}}\vec{\sigma}\chi^{c}_{s'_{2}} = -\frac 1 2 (\chi_{s'_2}^{\dagger}\vec \sigma\chi_{s_{2}}) \,,
\end{align}
where $\vec{\sigma}$ are the Pauli matrices and we have dealt with charge conjugated Pauli spinors for antiparticles.

Setting the pieces together, we obtain the invariant amplitude
\begin{equation}
{\cal M}_{fi} = -\frac{4}{3} \frac{g^{2}}{q^{2}+i\epsilon} ({\cal A}+{\cal B}) \,,
\label{eq:amplitude1}
\end{equation}
where
\begin{align}
{\cal N} &= \sqrt{\frac{E_{p'_{A}}+m_{1}}{2m_{1}}} \sqrt{\frac{E_{p_{A}}+m_{1}}{2m_{1}}} \sqrt{\frac{E_{p'_{B}}+m_{2}}{2m_{2}}} \sqrt{\frac{E_{p_{B}}+m_{2}}{2m_{2}}} \nonumber \\
&
= 1 + \frac{\vec{p\,}_{1}^{2}+\frac{1}{4}\vec{q\,}^{2}}{4m_{1}^{2}} + \frac{\vec{p\,}_{2}^{2}+\frac{1}{4}\vec{q\,}^{2}}{4m_{2}^{2}} + {\cal O}\left(\frac{1}{m_{i}^{3}}\right) \,,
\label{eq:amplitude2}
\end{align}
appearing in the term ${\cal A}$,
\begin{align}
{\cal A} &= {\cal N} \chi^{\dagger}_{s'_{1}} \left[ 1 + \frac{\vec{p\,}_{1}^{2} - \frac{1}{4}\vec{q\,}^{2} - i\vec{\sigma}\cdot (\vec{q}\times\vec{p}_{1})}{(E_{p'_{A}}+m_{1})(E_{ p_{A}}+m_{1})} \right] \chi_{s_{1}} \chi^{c\dagger}_{s_{2}} \left[ 1 + \frac{\vec{p\,}_{2}^{2} - \frac{1}{4}\vec{q\,}^{2} - i\vec{\sigma}\cdot (\vec{q}\times\vec{p}_{2})}{(E_{p'_{B}}+m_{2})(E_{p_{B}}+m_{2})} \right] \chi^{c}_{s'_{2}} \nonumber \\
&
= {\cal N}  \left[ 1 + \frac{\vec{p\,}_{1}^{2} - \frac{1}{4}\vec{q\,}^{2} - 2i\vec{s}_{1}\cdot (\vec{q}\times\vec{p}_{1})}{(E_{p'_{A}}+m_{1})(E_{p_{A}}+m_{1})} \right] \left[ 1 + \frac{\vec{p\,}_{2}^{2} - \frac{1}{4}\vec{q\,}^{2} + 2i\vec{s}_{2}\cdot (\vec{q}\times\vec{p}_{2})}{(E_{p'_{B}}+m_{2})(E_{p_{B}}+m_{2})} \right] \nonumber \\
&
= {\cal N}  \left[ 1 + \frac{\vec{p\,}_{1}^{2} - \frac{1}{4}\vec{q\,}^{2} - 2i\vec{s}_{1}\cdot (\vec{q}\times\vec{p}_{1})}{4m_{1}^{2}} + \frac{\vec{p\,}_{2}^{2} - \frac{1}{4}\vec{q\,}^{2} + 2i\vec{s}_{2}\cdot (\vec{q}\times\vec{p}_{2})}{4m_{2}^{2}} + {\cal O}\left(\frac{1}{m_{i}^{3}}\right) \nonumber \right] \\
&
=  \left[ 1 + \frac{2\vec{p\,}_{1}^{2} - 2i\vec{s}_{1}\cdot (\vec{q\,}\times\vec{p}_{1})}{4m_{1}^{2}} + \frac{2\vec{p\,}_{2}^{2} + 2i\vec{s}_{2}\cdot (\vec{q\,}\times\vec{p}_{2})}{4m_{2}^{2}} + {\cal O}\left(\frac{1}{m_{i}^{3}}\right) \right] \,.
\label{eq:amplitude3}
\end{align}
The term ${\cal B}$ is given by (the minus sign comes from the Minkowski space):
\begin{align}
{\cal B} &= - \chi^{\dagger}_{s'_{1}} \left[\frac{2\vec{p}_{1} - i\vec{\sigma}\times\vec{q}}{2m_{1}}\right] \chi_{s_{1}} \chi^{c\dagger}_{s_{2}} \left[\frac{2\vec{p}_{2} - i\vec{\sigma}\times\vec{q}}{2m_{2}}\right] \chi^{c}_{s'_{2}} + {\cal O}\left(\frac{1}{m_{i}^{3}}\right) \nonumber \\
&
=-\frac{1}{4m_{1}m_{2}} \left[(2\vec{p}_{1} - 2i \vec{s}_{1}\times\vec{q}) (2\vec{p}_{2} + 2i\vec{s}_{2}\times\vec{q}) \right] + {\cal O}\left(\frac{1}{m_{i}^{3}}\right) \nonumber \\
&
=-\frac{1}{4m_{1}m_{2}} \left[(4\vec{p}_{1}\cdot\vec{p}_{2} + 4i \vec{p}_{1}\cdot(\vec{s}_{2}\times\vec{q}) - 4i \vec{p}_{2} \cdot (\vec{s}_{1}\times \vec{q}) + 4 (\vec{s}_{1}\times\vec{q}) (\vec{s}_{2}\times\vec{q})\right] + {\cal O}\left(\frac{1}{m_{i}^{3}}\right) \nonumber \\
&
=-\frac{1}{4m_{1}m_{2}} \left[4\vec{p}_{1}\cdot\vec{p}_{2} + 4i \vec{s}_{2}\cdot(\vec{q}\times\vec{p}_{1}) - 4i  \vec{s}_{1} \cdot (\vec{q}\times \vec{p}_{2}) + 4 (\vec{q}^{2}(\vec{s}_{1}\cdot \vec{s}_{2})-(\vec{q}\cdot \vec{s}_{1})(\vec{q}\cdot \vec{s}_{2}))\right] + {\cal O}\left(\frac{1}{m_{i}^{3}}\right) \nonumber \\
&
=\frac{1}{4m_{1}m_{2}} \left[-4\vec{p}_{1}\cdot\vec{p}_{2} - 4i \vec{s}_{2}\cdot(\vec{q}\times\vec{p}_{1}) + 4i  \vec{s}_{1} \cdot (\vec{q}\times \vec{p}_{2}) - 4 \vec{q}^{2}(\vec{s}_{1}\cdot \vec{s}_{2}) + 4(\vec{q}\cdot \vec{s}_{1})(\vec{q}\cdot \vec{s}_{2}))\right] + {\cal O}\left(\frac{1}{m_{i}^{3}}\right) \,.
\label{eq:amplitude4}
\end{align}
To find the potential
corresponding to ${\cal M}_{fi}$, we recall that, for the nonrelativisic case,
the scattering amplitude is given by
\begin{eqnarray}
    \langle \vec p'_A \vec p'_B |S|\vec p_A \vec p_B \rangle  = \delta^{(3)} (\vec P_f-\vec P_i) \bigg[ \delta^{(3)}(\vec p'-\vec p) - 2\pi i \delta(E'-E) T(\vec p'_A \vec p'_B;\vec p_A \vec p_B) \bigg]
\end{eqnarray}
where $\vec P_i$ ($\vec P_f$) is the intial (final) total three momentum, $\vec p$ ($\vec p'$) is the initial (final) relative momentum and $E$ ($E'$) the initial (final) total energy.
In the relativistic case we have
\begin{eqnarray}
    \langle \vec p'_A \vec p'_B |S|\vec p_A \vec p_B \rangle  = \delta^{(3)} (\vec p'_A-\vec p_A) \delta^{(3)}(\vec p'_B-\vec p_B) + i (2\pi)^4  \delta^{(4)}(P_f-P_i) 
    \frac{1}{(2\pi)^6} \left(\frac{m_1^2m_2^2}{E_{p'_A}E_{p'_B}E_{p_A}E_{p_B}}\right)^{1/2}{\mathcal M}_{fi}(\vec p'_A \vec p'_B;\vec p_A \vec p_B)
\end{eqnarray}
where $P_i$ ($P_f$) is the initial (final) four-momentum which time component gives the energy conservation. 
At first order the scattering matrix is the potential in momentum space so
\begin{eqnarray}
      \tilde{V}(\vec{q};\vec{p}_{1},\vec{p}_{2})=T(\vec p'_A \vec p'_B;\vec p_A \vec p_B)
\end{eqnarray}
and comparing the two equations we find
\begin{equation}
\tilde{V}(\vec{q};\vec{p}_{1},\vec{p}_{2}) = - \frac{1}{(2\pi)^{3}} \frac{m_{1}m_{2}}{\sqrt{E_{p_{A}}}\sqrt{E_{p'_{A}}}\sqrt{E_{p_{B}}}\sqrt{E_{p'_{ B}}}} {\cal M}_{fi} \,.
\end{equation}
Expanding the energies the relation reads
\begin{equation}
\tilde{V}(\vec{q};\vec{p}_{1},\vec{p}_{2}) = - \frac{1}{(2\pi)^{3}} \left( 1 - \frac{\vec{p\,}_{1}^{2}}{2m_{1}} - \frac{\vec{q\,}^{2}}{8m_{1}} - \frac{\vec{p\,}_{2}^{2}}{2m_{2}} - \frac{\vec{q\,}^{2}}{8m_{2}} \right) {\cal M}_{fi}.
\label{eq:amplitude5}
\end{equation}

Combining Eqs.~(\ref{eq:amplitude1}),~(\ref{eq:amplitude2}),~(\ref{eq:amplitude3}),
~(\ref{eq:amplitude4}) and~(\ref{eq:amplitude5}) leads to (up to order ${\cal
O}(m_{i}^{-1}m_{j}^{-1})$):
\begin{align}
\tilde{V}(\vec{q};\vec{p}_{1},\vec{p}_{2}) &= +\frac{1}{(2\pi)^{3}} \frac{4}{3}\frac{g^{2}}{q^{2}+i\epsilon} \left[  1 + \frac{-\frac{1}{2}\vec{q\,}^{2} - 2i\vec{s}_{1}\cdot (\vec{q}\times\vec{p}_{1})}{4m_{1}^{2}} + \frac{-\frac{1}{2}\vec{q\,}^{2} + 2i\vec{s}_{2}\cdot (\vec{q}\times\vec{p}_{2})}{4m_{2}^{2}} \right. \nonumber \\
&
\left. +\frac{1}{4m_{1}m_{2}} \left(-4\vec{p}_{1}\cdot\vec{p}_{2} + 4i  \vec{s}_{1} \cdot (\vec{q}\times \vec{p}_{2}) - 4i \vec{s}_{2}\cdot(\vec{q}\times\vec{p}_{1}) - 4 \vec{q\,}^{2}(\vec{s}_{1}\cdot \vec{s}_{2}) + 4(\vec{q}\cdot \vec{s}_{1})(\vec{q}\cdot \vec{s}_{2}))\right)
\right].
\end{align}
Since we want to obtain a nonrelativistic potential we have to expand the
$4$-vector in the propagator, \emph{i.e.}
\begin{equation}
\frac{1}{q^{2}} \approx - \frac{1}{\vec{q\,}^{2}} - \frac{q_{0}^{2}}{\vec{q\,}^{4}} \quad\text{with}\quad
q_{0}^{2} = (E_{p'_{A}}-E_{p_{A}}) (E_{p_{B}}-E_{p'_{B}}) = \frac{(\vec{p}_{1}\cdot\vec{q}\,)(\vec{p}_{2}\cdot\vec{q}\,)}{m_{1}m_{2}} + {\cal O}\left(\frac{1}{m_{i}^{3}}\right) \,,
\end{equation}
and thus
\begin{equation}
\frac{1}{q^{2}} = -\frac{1}{\vec{q\,}^{2}} - \frac{1}{\vec{q\,}^{4}}\frac{(\vec{p}_{1}\cdot\vec{q}\,)(\vec{p}_{2}\cdot\vec{q}\,)}{ m_{1}m_{2}} + {\cal O}\left(\frac{1}{m_{i}^{3}}\right) \,,
\end{equation}
resulting in
\begin{align}
&
 \tilde{V}^{\rm Breit}(\vec{q};\vec{p}_{1},\vec{p}_{2}) = -\frac{1}{(2\pi)^{3}} \frac{4}{3}\frac{g^{2}}{\vec{q\,}^{2}} \left[  1 + \frac{-\frac{1}{2}\vec{q\,}^{2} - 2i\vec{s}_{1}\cdot (\vec{q}\times\vec{p}_{1})}{4m_{1}^{2}} + \frac{-\frac{1}{2}\vec{q\,}^{2} + 2i\vec{s}_{2}\cdot (\vec{q}\times\vec{p}_{2})}{4m_{2}^{2}} +\frac{4}{\vec{q\,}^{2}}\frac{(\vec{p}_{1}\cdot\vec{q}\,)(\vec{p}_{2}\cdot\vec{q}\,)}{4m_{1}m_{2}} \right.
\nonumber \\
&
\left. + \frac{1}{4m_{1}m_{2}} \left(-4\vec{p}_{1}\cdot\vec{p}_{2} + 4i  \vec{s}_{1} \cdot (\vec{q}\times \vec{p}_{2}) - 4i \vec{s}_{2}\cdot(\vec{q}\times\vec{p}_{1}) - 4 \vec{q\,}^{2}(\vec{s}_{1}\cdot \vec{s}_{2}) + 4(\vec{q}\cdot \vec{s}_{1})(\vec{q}\cdot \vec{s}_{2}))\right) \right].
\label{eq:VBreitinq}
\end{align}

The potential in coordinate space is given by the Fourier transformation\footnote{The operator in coordinate space is given by
\begin{eqnarray}
    \langle \vec r'_A \vec r'_B | V | \vec r_A \vec r_B \rangle &=& \frac{1}{(2\pi)^6} \int d^3 p_A d^3 p_B d^3p'_A d^3 p'_B 
    e^{i\vec p'_A\cdot \vec r'_A+i\vec p'_B\cdot \vec r'_B-i\vec p_A\cdot \vec r_A-i\vec p_B\cdot \vec r_B}
    \langle \vec p'_A \vec p'_B | V | \vec p_A \vec p_B \rangle
    \nonumber \\
    &=& \frac{1}{(2\pi)^6} \int d^3 q_1 d^3 q_2 d^3p_1 d^3 p_2 
    e^{i\vec q_1\cdot \frac{\vec r'_A+\vec r_A}{2} +i\vec q_2\cdot \frac{\vec r'_B+\vec r'_B}{2}+i\vec p_1 \cdot (\vec r'_A-\vec r_A)+i\vec p_2 \cdot (\vec r'_B-\vec r_B)}
    \langle \vec p'_A \vec p'_B | V | \vec p_A \vec p_B \rangle
\end{eqnarray}
with $\vec q_1 = \vec p'_A-\vec p_A$, $\vec q_2 = \vec p'_B-\vec p_B$, $\vec p_1=\frac{\vec p'_A+\vec p_A}{2}$ and $\vec p_2=\frac{\vec p'_B+\vec p_B}{2}$. For potentials 
\begin{eqnarray}
    \langle \vec p'_A \vec p'_B | V | \vec p_A \vec p_B \rangle &=& \delta^3(\vec p'_A+\vec p'_B-\vec p_A-\vec p_B) V(\vec q,\vec p_1,\vec p_2)
\end{eqnarray}
we have 
\begin{eqnarray}
    \langle \vec r'_A \vec r'_B | V | \vec r_A \vec r_B \rangle
    &=& \frac{1}{(2\pi)^6} \int d^3 q  d^3p_1 d^3 p_2 
    e^{i\vec q\cdot \left[\frac{\vec r'_A+\vec r_A}{2} - \frac{\vec r'_B+\vec r'_B}{2}\right]+i\vec p_1 \cdot (\vec r'_A-\vec r_A)+i\vec p_2 \cdot (\vec r'_B-\vec r_B)}
    V(\vec q,\vec p_1,\vec p_2)
\end{eqnarray}
where we have made the integral in $\vec q_2$ and rename $\vec q_1$ as $\vec q$. Now we can write
\begin{eqnarray}
    \langle \vec r'_A \vec r'_B | V | \vec r_A \vec r_B \rangle
    &=& \int d^3 q   
    e^{i\vec q\cdot \left[\frac{\vec r'_A+\vec r_A}{2} - \frac{\vec r'_B+\vec r'_B}{2}\right]}
    V(\vec q,i-i\vec \nabla^{r_A},-i\vec \nabla^{r_B})
    \frac{1}{(2\pi)^6} \int d^3p_1 d^3 p_2 e^{i\vec p_1 \cdot (\vec r'_A-\vec r_A)+i\vec p_2 \cdot (\vec r'_B-\vec r_B)}
    \nonumber \\ &=&
    \int d^3 q   
    e^{i\vec q\cdot \vec r}
    V(\vec q,-i\vec \nabla^{r_A},-i\vec \nabla^{r_B})
    \delta^3(\vec r'_A-\vec r_A)\delta^3 (\vec r'_B-\vec r_B)
\end{eqnarray}
where in the exponential we have used the relations given by the $\delta$ functions $\vec r'_A=\vec r_A$ and $\vec r'_B=\vec r_B$ and define $\vec r\equiv\vec r_A-\vec r_B$.
Notice that
\begin{eqnarray}
    \langle \vec r_A \vec r_B | V | \Psi \rangle &=& \int d^3 r''_A d^3 r''_B
    \langle \vec r_A \vec r_B | V | \vec r''_A \vec r_B \rangle \Psi(\vec r''_A,\vec r''_B) =
    V(\vec q,-i\vec \nabla^{r_A},-i\vec \nabla^{r_B}) \Psi(\vec r_A,\vec r_B)\,.
\end{eqnarray}}
\begin{equation}
V(\vec{r};\vec{p}_{1},\vec{p}_{2}) = \int d^{3}q\, e^{-i\vec{q}\cdot\vec{r}}\,
\tilde{V}(\vec{q};\vec{p}_{1},\vec{p}_{2}) \,,
\end{equation}
where $\vec p_1 = -i \vec \nabla^{r_A}$ and $\vec p_2 = -i \vec \nabla^{r_B}$ and all operators $\vec p_1$ and $\vec p_2$ are on the right.
The Fourier transform of the various terms in Eq.~(\ref{eq:VBreitinq}) is given by
\begin{align}
W &= \frac{1}{(2\pi)^{3}} \int d^{3}q\, e^{-i\vec{q}\cdot\vec{r}}\, \frac{1}{\vec{q\,}^{2}} = \frac{1}{4\pi r}, \\
W_{1} &= \frac{1}{(2\pi)^{3}} \int d^{3}q\, q_{j}\, e^{-i\vec{q}\cdot\vec{r}}\, \frac{1}{\vec{q\,}^{2}} = i\nabla_{j}W = i \frac{r_{j}}{r} W', \\
W_{2} &= \frac{1}{(2\pi)^{3}} \int d^{3}q\, e^{-i\vec{q}\cdot\vec{r}}\, \vec{q\,}^{2}\, \frac{1}{\vec{q\,}^{2}} = -\Delta \int \frac{d^{3}q}{(2\pi)^{3}}\, e^{-i\vec{q}\cdot\vec{r}}\, \frac{1}{\vec{q\,}^{2}} = -\Delta W = \delta^{(3)}(\vec{r}\,), \\
W_{3} &= \frac{1}{(2\pi)^{3}} \int d^{3}q\, q_{i}\, q_{j}\, e^{-i\vec{q}\cdot\vec{r}}\, \frac{1}{\vec{q\,}^{2}} = -\nabla_{i}\nabla_{j}W = -\left[W''-\frac{1}{r}W'\right] \left[\frac{r_{i}r_{j}}{r^{2}}-\frac{1}{3}\delta_
{ij} \right] - \frac{1}{3}\delta_{ij}\Delta W, \\
U &= \frac{1}{(2\pi)^{3}} \int d^{3}q\, e^{-i\vec{q}\cdot\vec{r}}\,
\frac{1}{\vec{q\,}^{4}} = -\frac{r}{8\pi}, \\
U_{1} &= \frac{1}{(2\pi)^{3}} \int d^{3}q\, q_{i}\, q_{j}\, e^{-i\vec{q}\cdot\vec{r}}\, \frac{1}{\vec{q\,}^{4}} = -\nabla_{i}\nabla_{j}U =
-\left[U''-\frac{1}{r}U'\right]\left[\frac{r_{i}r_{j}}{r^{2}}-\frac{1}{3}\delta_
{ij} \right] - \frac{1}{3}\delta_{ij}\Delta U,
\end{align}
with
\begin{equation}
\begin{split}
\Delta\frac{1}{r} &= -4\pi\delta^{(3)}(\vec{r}\,), \\
W'&= -\frac{1}{4\pi r^{2}}, \quad W''= \frac{1}{2\pi r^{3}}, \quad \Delta
W = -\delta^{(3)}(\vec{r}\,), \\
U' &= -\frac{1}{8\pi}, \quad U''=0, \quad \Delta U = -\frac{1}{4\pi r} \,, \\
\nabla_{i}\nabla_{j} W &= \left[W'' - \frac{1}{r}W' \right]
\left[\frac{r_{i}r_{j}}{r^{2}}-\frac{1}{3}\delta_{ij}\right] +
\frac{1}{3}\delta_{ij}\Delta W \,.
\end{split}
\end{equation}
Collecting all the pieces, we obtain the Breit quark-antiquark interaction in coordinate space:
\begin{align}
V^{\rm Breit}(\vec{r};\vec{p}_{1},\vec{p}_{2}) &=
-\frac{4\alpha_{s}}{3r} + \frac{2\pi\alpha_{s}}{3} \delta^{(3)}(\vec{r}\,) \left(
\frac{1}{m_{1}^{2}} + \frac{1}{m_{2}^{2}} \right) +
\frac{2\alpha_{s}}{3m_{1}m_{2}}\left[\frac{\vec{p}_{1}\cdot\vec{p}_{2}}{r} +
\frac{(\vec{r}\cdot\vec{p}_{1})(\vec{r}\cdot\vec{p}_{2})}{r^{3}} \right] \nonumber \\
&
+\frac{4\alpha_{s}}{3m_{1}m_{2}} \left[\frac{8\pi}{3} \delta^{(3)}(\vec{r}\,)
(\vec{s}_{1}\cdot\vec{s}_{2}) +
\frac{3(\vec{s}_{1}\cdot\hat{r})(\vec{s}_{2}\cdot\hat{r})-(\vec{s}_{1}\cdot
\vec{s}_{2})}{r^{3}} \right] \nonumber \\
&
+\frac{2\alpha_{s}}{3r^{3}} \left[
\frac{(\vec{r}\times\vec{p}_{1})\cdot\vec{s}_{1}}{m_{1}^{2}} -
\frac{(\vec{r}\times\vec{p}_{2})\cdot \vec{s}_{2}}{m_{2}^{2}} +
\frac{2}{m_{1}m_{2}} \left( (\vec{r}\times\vec{p}_{1})\cdot\vec{s}_{2}
-(\vec{r}\times\vec{p}_{2})\cdot\vec{s}_{1} \right) \right] \,.
\label{eq:VBreitinr}
\end{align}
with $\alpha_s=\frac{g^2}{4\pi}$ and
\begin{eqnarray}
    (\vec r \cdot \vec p_1)(\vec r \cdot \vec p_2) \equiv -r_i r_j \nabla_{i}^{r_A} \nabla_{j}^{r_B}
\end{eqnarray}
is understood.

If one wants to obtain the quark-antiquark interaction in operator form one only needs to change the factor $\frac 4 3$ by a $\lambda^a_1 (\lambda^a_2)^T$ operator and the expectation values of
the spin operators $\vec s_i$ by the operators $\frac 1 2 \vec \sigma_i$. The quark-quark interaction is the same changing $\lambda^a_1 (\lambda^a_2)^T$ by $-\lambda^a_1 \lambda^a_2\equiv -\lambda_1 \cdot \lambda_2$, since in our
convention the antiquark current is the same as the quark current and the minus sign comes from the different number of interchange of fields in the normal product (Feynman rule for fermions). So for quark-quark interaction we have
\begin{align}
V^{\rm Breit}(\vec{r};\vec{p}_{1},\vec{p}_{2}) &=
\frac 1 4 \vec \lambda_1 \cdot \vec \lambda_2 \alpha_s \bigg\{\frac{1}{r} - \frac{\pi}{2} \delta^{(3)}(\vec{r}\,) \left(
\frac{1}{m_{1}^{2}} + \frac{1}{m_{2}^{2}} \right) -
\frac{1}{2m_{1}m_{2}}\left[\frac{\vec{p}_{1}\cdot\vec{p}_{2}}{r} +
\frac{(\vec{r}\cdot\vec{p}_{1})(\vec{r}\cdot\vec{p}_{2})}{r^{3}} \right] \nonumber \\
&
-\frac{1}{m_{1}m_{2}} \left[\frac{2\pi}{3} \delta^{(3)}(\vec{r}\,)
(\vec{\sigma}_{1}\cdot\vec{\sigma}_{2}) +
\frac{3(\vec{\sigma}_{1}\cdot\hat{r})(\vec{\sigma}_{2}\cdot\hat{r})-(\vec{\sigma}_{1}\cdot
\vec{\sigma}_{2})}{4r^{3}} \right] \nonumber \\
&
-\frac{1}{2r^{3}} \left[
\frac{(\vec{r}\times\vec{p}_{1})\cdot\vec{\sigma}_{1}}{2m_{1}^{2}} -
\frac{(\vec{r}\times\vec{p}_{2})\cdot \vec{\sigma}_{2}}{2m_{2}^{2}} +
\frac{1}{m_{1}m_{2}} \left( (\vec{r}\times\vec{p}_{1})\cdot\vec{\sigma}_{2}
-(\vec{r}\times\vec{p}_{2})\cdot\vec{\sigma}_{1} \right) \right] \bigg\} \,.
\label{eq:VBreitinrqq}
\end{align}
This can be also seen as the quark-antiquark or antiquark-antiquark interaction considering the dual representation in color for antiquarks. We have that between color singlet states
\begin{eqnarray}
    \langle \vec \lambda_i \cdot \vec \lambda_j \rangle &=& -\frac{4}{3} \quad {\rm mesons}
    \nonumber \\
    \langle \vec \lambda_i \cdot \vec \lambda_j \rangle &=& -\frac{8}{3} \quad {\rm baryons}
\end{eqnarray}
This is why sometimes in the literature we find the relation $V_{qq}=2 V_{q\bar q}$ assuming that are spectation values in color singlets for baryons and mesons respectively.

Following this detailed derivation of the one-gluon exchange interaction between quark and antiquark, which appears to govern short interquark distances, one might get the impression that our understanding of QCD is nearly complete. However, nothing could be further from the truth. We still do not fully understand why quarks and gluons, the fundamental fields of QCD, do not manifest in the theory’s spectrum as asymptotic particle states. This gap highlights our broader lack of knowledge about the behavior of non-Abelian gauge field theories in general, and QCD specifically, at large distance scales.

The word confinement in the context of hadron physics is originally referred to the fact that quarks and gluons appear to be trapped inside hadrons, from which they cannot escape. While the problem of confinement remains unsolved, certain general features of the confining force are understood. Lattice-QCD studies have shown that multi-gluon exchanges create an attractive, linearly rising potential proportional to the distance between infinitely heavy quarks~\cite{Bali:2005fu}. Additionally, the spontaneous creation of light-quark pairs from the QCD vacuum plays a key role in the dynamics of strong interactions, potentially contributing to the screening of this rising potential at low momenta and eventually to the breaking of the quark-antiquark binding string~\cite{Bali:2005fu}. These insights have been integrated into any QCD-based quark model. For example, a potential of the form:
\begin{equation}
V_{\rm CON}(\vec{r}\,) = a_{c} \big( 1-e^{-\mu_{c}r} \big) \,,
\label{eq:VCONC}
\end{equation}
where $a_{c}$ and $\mu_{c}$ are parameters, exhibits a linear behavior at short distances, characterized by an effective confinement strength $\sigma = a_{c}\mu_{c}$, and approaches a constant value at large distances, with the threshold given by $V_{\rm thr} = a_{c}$. It can be assumed that no hadronic states exist for energies above this threshold. At this point, the system undergoes a transition from a color string configuration between static color sources to a pair of static hadrons, as the color string breaks, leading to the most likely decay into hadrons.

%%%%%%%%%%%%%%%%%%%%%%%%%%%%%%%%%%%%%%%%%%%%%%%%%%%%%%%%%%%%%%%%%%%%%%%%%%%%%%%%%%%%%%%%%%%%%%%

\section{Spectrum of Baryons and Mesons in the naive quark model}

As we have shown, in order to understand the magnetic moments of the nucleons, we need large effective quark masses and one could justify, as a first approximation, that baryons are simply  non-relativistic  bound states from gluon exchange interactions. 
As we will see, this naive picture works surprisingly well although we do not understand why it works as well as it does.

Many studies were made based on the ideas of De Rujula,
Georgi, and Glashow ~\cite{PhysRevD.12.147} (see ~\cite{Capstick:2000qj} for a review). These models have sometimes been called quark potential models or non-relativistic quark models but we prefer to refer to them as naive constituent quark models.

The most detailed and phenomenologically successful of these first models is the one developed by Isgur and Karl and their 
collaborators~\cite{Isgur:1977ef, PhysRevD.18.4187,PhysRevD.19.2653, Isgur:1979be, Isgur:1979ee, Koniuk:1979vy, Isgur:1981yz}. 
The model has general characteristics that are also shared by other models, so we will focus on it as an example.

Isgur and Karl assume a non relativistic kinetic energy $T$ and a quark-quark interaction based on the one-gluon-exchange (OGE) of De Rújula {\it et al.}  with a hamiltonian~\cite{PhysRevD.18.4187,PhysRevD.19.2653}

\begin{equation}
{\mathcal H}=\sum_i {\left( m_i+\frac{p_i^2}{2m_i} \right)}+\sum_{i<j}{\left( V^{ij}+H^{ij}_{hyp} \right )}
\label{hamIK}
\end{equation}
The potential $V^{ij}$ should includes the chromoelectric one gluon exchange piece but in practice it is replaced by an harmonic oscillator potential $K(r_i-r_j)^2/2$ which plays the role of the confinement potential plus, sometimes, an unspecified anharmonic $U_{ij}$ which is treated perturbatively. This term is some unknown potential which should incorporate a short-range attractive potential
(of Coulomb-type) and deviations
of the long-range part of the potential from the
harmonic-oscillator form. The $H^{ij}_{hyp}$ term corresponds to the chromomagnetic piece of the OGE interaction given by
\begin{equation}
H^{ij}_{hyp}=-\frac{\alpha_s}{4}(\vec \lambda_i\cdot\vec \lambda_j)\left (\frac{2\pi}{3m_im_j}{\vec \sigma_i\cdot\vec \sigma_j}\delta^{(3)}(r_{ij})+ \frac{1}{4m_im_j}\frac{1}{r^3_{ij}} \left[\frac{3(\vec \sigma_i\cdot \vec r_{ij})(\vec \sigma_j\cdot \vec r_{ij})}{r^2_{ij}}-\vec \sigma_i\cdot\vec \sigma_j \right] \right)
\label{eq:ham}
\end{equation}
where  $\vec \sigma_i$ is the spin operator of the $\it i$-th quark.
Notice that in the one gluon exchange interaction there are also momentum dependent terms that were not included by Isgur and Karl.

To solve the Schr\"odinger equation for baryons one introduces the Jacobi coordinates 
\begin{equation}
\vec \rho=\frac{\vec r_1-\vec r_2}{\sqrt{2}} \; \;\; \vec \lambda=\frac{1}{\sqrt{6}}(\vec r_1+\vec r_2-2\vec r_3)
 \; \;\; \vec R_{cm} = \frac{m(\vec r_1+\vec r_2) + m' \vec r_3}{2m+m'}
\end{equation}
considering two equal masses for quarks 1 and 2 and a different mass for quark 3.
This change of variables separate the Hamiltonian Eq.(\ref{hamIK}) in two independent harmonic oscillators when $H_{hyp}$ and $U$ are set to zero given by
\begin{eqnarray}
    H= \frac{P_{cm}^2}{2M} + \frac{P_{\rho}^2}{2m_\rho} + \frac{P_{\lambda}^2}{2m_\lambda} + \frac 3 2 K (\rho^2+\lambda^2)
\end{eqnarray}
with $M=2m+m'$, $m_\rho=m$ and $m_\lambda=\frac{3mm'}{2m+m'}$. Notice that $3(\rho^2+\lambda^2)=\sum_{i<j}(\vec r_i-\vec r_j)^2$ and so is simetric to the exchange of particles.
The energies are given by
\begin{eqnarray}
    E_{n_\rho l_\rho n_\lambda l_\lambda} =  \omega_\rho (2 n_\rho + l_\rho + \frac 3 2) +  \omega_\lambda (2 n_\lambda + l_\lambda + \frac 3 2)
\end{eqnarray}
with $\omega_\rho=\sqrt{\frac{3K}{m_\rho}}$ and $\omega_\lambda=\sqrt{\frac{3K}{m_\lambda}}$ (with $\omega_\rho=\omega_\lambda$ for $m'=m$).
Then, the baryon wave function can be written as solutions of two three dimensional harmonic oscillator eingenstates $\Psi_{NLM}(\vec{\rho},\vec{\lambda})$ with 
\begin{equation}
\Psi_{NLM}(\vec{\rho},\vec{\lambda})=C_N P_N(\vec \rho,\vec \lambda)e^{-\frac{1}{2}(\alpha_\rho^2\rho^2+\alpha_\lambda^2\lambda^2)}
\end{equation}
where $\alpha_\rho^2=\sqrt{3Km_\rho}$, $\alpha_\lambda^2=\sqrt{3Km_\lambda}$, $C_N$ is a normalization factor and $P_N$ a polynomial of degree $N$.
By combining these spatial wave functions with the corresponding quark spin, flavor and color wave functions to obtain totally antisymmetric functions one can generate the baryon spectrum at first order.
Notice that this could be done for baryons with $u$, $d$ and $s$ quarks considering $m_u=m_d$. For example, for the $\Delta^{++}$ baryon with spin component $m$ we would have that the state is
\begin{eqnarray}
    |\Psi_{\Delta^{++}m} \rangle &=& C e^{-\frac{1}{2}\alpha^2(\rho^2+\lambda^2)} |\chi_{\frac 3 2 m} \rangle | \phi_{\frac 3 2 \frac 3 2} \rangle\, \xi_c[1^3]
\end{eqnarray}
with the definitions given by Eqs.(\ref{j32}) and (\ref{DelatF}) and $\xi_c[1^3]=\frac{1}{\sqrt 6} \epsilon_{ijk} |q_{c_i} q_{c_j}q_{c_k}\rangle$ representing the totally antisymmetric singlet color wave function.
For the $p$ with spin up
\begin{eqnarray}
    |\Psi_{p \uparrow} \rangle &=& C e^{-\frac{1}{2}\alpha^2(\rho^2+\lambda^2)} |p \uparrow \rangle\, \xi_c[1^3]
    \label{pup}
\end{eqnarray}
with the definition Eq.(\ref{eq:dipole1}). In both cases $\alpha=\sqrt{3Km_u}$ ($m'=m=m_d=m_u$) and the states would be degenerate at first order.
Once the order zero states are built, one uses perturbation theory to include the effects of $H_{hyp}$ and $U$.

It is important to note that in the harmonic oscillator model ($H_{hyp}=U=0$) the states of each multiplet are degenerated. This approximation works for the negative-parity $P$-wave baryons because all of them are contained in a single SU(6) multiplet (a 70-plet). However, in the case of the positive parity $P$-wave sector, states are classified in five different multiplets. The harmonic oscillator model would make all the SU(6) multiplets degenerate with energy $2\hbar\omega$ being $\omega=\sqrt{\frac{3K}{m}}$. The experimental spectrum does not show this degeneracy, then one has to conclude that the confining forces are not purely harmonic.

To solve this problem Isgur and Karl introduces the anharmonic potential $U(r_i-r_j)$ which is treated as a perturbation. Defining the momenta $a_1$, $a_2$, $a_4 $ of the $U(r_i-r_j)$ potential by
\begin{equation}
\left\{a_0,a_2,a_4 \right\}= 3\frac{\alpha^3}{\pi^{3/2}}\int{d^3\rho \left\{1,\alpha^2\rho^2,\alpha^4\rho^4  \right\} U(\sqrt{2}\rho) e^{-\alpha^2\rho^2}}
\end{equation}

One can shown that~\cite{PhysRevD.19.2653} the positive parity $P$ wave states split in a pattern which is independent of the functional form of the potential $U(r_i-r_j)$. The ground state and the first negative parity excitation are also modified giving the energies
\begin{eqnarray}
E[56,0^+]&=&E_0
\nonumber \\ 
E[56', 0^+]&=&E_0+2\Omega-\Delta
 \nonumber \\
E[70,0^+]&=&E_0+2\Omega-\Delta/2
\nonumber \\
E[56,2^+]&=&E_0+2\Omega-2 \Delta/5
\nonumber \\
E[70,2^+)]&=&E_0+2\Omega-\Delta/5
\nonumber \\
E[20,1^+)]&=&E_0+2\Omega
\end{eqnarray}
where $E_0=3\hbar\omega+a_0$, $\Omega=\hbar \omega -a_0/2+a_2/3$ and $\Delta=-5a_0/4+5a_2/3-a_4/3$.

Isgur and Karl~\cite{PhysRevD.19.2653} makes a rough qualitative
judgment (in their own words) to obtain the values $E_0=1150$, $\Omega=440$ and $\Delta=440$ (note that our definitions are slightly different from the original in Ref~\cite{PhysRevD.19.2653}).

In this way Isgur and Karl were able to make the first radial excitation $[56',O^+]$ low-lying (corresponding to
the Roper resonance) and makes the unobserved $[20, 1^+]$ multiplet lie very high.

The hyperfine interaction is diagonalized in the basis of the harmonic oscillator wave functions, being its strength $\alpha_s$ determined by fitting the $N\Delta$ mass splitting
\begin{equation}
\delta=\frac{4^3\alpha_s \alpha^3}{3\sqrt{2\pi}m_u^2}\,.
\end{equation}
The resulting description of the spectrum is quite good. For further details one has to resort to the original papers.

Despite the main features of the low lying baryon spectra are reasonably well reproduced by the model, it has several flaws. First of all, the bound states of light quarks shows a ratio $p/m$ of the order of unity. Then, the use of a non relativistic approximation is questionable. It is true that since this is a phenomenological model, some of the relativistic corrections may be hidden in the adjustable parameters, but a study of the relativistic corrections should be incorporated into the model.

Another inconsistency is that the incorporation of the spin-orbit piece in the potential destroy the agreement with the experimental data. 
Also it is questionable the role played by the anharmonicity in the confinement potential.
Then, one can conclude the model is a preliminary, and interesting attempt to describe the baryon spectrum.

The work of Isgur and Karl was continued with a relativezed model, first applied to meson spectroscopy by Godfrey and Isgur~\cite{Godfrey:1985xj}  and later to baryons~\cite{Capstick:1985xss}. Both approaches start with a Hamiltonian given by 
\begin{equation}
{\mathcal H}=\sum_i {\sqrt{p^2_i+m^2_i}+V}
\end{equation}
where $V$, besides the OGE interaction, includes a linear confinement interaction, instead the quadratic one used by Isgur and Karl, and a Thomas precession spin-orbit term coming from confinement. The model is relativized by: (1) the interquark coordinate $r$  is smeared out over distances of the order of the inverse quark masses. In practices this modification is implemented by convoluting all the potential with the smearing function $\rho_{ij}(r_i-r_j)=\sigma_{ij}^3e^{-\sigma_{ij}^2 (r_i-r_j)^2}/\pi^{3/2}$
where $\sigma_{ij}$ is a two parameter function of the quark masses~\cite{Godfrey:1985xj}; (2) The convoluted potentials $\tilde{V}_k(r_{ij})$, with the exception of confinement, are modified according to
\begin{equation} 
\frac{\tilde{V}_k(r_{ij})}{m_im_j}\to \left(\frac{m_im_j}{E_iE_j}\right)^{1/2+\epsilon_k}\frac{V_k(r_{ij})}{m_im_j}\left(\frac{m_im_j}{E_iE_j}\right)^{1/2+\epsilon_k}
\end{equation}
where $\epsilon_k$ is a parameter which is expected to be small.

The definition of the model is completed by introducing a phenomenological annihilation potential and parametrizing the behavior of the running gluon coupling constant by the convenient form
\begin{equation}
\alpha_s(Q^2)=\sum_k\alpha_k e^{Q_2/4\gamma_k^2}
\end{equation}
where $\alpha_k$ and $\gamma_k$ are parameters fitted  to a QCD curve for $\alpha_s(Q^2)$.

Godfrey and Isgur performed a detailed study calculating first the meson spectrum and then performing an
extensive analysis of meson couplings. They conclude that all the mesons properties, from the pion to the upsilon, can be described in a unified framework. In fact the paper is still today used as a template to decide if a experimental signal  can by a $q\bar q$ meson or a more complicated structure.

To describe the baryons, in addition to using the relativized model of Godfrey and Isgur, Capstick and Isgur introduce a confinement potential based on the flux tube model of low energy  QCD~\cite{ISGUR1983247,Capstick:1985xss,PhysRevD.27.233}. The model 
is based on the idea that gluons should rearrange themselves into flux tubes which, in the heavy
quark limit, adjust their configuration instantaneously in response to quark motion. The lowest lying
$qqq$ flux tube configuration is either a symmetric Y string configuration or, if one of the angles
formed by the $qqq$ triangle is greater than 120°, two
strings joining the three quarks. 

The string potential in terms of $\rho$, $\lambda$ and $\cos\theta=\frac{\vec{\rho}\cdot\vec{\lambda}}{\rho\lambda}$ is then~\cite{Capstick:1985xss}
\begin{eqnarray}
V_{string}&=&C_{qqq}+b \sqrt{\frac{3}{2}}\left[\rho^2+\lambda^2+2\rho\lambda\sqrt{1-\cos^2\theta}\right]^{1/2} \,\text{if all angles}~~\theta_{ijk}<120º 
\nonumber \\
V_{string}&=&C_{qqq}+b(r_{12}+r_{13}), \theta_{312}>120º,
\nonumber \\
V_{string}&=&C_{qqq}+b(r_{12}+r_{23}), \theta_{123}>120º,
\nonumber \\
V_{string}&=&C_{qqq}+b(r_{13}+r_{23}), \theta_{132}>120º,
\end{eqnarray}
where $C_{qqq}$ is an overall energy shift and
\begin{eqnarray}
r_{12}&=&\sqrt{2}\rho,
\nonumber \\
r_{13}&=&\frac{1}{\sqrt{2}}(\rho^2+3\lambda^2+2\sqrt{3}\rho\lambda \cos\theta)^{1/2}
\nonumber \\
r_{23}&=&\frac{1}{\sqrt{2}}(\rho^2+3\lambda^2-2\sqrt{3}\rho\lambda \cos\theta)^{1/2}
\end{eqnarray}

To simplify the calculation, Capstick and Isgur divide the $V_{string}$ into a two-body piece and a three-body piece.
\begin{equation}
V_{string}=C_{qqq}+fb\sum_{i<j}{r_{ij}}+V_{3b}
\end{equation}

Where 
\begin{equation}
V_{3b}=b\left[\sum_{i=1}^3|r_i-r_4|-f\sum_{i<j}r_{ij}\right]
\end{equation}

The parameter $f$ is chosen in order to minimize the expectation value  of the $V_{3b}$ in the harmonic oscillator ground state of the baryon system. 
Then one can treat the two body part of $V_{string}$ exactly and compute  the $V_{3b}$ part perturbatively.

Concerning the results, these relativizing version of  quark model leads to a picture of the baryon spectrum that is very similar to the non relativistic version of Ref.~\cite{PhysRevD.18.4187,PhysRevD.19.2653}. However, they provide a demonstration that its is possible describe in a single consistent framework both mesons and baryons  without any change of the used parameters. Furthermore,  it explain one of the unpleasant features of the non relativistic version: baryon spin-orbit splittings
appear to be much smaller than expected from their one gluon-exchange matrix elements. The cure of this problem is mainly twofold. One is the cancellation of the one-gluon exchange spin orbit by the Thomas precession term coming from the confinement interaction. The second is related with the modification of the $\delta$ piece of the one gluon exchange hyperfine interaction due to the smearing of the quark wave function. This modification produces hyperfine splittings which are less strong that the one of the non relativistic version of the model. 

As conclusion, all these works represents a first attempt to show that mesons and baryons can be reasonably well understood in terms of quarks and gluon degrees of freedom.

\section{Dynamical quark mass and boson exchanges}

The constituent quark model we have described so far is based on the assumption that light quarks have an effective mass of the order of $300$ MeV, much larger than the masses appearing in the QCD lagrangian. In this section we will try to explain how these masses appear and what are the consequences that follows from this fact. Details about spontaneously broken symmetries in Quantum Field Theory can be found in other chapters.

In Quantum Field Theory mass generation is associated with the phenomena of the spontaneous breaking of a certain symmetry. To show how this comes about, let's assume that $U$ is an element of a symmetry group which leaves invariant a certain Hamiltonian $H$ 
\begin{equation}
UHU^+=H
\end{equation}
and let $|A\rangle$ and $|B\rangle$ two eigenstates of $H$ connected by $U$
\begin{equation}
U|A\rangle=|B\rangle\,.
\end{equation}
Then we have
\begin{equation}
\label{deg}
E_A=\langle A|H|A\rangle=\langle B|H|B\rangle=E_B
\end{equation}
Then the symmetry of the hamiltonian is manifest in a degeneration of the eigenstates connected by such symmetry.
However the former results is only true if the ground state is invariant under the symmetry transformation. In fact, if $|A\rangle$ and $|B\rangle$ are related with the ground states $|0\rangle$ through certain creation operators $\psi_A$ and $\psi_B$
\begin{equation}
|A\rangle=\psi_A |0\rangle~~,~~|B\rangle=\psi_B |0\rangle
\end{equation}
verifiyig
\begin{equation}
\label{trans}
U\psi_A U^+=\psi_B
\end{equation}
Eq.~(\ref{deg}) follows only if $U|0\rangle=|0\rangle$.

If the ground state is not invariant under the symmetry transformation the degeneration of the energy levels does not hold. This situation is refereed as a spontaneous symmetry breaking (SSB). It is worth to notice that even though the symmetry is not manifest in the degeneration of the energy levels, the Hamiltonian is still invariant under $U$.

There is another signal of the SSB. For $U=e^{i\epsilon^aQ^a}$ where $Q^a$ are the  group generator and $\epsilon^a$ the continuous group parameters, $U|0\rangle=|0\rangle$ can be expressed by the statement that the group generators does not annihilate the vacuum
\begin{equation}
\label{Q}
Q^a|0\rangle\neq 0
\end{equation} 
An equivalent statement to Eq.~(\ref{Q}) is that certain field operators have nonvanishing vacuum expectation values
\begin{equation}
\langle0|\phi_j|0\rangle\neq 0
\end{equation}
which follows from writing Eq.(\ref{trans}) for the operators $\phi_i$ in terms of $Q^a$
\begin{equation}
\left[Q^a, \phi_i\right]=it^a_{ij}\phi_j
\end{equation}
and 
\begin{equation}
\langle0|\left[Q^a, \phi_i \right]|0\rangle=it^a_{ij}\langle0|\phi_j|0\rangle\neq 0
\end{equation}

In the case of the QCD Lagrangian with massless quarks, there is a global symmetry, given by Chiral transformations
\begin{equation}
\psi\rightarrow \psi'=e^{i\alpha^a \lambda^a \gamma_5} 
\end {equation}
Where $\alpha^a$ are the parameters of the transformation and $\gamma_5$ is the Dirac gamma matrix. 
Chiral symmetry instead of rotating separately the 2-component Weyl spinors, make independent vector and axial
rotations of the full 4-component Dirac spinors. Meanwhile, the axial transformation mixes
states with different $P$-parities. Therefore, if that symmetry remains exact, it implies the existence of parity doublets. For example the vector $\rho$ meson $\rho(770) 1^-$ would be degenerated with the axial $a_1$ meson $a_1(1260) 1^+$ or the nucleon $N(940) \frac{1}{2}^+$ would be degenerated with the resonance $N(1535) \frac{1}{2}^-$. Since this is not the case one have to conclude that chiral symmetry is spontaneously broken in nature.

The second signal of SSB is the quark condensate $\langle\bar{\psi}\psi\rangle\approx-(250 MeV)^3$. By definition this is the quark Green function. If the quark propagator has only a slash term $\slash\!\!\!k$ in the denominator  (as correspond to the propagator of a massless particle), the trace over the spinor indices implied that the loop is identically zero. Therefore, chiral symmetry breaking require that quarks develops a dynamical mass so that
\begin{equation}
\langle\bar{\psi}\psi\rangle=-N_c\int\frac{d^4k}{(2\pi)^4i}{\rm Tr}\bigg[\frac{Z(k)}{M(k)-\slash\!\!\!k} \bigg]
\end{equation}
How the spontaneously breaking of the chiral symmetry comes about is still a topic of discussion. 

Without identifying the cause of spontaneous symmetry breaking, Manohar and Georgi~\cite{Manohar:1983md} developed a model in which they assume that the two non-perturbative phenomena which characterizes QCD  are confinement and chiral symmetry breaking. They argue that there are no reason for the chiral symmetry scale $\Lambda_{\chi}\approx 1$ GeV and the confinement scale $\Lambda_{QCD}\approx 100-300$ MeV to be the same. Thus, the effective Lagrangian in the region between $\Lambda_{\chi}$ and $\Lambda_{QCD}$ will include as fundamental fields quark and gluon fields since they are colorful particles not bound into color-singlet
hadrons at such short distances. Furthermore the quark-gluon interaction will be
described by an SU(3) color gauge theory. 
Since the SU$_L$(3) × SU$_R$(3) global chiral symmetry is spontaneously broken, there is also an octet of pseudoscalar Goldstone
bosons, which complete the quark-quark interaction. The model leads to a consistent picture which qualitative explains the successes of the nonrelativistic quark model.

Shuryak~\cite{Shuryak:1981ff,Shuryak:1982dp} proposed a model of the QCD vacuum based in the hypothesis that the main non-perturbative gluon field fluctuations are of the instanton type and the QCD vacuum can be describes as a instanton liquid.\footnote{ Instantons are certain configurations of the Yang–Mills potential
satisfying the equations of motion $ d^{ab}_{\mu}F^b_{\mu \nu}$ in euclidean space, i.e. in imaginary time.}. This hypotesis has been suported by calculations on the lattice~\cite{Chu:1994vi, Faccioli:2003qz} which provides a strong evidence that
instantons and antiinstantons  are the only non-perturbative gluon configurations surviving after a sufficient smearing of the quantum gluon fluctuations. 
 
The model was developed later on by Diakonov {\it et al.}~\cite{Diakonov:1983hh} and explains in a simple and elegant way the spontaneous
breaking of chiral symmetry (see ref~\cite{Diakonov:1995ea, DIAKONOV19961} for a review).

When quarks are introduced in the instanton vacuum its propagator gets modified by interacting with the fermionic zero modes of the individual instantons. Averaging  over the instanton ensemble Diakonov {\it et al.}~\cite{Diakonov:1983hh} has derived a closed equation for the averaged quark propagator, which has the form of a massive propagator with a momentum-dependent
dynamical mass
\begin{equation}
S(p)=\frac{\gamma^{\mu}p_{\mu}+iM(p^2)}{p^2+M^2(p^2)} 
\end{equation}
where the quark momentum dynamical mass is related with the instanton density, $N/V$, and  the average instanton size, $\bar{\rho}$, as
\begin{equation}
M(p^2)=~\sqrt{\frac{\pi^2N\bar{\rho}^2}{VN_c}}F(\bar{\rho} p)
\end{equation}
where $F(z)$ is a combination of the modified Bessel functions, which is equal to 1 at $z=0$ and decreases rapidly with the momentum measured in units of the inverse of instanton size. 
Ref.~\cite{DIAKONOV19961} estimated that the value of the dynamical mass at zero momentum is around $350$ MeV/$c^2$ which agrees with the values used in the non relativistic quark model.

In the domain of the low momenta $k\ll 1/\bar{\rho}$, QCD reduces to a remarkably simple no trivial theory of massive quarks interacting with Goldstone bosons. It is given by the partition function~\cite{DIAKONOV19961} 
\begin{eqnarray}
  Z=\int {\mathcal D} \pi^A \int {\mathcal D} \psi^+
  {\mathcal D} \psi \exp \left\{ \int d^{4} x \,
  \psi^+ (x) \left[ i \gamma^\mu \partial_\mu + i M
  e^{i\frac{\lambda _a}{f_{\pi }}\phi ^{a}\gamma _5}
  \right] \psi (x) \right\}
\end{eqnarray}
from which an effective Lagrangian, invariant under chiral rotations can be derived as 
\begin{equation}
\label{lagrangian}
{\mathcal L}=\overline{\psi }(i\, {\slash\!\!\! \partial} -M(q^{2})U^{\gamma_{5}})\,\psi 
\end{equation}
where  $\psi$ is the quark spinor $U^{\gamma_5}=e^{i\frac{\lambda _a}{f_{\pi }}\phi ^{a}\gamma _5}$ is the Goldstone boson fields matrix and $M(q^2)$ is the dynamical
mass. We will analyze the consequences of the effect in the hadron spectra in the next section.

\section{Applications to the hadron spectra}
\label{sec7}

In this section we do not intend to make an exhaustive analysis of all papers published on the subject rather we will limit ourselves to reporting those works that can shed light on the contribution of chiral interaction to the understanding of the hadron spectrum. As a general comment we refer to the original papers to a more detailed description of the topics we are interested in.

The  one-gluon exchange color-magnetic interaction, which has been used in earlier attempts to describe the baryon spectra provide many of the qualitative and some of the quantitative features of the fine structure
of the baryon spectra. However, several important  features lacks of a convincing  explanation.
The first shortcoming of these type of models is the large spin-orbit interaction predicted by the one gluon exchange interaction in conflict with the empirical evidences. This problem can be mitigated by a) decreasing this interaction by convoluting the
potentials with a smearing function as suggested by relativistic dynamics~\cite{Capstick:2000qj} and b) partial canceling  the spin-orbit term with the spin-orbit interaction associated with Thomas precession in the confining potential.

A second problem is related with the color operator structure $(\vec \sigma_i\cdot \vec \sigma_j)(\vec \lambda_i\cdot \vec \lambda_j)$ of the one-gluon exchange interaction. The one-gluon exchange interaction is attractive in color-spin symmetric quark pair states and repulsive in antisymmetric ones. Therefore, the $\frac{1}{2}^+N(1440)$ resonance, which belongs to the N = 2 band, should have higher mass than the $\frac{1}{2}^-N(1535)$ (N=l) as both have the same mixed color-spin symmetry. Similar examples can be found in the $\Delta$ and hyperon spectra. These problems can be solved if the effects of the one gluon exchange is totally or partially replace by the one-pion exchange with a operator structure $(\vec \sigma_i\cdot \vec \sigma_j)(\vec \tau_i\cdot \vec \tau_j)$.

Among the quark models which incorporate chiral symmetry breaking, one finds mainly two different approaches developed by the groups of Graz (G model)~\cite{Glozman:1995fu} and Salamanca-T\"ubingen
(ST model)~\cite{PhysRevC.62.034002,Fernandez_1993}.

The G-model works in the spirit of the two-scale picture of Manohar and Georgi~\cite{Manohar:1983md} assuming that beyond the chiral symmetry breaking scale but within the confinement scale, a baryon should be considered as a system of three massive constituent quarks with an effective quark-quark interaction given by a central confining part, assumed to be harmonic or linear, and a chiral interaction that is mediated by the octet of pseudoscalar Goldstone bosons (GBE interaction) between the constituent quarks~\cite{PhysRevD.58.094030}
\begin{equation}
{\mathcal V_{\chi}}=\left[\sum_{a=1}^3 V_{\pi}(r_{ij})\lambda^a_i\lambda^a_j+\sum_{a=4}^7 V_{\kappa}(r_{ij})\lambda^a_i\lambda^a_j+\sum_{a=8}^8 V_{\eta}(r_{ij})\lambda^a_i\lambda^a_j + \frac{2}{3} V_{\eta'}(r_{ij})\right]
(\vec \sigma_i\cdot \vec \sigma_j)
\end{equation}
being $\lambda^a$ the flavor matrices and
\begin{equation}
{\mathcal V_{\gamma}}=\frac{g^2_{\gamma}}{4\pi}\frac{1}{3}\frac{1}{4m_im_j}\left\{\mu^2_{\gamma}\frac{e^{-\mu_{\gamma}r_{ij}}}{r_{ij}}-4\pi\delta(r_{ij})\right\}
\end{equation}
 with $\gamma=\pi$, $K$, $\eta$ and $\eta'$.
Due to the difficulties of the  color interaction with the level ordering, the model neglects the one gluon exchange interaction.

The GBE interaction has been derived only for point-like particles but the constituent quarks and pseudoscalar mesons have finite size. Therefore, it is quite natural to assume that at distances $r<r_0$, where $r_0$ is related to the constituent quark and pseudoscalar meson sizes, the GBE interaction should be attenuated. The  G-model uses two different strategies, either one uses a step function~\cite{Glozman:1996wq} or uses a suitable subtraction~\cite{Glozman:1997ag}
\begin{equation}
{\mathcal V_{\gamma}}=\frac{g^2_{\gamma}}{4\pi}\frac{1}{3}\frac{1}{4m_im_j}\left\{\mu^2_{\gamma}\frac{e^{-\mu_{\gamma}r_{ij}}}{r_{ij}}-
\lambda^2_{\gamma}\frac{e^{-\lambda_{\gamma}r_{ij}}}{r_{ij}}\right\}
\end{equation}
where
\begin{equation}
\lambda_{\gamma}=\lambda_0+\kappa\mu_{\gamma}
\end{equation}
To try to keep the number of free parameters as small as possible  it is assumed the same single octet-quark coupling constant $g_8^2/4\pi$ for all octet mesons. The value of this constant is extracted from the pion-nucleon coupling constant. The singlet coupling constant $g_0$ is scaled from the $g_8$ as $g_0=\epsilon g_8$. Taken the quark masses as $m_q=340$ and $m_s=500$ and the mesons masses at the experimental values, the only free parameters of the model are $\epsilon$, $\lambda_0$ and $\kappa$ and those involved in the confinement potential.

The ST-model was first developed to describe the $NN$ interaction in~\cite{Fernandez_1993}. It is based in a quark-quark potential derived from the Diakonov Lagrangian~\cite{Diakonov:1995zi}
\begin{equation}
\label{lagrangian}
{\mathcal L}=\overline{\psi }(i\, {\slash\!\!\! \partial} -M(q^{2})U^{\gamma_{5}})\,\psi 
\end{equation}
As we have seen in the previous section, in this model the dynamical mass $M(q^{2})$ vanishes at large momenta and it is frozen at low 
momenta, for a value around 300 MeV. 
To simulate this behavior one parametrize the dynamical mass as
\begin{equation}
M(q^{2})=m_{q}F(q^{2})\,, 
\end{equation}
where $m_{q}\simeq 300$ MeV, corresponds to the constituent quark mass and
\begin{equation}
F(q^{2})=\left[ \frac{\Lambda _{\chi}^{2}}{\Lambda _{\chi}^{2}+q^{2}}%
\right] ^{\frac12} \, .
\end{equation}
acts as a form factor of the boson exchange interactions. The cut-off $\Lambda _{\chi}$ fix the chiral symmetry breaking scale. 

Instead the formulation of~\cite{Fernandez_1993} we will describe the update version of Ref.~\cite{Vijande_2005}
which allows an straightforward generalization to the heavy quark sector.
In this new version the Goldstone boson fields matrix $U^{\gamma_5}$ is expanded in terms of boson fields as
\begin{equation}
U^{\gamma_5}= 1+\frac{1}{f_{\pi}}\gamma^5\lambda^a \pi^a-\frac{1}{2f^2_{\pi}} \pi^a \pi^a+...
\end{equation}
with $a=1,\ldots,8$.
Inserting this expansion in Eq.(\ref{lagrangian}) one obtains
\begin{equation}
{\mathcal L}=\overline{\psi }\left(i\, {\slash\!\!\! \partial} -m_qF(q^{2})\left\{1+\frac{1}{f_{\pi}}\gamma^5\lambda^a \pi^a-\frac{1}{2f^2_{\pi}} \pi^a \pi^a+...\right\}\right)\psi
\end{equation}
with $g_{ch}\approx m_q/f_{pi}$
The second term generates the constituent quark mass and the third one gives rise to three pseudoscalar one-boson
exchange interaction between quarks, the one-pion exchange $\pi^a\equiv \vec{\pi}$, the one-kaon exchange $\pi^a\equiv K^a$ and one-eta exchange $\pi^a\equiv\eta_8$. The main contribution of the third term comes from the exchange of two correlated pions that can be simulated by means of a scalar $\sigma$ exchange potential. The two-kaon and two-eta exchanges are very short ranged and can see safely be neglected. 
The different terms of the potential contains central and tensor, or central and spin-orbit contributions. 

The chiral coupling constant $g_{ch}$ is determined from the $\pi NN$ coupling constant. The mass of the pseudoscalar boson is fixed from its experimental values. The mass of the sigma meson is provided by the PCAC relation~\cite{PhysRevD.26.239} $m_{\sigma}^2=(2m_q)^2+m_{\pi}^2$.
$\Lambda_{\chi}$ is in the range 600 MeV and 1 GeV in order to reproduce the adequate
cut-off due to the instanton fluctuation size. One can estimate the
$N\Delta$ mass difference due to the OPE interaction and ranges between 150
and 200 MeV for these cut-off values. These values are far from the experimental one (300 MeV), meaning that
the non-perturbative contributions are not enough to reproduce it.
The rest of the mass difference must have its origin in a perturbative process. This effects are taken into account trough the one-gluon-exchange potential. 
Therefore, $\Lambda_{\chi}$ controls the ratio between the contribution of the
pion and the gluon. To fix more accurately the value of this parameter one has to look for a
process where the contribution of one of the two exchanges is suppressed. A process driven
by the tensor part of the interaction would be adequate for our purposes, since we know
that the tensor interaction coming from the OGE is negligible in comparison with the one
coming from the OPE. Thus the value of $\Lambda_{\pi}$ has been fixed in Ref.~\cite{Fernandez:1992xs}
studying the reaction $pp\to n\Delta^{++}$,
a process driven by the tensor interaction. The value of $\Lambda_K$ is taken as 
$\Lambda_K\approx\Lambda_{\pi}+m_s$ to include the large strange current quark mass.

The one-gluon exchange potential is obtain as the Breit interaction Eq.(\ref{eq:VBreitinr}), although some momentum dependent terms are usually neglected.
This potential presents a contact term which is regularized in a suitable way
\begin{equation}
\delta(r)\to \frac{1}{4\pi r_0^2(\mu)}\frac{e^{-r/r_0(\mu)}}{r}
\end{equation}
where $r_0(\mu)=\hat{r_0}/\mu$ scaling with the reduced mass $\mu$.

In order to cover a wide energy range to describe light, strange and heavy mesons, the quark gluon coupling constant is scaled with the  reduced mass $\mu$ of the interacting $qq$ pair as
\begin{equation}
\alpha_{s}(\mu)=\frac{\alpha_{0}}{\ln\left( 
\frac{\mu^{2}+\mu_{0}^{2}}{\Lambda_{0}^{2}} \right)},
\label{eq:alpha_s}
\end{equation}
where $\mu$ is the reduced mass of the $qq$ pair and $\alpha_{0}$,
$\mu_{0}$ and $\Lambda_{0}$ are parameters of the model determined by a global
fit to the hyperfine splittings from the light to the heavy quark sector.

The quark-quark interaction also includes an screened confining potential which, as mentioned at the end of Sec.\ref{sec4}, simulate at some distances the spontaneous creation of $q\bar q$ pairs
\begin {equation}
V_{\rm CON}^{\rm C}(\vec{r})=\left[ -a_{c}(1-e^{-\mu_{c}r})+\Delta \right] 
(\vec{\lambda}_{1}\cdot\vec{\lambda}_{2}),
\end{equation}
where $\Delta$ is a global constant fixing the origin of energies.

Both models has been applied to the description of the baryon spectra. In Ref.~\cite{Glozman:1996wq}  the group of Graz describe with a rigorous three-body Faddeev calculations the masses of all 14 lowest states in the $N$ and $\Delta$ spectra using the Goldstone-boson-exchange interaction plus linear confinement.
The whole set of lowest $N$ and $\Delta$ states is quite correctly reproduced. In the most unfavorable
cases deviations from the experimental values do not
exceed $3\%$. In addition all level orderings are correct.
In particular, the positive-parity state $N(1440)$ (Roper
resonance) lies below the pair of negative-parity states
$N(1535)$ and $N(1520)$ . The same is true in the $\Delta$ spectrum
with $\Delta(1600)$ and the pair $\Delta( 1620)-\Delta( 1700)$.
Thus with this type of interactions the long-standing problem of baryon spectroscopy
seems to be resolved.

In Ref.~\cite{Glozman:1997ag} the study is extended to all the light and strange baryons. The authors use a semi-relativistic
chiral constituent quark model based on the following
three-quark Hamiltonian:
\begin {equation}
{\mathcal H}=\sum_{i=1}^3 \sqrt{\vec{p}_i^2+m_i^2}+\sum_{i<j=1}^3 V_{ij}
\end{equation}
where the relativistic form of the kinetic-energy operator is
employed and the interaction part is
\begin{equation}
V_{ij}=V_{\chi}+Cr_{ij}+V_0
\end {equation}
The Schr\"odinger equation  corresponding to the three-quark  Hamiltonian is
solved using the stochastic
variational method~\cite{PhysRevA.53.1907}. 
Again a quite satisfactory description of the spectra of all low-lying light and strange baryons is achieved. 
In the $\Lambda$ and $\Sigma$ spectra the positive-parity fall below the negative  excitations, resolving the problem as in the non-strange sector.

In a series of papers~\cite{PhysRevC.64.058201, PhysRevC.68.035207, PhysRevC.72.025206, Garcilazo:2007eh} the ST model has been applied to the baryon spectrum. All the papers solve the three body problem using the Faddeev method in momentum space. They do not consider the non-central contributions arising from the different terms of the interaction because experimentally there are no evidence for important effects of this interaction pieces on the baryon spectra. Ref.~\cite{PhysRevC.64.058201} shows that using a set of parameters that allows to understand the $NN$ phenomenology~\cite{PhysRevC.62.034002} one obtains a quite reasonable description of the baryon spectrum. However to get the correct level ordering it is necessary to use a stronger pionic interaction at the expense of loosing the correct description of the $NN$ system. In Ref.~\cite{PhysRevC.68.035207} the same problem is addressed using  relativistic kinematics. The relative position of the negative and parity low energy states is determined basically by the relativistic kinematics together with the one-pion exchange interaction. The $\sigma$ exchange improve the description. The one gluon exchange, when properly taken into account, does not destroy the reproduction of the experimental data. 

As explained above, the inclusion of OGE allows to extend the calculation to the heavy quark sector. In the case of heavy quarks ($c$ or $b$) chiral symmetry is explicitly broken and therefore  boson exchange does not contribute and the only interactions are confinement and OGE. In Ref.~\cite{Garcilazo:2007eh} Garcilazo {\it el al.} compute the spectrum of $\Lambda_i$, $\Sigma_i$, $\Xi_i$, $\Omega_i$ ($i=c$ or $b$). Comparing the gross structure of the experimental spectrum and the theoretical predictions the authors found an overall good agreement. 

In this work the interplay between the pseudoscalar and OGE is clearly illustrated by studying the different behavior of the $\Sigma_i(3/2)^+-\Lambda_i(1/2^+)$ and the $\Sigma_i(3/2)^+-\Sigma_i(1/2^+)$ mass differences. In the $\Lambda$ the two light quarks are coupled to spin $0$ while in the $\Sigma$ are coupled to spin $1$. This fact makes that in the $\Sigma$ the pseudo-scalar and OGE interactions between the two light quarks almost cancel and the attraction is provided by the interaction between the light and heavy quark which is given only by the OGE potentials.

The ST model has been also applied to the meson spectra. Apart for an early work which shows how the model is able to reproduce the light meson spectra~\cite{PhysRevC.59.428}, Vijande {\it et al.}~\cite{Vijande_2005} performed a complete study of the meson spectra from the light $q\bar q$ states to the $b\bar b$ mesons with the same interaction which is obtained from the interaction using in the study of the baryon spectra thorough the transformation 
\begin{equation}
V_{q\bar q}=\sum_{\alpha} (-1)^{G_{\alpha}} V_{qq}(\alpha)
\end{equation}
where $G_{\alpha}$ is the $G$ parity of the exchanged field $\alpha$.

An overall  reasonable description of the well established $q\bar q$ states is obtained for all flavors. This calculation can be used as a starting point to compare the new mesons, whose nature is still unknown and some of them are in conflict with the naive quark model expectations. As an example, one has the case of the light scalar sector,  with  results that suggest that this states cannot be pure $q\bar q$ states. Other cases are the open charm mesons where the $D^*(2640)$ and $D_{Sj}(2573)$ seem to be compatible with $c\bar n$ and $c\bar s$ mesons whereas the $D^*_{Sj}(2317)$ seems to have a completely different structure. The study of these more complex structures that may include $qq\bar q\bar q$ states  or molecular structures is addressed in another section of this chapter.

The study of the meson spectrum is completed with the study of different meson decay modes. Besides the electromagnetic decays, the strong decays of a meson into two mesons are described in Refs.~\cite{Bonnaz:2001hz,Segovia:2013wma, Segovia:2016xqb} from light to heavy mesons. The decay  model used is the $^3P_0$ model. This model assumes a creation of a $q\bar q$ pair from the
vacuum with vacuum quantum numbers $J^{PC}=0^{++}$. It has only one parameter, the strength $\gamma$ of the decay interaction, which is regarded as a free constant. In Ref.~\cite{Segovia:2012cd}  a scale-dependent strength as a function of the reduced mass of the
quark–antiquark pair of the decaying meson is proposed  to achieve a global description of the strong decays
\begin{equation}
\gamma(\mu)=\frac{\gamma_0}{log\frac{\mu}{\mu_0}}
\end{equation}
This scale dependent strength allows to extend the results from the light to the charm and bottom sectors.
The results predicted by the $^3P_0$ model with the suggested running of the $\gamma$ parameter are in a global agreement with the experimental data.

The conclusion of this section is that models which incorporate the consequences of chiral symmetry breaking  and, in some cases, perturbative aspects of QCD thorough the OGE are able to give a reasonable description of the meson and the baryon spectrum together with a correct description of the baryon-baryon phenomenology.

\section{The 2003 revolution: The X(3872), a new spectroscopy}

The research on heavy hadron spectroscopy aims to reproduce the masses and properties of the mesons and baryons containing one or several heavy quarks, $Q\equiv\{c,b\}$. This quest started in 1974 with the discovery of the $J/\psi$ resonance in Brookhaven National Laboratory~\cite{PhysRevLett.33.1404} and Stanford Linear Accelerator~\cite{PhysRevLett.33.1406}, a particle rapidly assigned to the $c\bar c$ $1^3S_1$ state of the quark model which confirmed the existence of the charm quark predicted some years earlier~\cite{PhysRevD.2.1285}. The theoretical description of the $J/\psi$ resonances and other states detected afterwards, such as the $\Upsilon(1S)$ in 1977 at Fermilab~\cite{PhysRevLett.39.252}, were well understood within naive quark models.
As an example, for the charm and bottom meson sector, potential models have been very successful at describing the $Q\bar Q$ spectra even at the early years~\cite{Eichten:1978tg,Eichten:1979ms,Gupta:1984um,Barnes:1996ff}.
Already in 1978, the Cornell model~\cite{PhysRevD.17.3090} allowed to understand the low-lying charmonium states by means of a non-relativistic approach of heavy quarks interacting via a Coulomb-like term (inspired by the one-gluon exchange potential) plus a linear confining potential.
Not only the Cornell model was able to describe the 11 known charmonium states at that time, but correctly predicted the energy range of further 15 states that were discovered in the following years~\cite{PhysRevD.21.203}. In fact, the success of such a simple model is not accidental, as it can be related with the QCD static potential~\cite{SUMINO2003173,Mateu:2018zym}.

Hence, up to the beginning of the XXI century it seemed that everything was sorted out in the heavy spectroscopy.
However, nature is more complex and, in 2003, the $X(3872)$ resonance, now labeled as $\chi_{c1}(3872)$ in the PDG, was discovered first by the Belle Collaboration~\cite{PhysRevLett.91.262001} and soon after confirmed by BaBar~\cite{PhysRevD.71.071103}, CDF~\cite{PhysRevLett.93.072001} and D0~\cite{PhysRevLett.93.162002} Collaborations. This state was identified in the $J/\psi\pi^+\pi^-$ invariant mass of the
$B^+\to K^+ \pi^+\pi^-J/\psi$ decay, where the two pions originate from a $\rho^0$ meson~\cite{CDF:2005cfq},
that's it, an isospin 1 final state. The isospin-0 decay channel $J/\psi\omega\to J/\psi\pi^+\pi^-\pi^0$ was also measured~\cite{Belle:2005lfc}, finding an almost equal ratio between these two decay modes,
\begin{align}
        \frac{{\mathcal B}(\chi_{c1}(3872)\to \pi^+\pi^-\pi^0 J/\psi)}{{\mathcal B}(\chi_{c1}(3872)\to \pi^+\pi^- J/\psi)}&=
        1.0 \pm0.4({\rm stat})\pm0.3({\rm syst})\,.
\label{X3872}
\end{align}
This large isospin breaking is not compatible with a pure $c\bar c$ state, so either the isospin is not conserved in the decay process or the $\chi_{c1}(3872)$ has a sizable $I=1$ component. This puzzling property was difficult to accommodate in the simple quark model picture. A key feature that can give us a hint is that this state is
strikingly near the $D^0\bar D^{*\,0}$ threshold, with a binding energy given by~\cite{Belle:2008fma,Guo:2019qcn,LHCb:2020fvo},
\begin{align}
m_{D^{*0}}+m_{\bar D^0} -m_{X(3872)}
&= \{1.1^{+0.6+0.1}_{-0.4-0.3},\,
\,{\rm MeV}\,,\, (0.00\pm0.18)\,{\rm MeV}\,,\, (0.07\pm0.12)\,{\rm MeV} \}.
\end{align}
Additionally, the nearest quark model state with $J^{PC}=1^{++}$ quantum numbers is the $\chi_{c1}(2P)$ but most models predict its mass at higher energies. The most reasonable explanations for this deviation are that the closeness of the $D\bar D^*$ threshold has an influence on the bare $\chi_{c1}(2P)$ charmonium state, renormalizing its mass, or that the $\chi_{c1}(3872)$ resonance is a dynamically generated state in the $D\bar D^*$ channel, that's it, a hadronic molecule mainly bound due to pion exchange.
Actually, the first suggestion that loosely bound states can exist near the charm thresholds was already envisaged in the 1970's~\cite{Nussinov:1976fg,DeRujula:1976zlg}.
Indeed, the coupling of bare $c\bar c$ states with meson-meson thresholds was evaluated for excited states in the original Cornell model~\cite{PhysRevD.17.3090}, although it was found to be irrelevant for the states analyzed then.
The discovery of the $\chi_{c1}(3872)$ recovered this idea and soon after plenty of works developed this molecular picture~\cite{Tornqvist:2004qy,Close:2003sg,Voloshin:2003nt}.

Since the discovery of the $\chi_{c1}(3872)$, the B-factories (such as BaBar, Belle and CLEO), $\tau$-charm facilities (CLEO-c, BESIII) and proton--(anti-)proton colliders (CDF, D0, LHCb, ATLAS, CMS) have discovered a large number of possible exotic charmonium- and bottomonium-like states, and also states compatible with multiquark structure, i.e., with minimum quark content of four and five particles. The characteristics of these exotic mesons, baryons, tetraquarks and pentaquarks are diverse and challenge their theoretical study. For a exhaustive description of the state of the art in heavy quarkonium physics the reader is referred to some recent reviews~\cite{Mai:2022eur, Chen:2022asf, Guo:2022kdi, Huang:2023jec}.

Nevertheless, one of the strongest hints that coupled channel effects must be taken into account in the naive quark model are the recently discovered multiquark-like states. The states with minimum four quark content are, e.g., the charmonium-like charged $Z_c(3900)$, $Z_c(4020)$~\cite{BESIII:2013ris,BESIII:2013ouc}, $Z_{cs}(3985)^-$ and $Z_{cs}(4220)^+$~\cite{BESIII:2020qkh,LHCb:2021uow}, the bottomonium-like $Z_b(10610)$ and $Z_b(10650)$~\cite{Belle:2011aa,Belle:2015upu} and the tetraquark-like structures such as the $T_{cc}(3875)^+$~\cite{LHCb:2021vvq,LHCb:2021auc}, the $T_{cs0}(2900)^0$, $T_{cs1}(2900)^0$~\cite{LHCb:2020pxc} and the $T_{\psi\psi}$ states~\cite{LHCb:2020bwg,CMS:2023owd,Zhang:2022toq,ATLAS:2022hhx,Xu:2022rnl}.
The pentaquark-like states include the so-called $P_c(4457)$, $P_c(4440)$, $P_c(4380)$ and $P_c(4312)$~\cite{LHCb:2015yax,LHCb:2019kea}, with $uudc\bar c$ minimum quark content, or the recent hidden-charm pentaquarks with strangeness, dubbed $P_{cs}(4459)$ and $P_{cs}(4338)$~\cite{LHCb:2020jpq,LHCb:2022ogu}, with $udsc\bar c$ minimum quark content.

The question whether these structures are actual compact tetraquarks or pentaquarks or they are dynamically generated as meson-meson or meson-baryon molecules is an intense field of research nowadays. In this section we will focus on the second option, as most of these states are close to meson-meson or meson-baryon thresholds, which can be seen as an indication that the coupled-channels effect can be relevant for their formation.
Thus, we will discuss how to extend the naive quark model so it incorporates the effect of close channels, focusing on the meson spectrum~\footnote{The extension to the baryon spectrum is more complex, but follows the same logic.}.

\subsection{Unquenching the quark model}

The first tentative extension to the description of the meson spectrum in naive quark model is the inclusion of the effect of meson-meson channels,
which forces to create a $q\bar q$ to couple the $q\bar q+ q\bar qq\bar q$ sectors.
This are the so-called \emph{unquenched or unitarized quark models}~\cite{Bijker:2009up,Heikkila:1983wd} which
incorporates this effect through hadron loops, with or without including the interaction between the two hadrons in the loop.
Herein we will briefly describe how they are implemented.

As already mentioned, first we need to create an additional $q\bar q$ pair. In principle, one can use the same hamiltonian that generates the meson spectrum to study the quark-pair creation mechanism.
However, there are phenomenological models such as the $^3P_0$ model, which assumes that a $q\bar q$ is created from the vacuum, that are much simpler and that work reasonably well to describe the hadron strong decays~\cite{Ackleh:1996yt}.

\begin{figure}[t!]
\centering
\includegraphics[width=.45\textwidth]{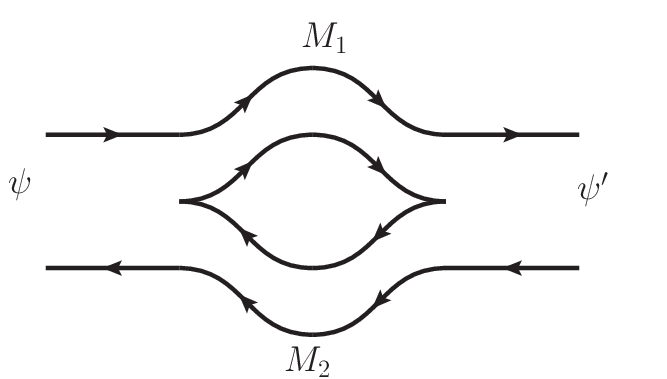}
\caption{\label{fig:loop} Schematic diagram of a meson-meson loop $q\bar q\to M_1M_2\to q\bar q$.}
\end{figure}

\begin{figure}[b!]
\centering
\includegraphics[width=.5\textwidth]{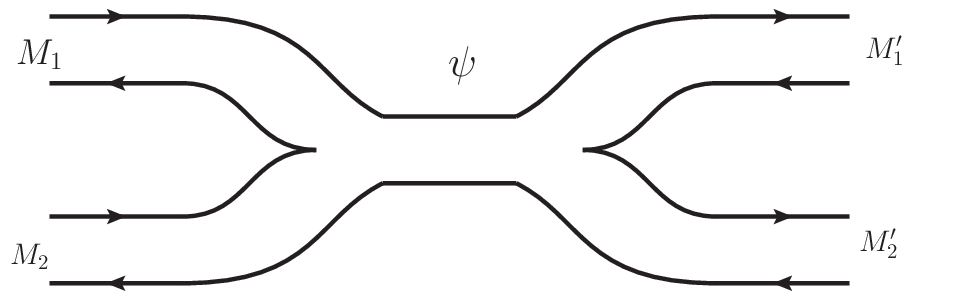}
\includegraphics[width=.35\textwidth]{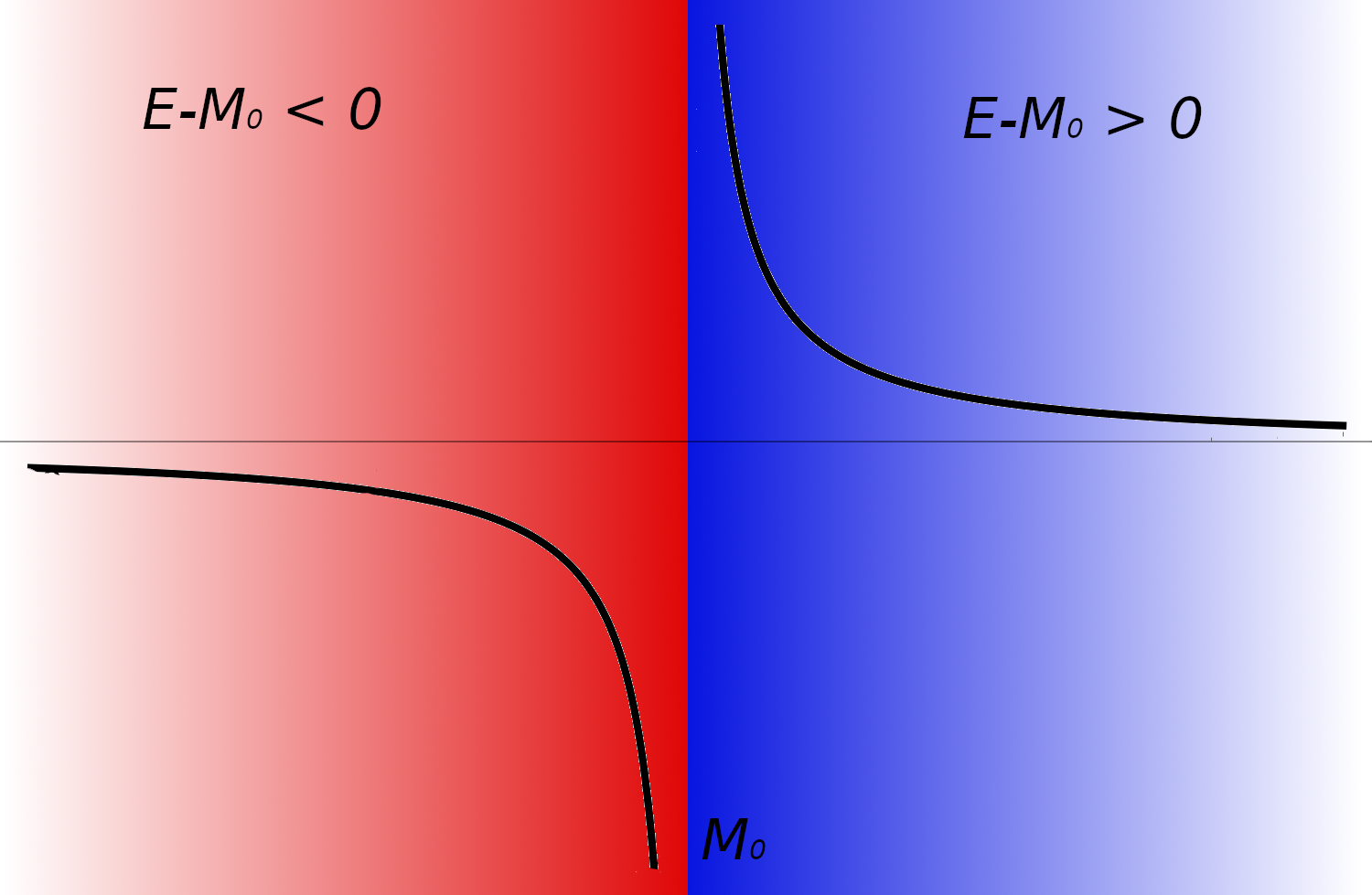}
\caption{\label{fig:Vcou} Left: Diagram of the effective potential of Eq.~\eqref{eq:couV} between two mesons due to the coupling with the meson spectrum, $M_1M_2\to\psi\to M_1'M_2'$. Right: Attractive (red) and repulsive (blue) regimes of $V_{\rm cou}$ for a given $\psi$ of mass $M_0$.}
\end{figure}

Regardless of the creation mechanism that is used, the one and two-hadron states gets connected, as shown in Fig.~\ref{fig:loop}. Depending on the case, this mixing
can be relevant, specially for mesons close to meson-meson thresholds, respectively.
Then, the physical state is no longer a simple $q\bar q$ but can be written as,

\begin{align}
| \Psi \rangle &= \sum_{i=1}^{N_\psi} c_i |\psi_i \rangle
 + \sum_{\alpha=1}^{N_\chi} \chi_\alpha(p)|\phi_{M_1}\phi_{M_2}\rangle_\alpha\,,
\end{align}
where $|\psi_i\rangle$ is the $i$-th naive $q\bar q$ quark model states of the spectrum
and $|\phi_{M_1}\phi_{M_2}\rangle_\alpha$ are two-meson states $M_1M_2$ with relative momentum $p$
and $\alpha$  quantum numbers. $N_\psi$ and $N_\chi$ are the number of $q\bar q$ and meson-meson channels
included in the calculation, respectively.
The Schr\"odinger equation for the composed state, $H | \Psi \rangle = E | \Psi \rangle$
can be written in terms of the relative wave function of the meson-meson channel,

\begin{align}\label{Hmult}
\sum_{\alpha=1}^{N_\chi} \int \left( H^{\alpha' \alpha}_{M_1M_2}(p',p) +
V^{\alpha'\alpha}_{\rm cou}(p',p)\right)\chi_\alpha(p) p^2 dp=E\chi_{\alpha'}(p')\,,
\end{align}
being $H_{\alpha' \alpha}^{M_1M_2}(p',p)$ the four-quark Hamiltonian, that encodes the kinetic energy of all quarks and interaction
between pairs of quarks, and $V_{\alpha'\alpha}^{\rm cou}$ is the interaction which is
generated by the coupling with the meson spectrum, shown in Fig.~\ref{fig:Vcou} and given by

\begin{align}\label{eq:couV}
V^{\alpha'\alpha}_{\rm cou}(p',p) = \sum_{i=1}^{N_\psi} \frac{h_{\alpha'i}(p')h_{i\alpha}(p)}{E-M_i}\,.
\end{align}
with $h_{i\alpha}(p)$ the amplitude that describes the $|\psi_i\rangle\to |\phi_{M_1}\phi_{M_2}\rangle_\alpha$ process.
 In Eq.(\ref{Hmult}) all internal degrees of freedom have been integrated out to get an equation for the relative motion using the Resonating Group Method (RGM)
that will be introduced in the last section.

The formalism can be extended to the complex plane as done in, e.g., Ref.~\cite{Ortega:2020tng,AHEP} using the Lippmann-Schwinger or the Bethe-Salpeter equation, but the relevant phenomenology
of the coupling can be already appreciated from the structure of $V_{\rm cou}$.
As an example, let's focus on one single meson-meson threshold $E_{\rm thr}$ and one single meson state $\psi$ with mass $M_0$. Then, the closeness and relative position of $M_0$ with respect to $E_{\rm thr}$ can add attraction or repulsion to $M_1M_2$. This is specially relevant at energies $E \sim E_{\rm thr}$ where one expects the hadronic molecules to appear.
Then, we will expect attraction for $E\sim E_{\rm thr} < M_0$, as $V_{\rm cou}<0$
and repulsion for $E\sim E_{\rm thr} > M_0$, being more intense when the closer the bare state is to the threshold ($M_0 \sim E_{\rm thr}$).
This feature can help to understand the nature of the $\chi_{c1}(3872)$ resonance~\cite{Ortega:2009zz,Cincioglu:2016fkm,Takizawa:2012hy}: The $D\bar D^*$+h.c. is almost bound due to the pion interaction, but the interaction is not strong enough to produce a molecular state. However, the bare $\chi_{c1}(2P)$ is around $60-80$ MeV above the $D\bar D^*$+h.c. threshold, so the $V_{\rm cou}$ is attractive. This coupling effect adds the necessary attraction to bind the $D\bar D^*$ system, generating the $\chi_{c1}(3872)$ as a $D\bar D^*$ molecule with a sizable $\chi_{c1}(2P)$ state. The bare $c\bar c$ state, on the other hand, gets dressed due to the coupling with the $D\bar D^*$ channel, acquiring width. This state could be the recently discovered $\chi_{c1}(4010)$~\cite{LHCb:2024vfz}.

So, summarizing, meson states above a meson-meson threshold help to bind the channel, while meson states below the threshold unbind it.
This complicates the study of the meson spectrum as the number of meson-meson thresholds grow and whether or not the meson-meson interaction is included in the hadron loop.

\subsection{Coupled channels}

As mentioned above, since 2003 it has become clear that the simple quark model is not sufficient to describe the heavy hadron spectra. The coupling of the meson and baryon spectra to
higher Fock components can partially solve this caveat. But, in some cases, as for the
tetraquark or pentaquark-like states, this inclusion is unavoidable, as the four and five quark
components are the minimum Fock spaces the system can access.
These multiquark states don't need to be colorless compact objects, but can still be resonances or bound states composed of two mesons or a meson and a baryon.

In this section we will give some examples of states that can be understand as dynamically generated objects of two hadrons due to the interaction of several channels.
The first signs that meson-baryon states can exist in nature dates back to the 1960's, with the discovery of the $\Lambda(1405)$ in the $\pi\Sigma$ invariant mass distribution of the $K^-p \rightarrow \pi\pi\pi\Sigma$ reaction at $1.15$ GeV~\cite{Dalitz:1959dn,Dalitz:1960du,Alston:1961zzd}.
The structure of this $J^P=\frac{1}{2}^-$ strange baryon~\cite{CLAS:2014tbc} is compatible with a quasi-bound $\bar KN$ state embedded in the $\pi\Sigma$ continuum.
Thus, the $\Lambda(1405)$ resonance was a pioneering example of an \emph{exotic} baryon, triggering a plethora of theoretical and experimental research that aimed to explore its inner structure.

The minimum quark content of the $\Lambda(1405)$ is $uds$, but this naive baryon structure can mix with $udsu\bar u$ or $udsd\bar d$. However, most of the studies ignore this coupling and are able
to describe the $\Lambda(1405)$ by performing a coupled-channels calculation of the $\bar KN$ and $\pi\Sigma$ channels. In this picture, the $\Lambda(1405)$ emerges from the interaction of these two channels, which can be explored, e.g., by using the Lippmann-Schwinger equation,

\begin{align}
\label{ec.14}
T^{\alpha'}_{\alpha}(z;p',p) =& V^{\alpha'}_{\alpha}(p',p)+\sum_{\alpha''}\int dp'' p''^{2} V^{\alpha'}_{\alpha''}(p',p'')\frac{1}{z-E_{\alpha''}(p'')}T^{\alpha''}_{\alpha}(z;p'',p),
\end{align}
where $\alpha$ represents the set of quantum numbers for a given partial wave $JLST$, $V$ is the full potential and $E_{\alpha}(q)$ is the non-relativistic energy for the momentum $q$.

The resonances, bound states and virtual states are, then, obtained from the poles of the on-shell $S$ matrix, which is calculated from the $T$-matrix in the non-relativistic kinematics,
\begin{align}
    S_\alpha^{\alpha'}(E) &= \delta_\alpha^{\alpha'}-2\pi\,i\,\sqrt{\mu_\alpha\mu_{\alpha'}k_\alpha k_{\alpha'}}T_\alpha^{\alpha'}(E;k_{\alpha'},k_\alpha),
\end{align}
with $k_\alpha$ the on-shell momentum of the meson-baryon system.

Most studies suggest a two-pole nature for the $\Lambda(1405)$ in the $\pi\Sigma$ nonphysical sheet, one associated to the $\bar KN$ and another to the $\pi\Sigma$ channel (see, e.g., Refs.~\cite{Hyodo:2011ur,Mai:2020ltx,Meissner:2020khl}). A similar state is found in the charm sector, the $\Lambda_c(2940)^+$~\cite{BaBar:2006itc},
which is consistent with a $D^*N$ molecular state in the $\pi\Sigma_c$ continuum.

Another interesting example of coupled-channels effect are the famous $P_c$ pentaquark states~\cite{LHCb:2015yax,LHCb:2019kea}, which are close to the
$\Sigma_c \bar D$, $\Sigma_c^* \bar D$ and $\Sigma_c \bar D^*$ thresholds. In this case, there is no possibility to couple with the baryon spectrum, as the minimum
quark content is $uudc\bar c$ and the energy is around $4.5$ GeV. The closeness of the poles to these meson-baryon thresholds
favours the hadron-hadron molecular interpretation, rather than a compact pentaquark state.
Similarly, many of the recent $T_{\psi\psi}$ states, such as the $T_{\psi\psi}(6200)$, $T_{\psi\psi}(6600)$ or the $T_{\psi\psi}(6900)$, tend to emerge near charmonium-charmonium thresholds such as the $J/\psi J/\psi$, $\eta_c\eta_c^\prime$ or the $J/\psi \psi^\prime$ threshold, respectively. Then, it is reasonable to investigate their properties in a coupled-channels formalism as it was done in Ref.~\cite{Ortega:2023pmr} based on a constituent quark model (CQM).

\subsection{Thresholds effects}

Mutual interactions between channels in a coupled-channels calculation usually results in the appearance of new states, either resonances or bound states, which emerge as poles in the $S$-matrix. However, another possible outcome of the coupled-channels calculations is that there are no bound states or resonances, but still the cross section shows enhancements due to the opening of a threshold.

Let's consider a single channel case. The effective range expansion for the $S$-matrix amplitude of a two-body scattering near threshold is given by~\cite{Dong:2020hxe},

\begin{align}
 f^{-1} &=\frac{1}{a}-i\,k+{\cal O}(k^4)
\end{align}
being $a$ the $S$-wave scattering length and $k$ the relative momentum.
Considering a non-relativistic approach, near the threshold $k=\sqrt{2\mu E}$, with $\mu$ the reduced mass and $E$ the
energy of the two-body system with respect to threshold. Then, we can write,

\begin{align}
 f^{-1} &=\frac{1}{a}-i\,\sqrt{2\mu E}\,.
\end{align}
The line shape will be proportional to $|f(E)|^2$, so,

\begin{align}\label{eq:cusp}
 |f(E)|^2 =\left\{\begin{array}{lll} \frac{1}{a^{-2}+2\mu E}&\mbox{for} &E\ge 0,\\
                   \frac{1}{(a^{-1}+\sqrt{-2\mu E})^2} & \mbox{for} & E<0.
                  \end{array}
\right.
\end{align}

\begin{figure}[t!]
\centering
\includegraphics[width=.5\textwidth]{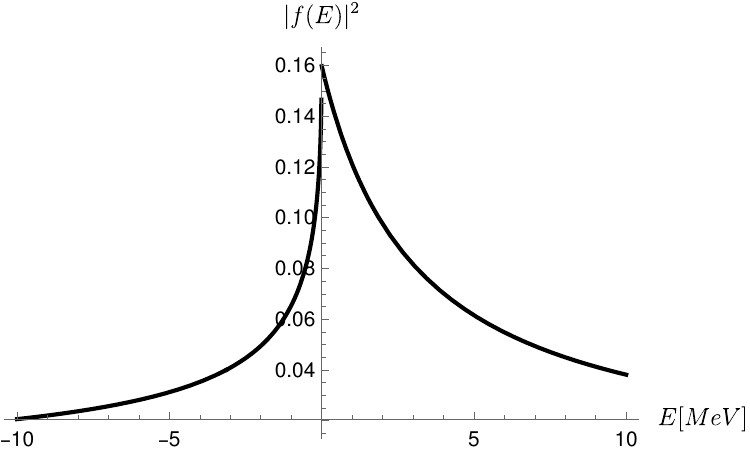}
\caption{\label{fig:cusp} Left: Example of a threshold cusp, Eq.~\eqref{eq:cusp}, with $\mu=1$ and $a=0.4$.}
\end{figure}

So, for $a>0$ (that's it, an attractive interaction but not strong enough to bind the two-body system) a peak appears at threshold ($E=0$) without any associated bound state, as shown in Fig.~\ref{fig:cusp}. Actually, as argued in Refs.~\cite{Dong:2020hxe,Guo:2014iya}, the threshold cusps may not appear
without the existence of a nearby pole (bound, virtual or resonance). In fact, the previous equation predicts a virtual state, a pole in the second Riemann sheet of the complex plane below the threshold, at energy $E=-(2\mu a^2)^{-1}$.

This effect has been used to explain the $X(4140)$ state, a $J^{PC}=1^{++}$ structure with hidden charm observed
in the $\phi J/\psi$ invariant mass spectrum~\cite{CDF:2011pep,Belle:2009rkh}.
Some studies have assigned this structure as a threshold cusp due to the opening of the $\phi J/\psi$ threshold,
associated to the presence of the $D_s\bar D_s^*$ channel. The residual $D_s\bar D_s^*$ interaction is strong enough to create a sudden rise of the line shape, but too weak to generate a bound or virtual state.

A more delicate example are the $Z_c$ and $Z_b$ charged resonances, which force to have a minimum quark content of four quarks.
The $Z_c(3900)$ was first discovered in $e^+e^-$ collisions by looking into the $\pi J/\psi$ and $D\bar D^*$ invariant mass distributions~\cite{BESIII:2013ris,Belle:2013yex,BESIII:2013qmu}.
Its heavy partner, the $Z_c(4020)$, was later discovered in the $\pi h_c$ invariant mass spectrum~\cite{BESIII:2013ouc}.
Both are $I^G(J^{PC})=1^+(1^{+-})$ resonances which have their bottom partners in the $Z_b(10610)$ and $Z_b(10650)$ resonances near the $B\bar B^*$ and $B^*\bar B^*$ thresholds, respectively~\cite{Belle:2011aa}.

The coupled-channels calculation of the $D^{*}\bar D^{(*)}$, $\pi J/\psi$ and $\rho \eta_c$ channels does not seem to generate a bound state that
describes the experimental line shapes. That is why threshold cusps have been used to explain strong energy dependencies near the $D^{(*)}\bar D^*$ and $B^{(*)}\bar B^*$ thresholds~\cite{Bugg:2011jr,Swanson:2014tra}, but the experimental peak is strong which points to the existence of a nearby pole.
More dedicated analysis of the BESIII data considering two-body coupled channels and triangle singularities~\cite{Albaladejo:2015lob,Pilloni:2016obd} shows the presence of virtual poles that can account for the experimental signal.

\begin{figure}[t!]
\centering
\includegraphics[width=.5\textwidth]{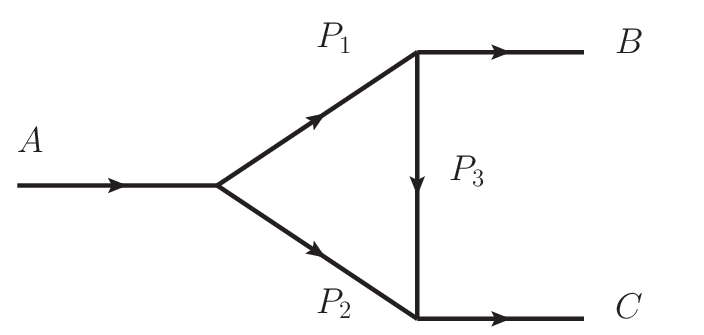}
\caption{\label{fig:triangle} Triangle diagram for $A\to BC$.}
\end{figure}

Triangle singularities appear when we have a loop of three particles which can be simultaneously on-shell, two of them having parallel momenta (see Fig.~\ref{fig:triangle}). Then, in the rest frame of particle $A$, the energies of the external particles are~\cite{Bayar:2016ftu,Abreu:2020jsl},

\begin{align}
 E_A&=P^0=m_A,\nonumber\\
 E_B&=k^0=\frac{m_A^2+m_B^2-m_C^2}{2m_A},\nonumber\\
 E_C&=P^0-k^0=\frac{m_A^2-m_B^2+m_C^2}{2m_A},\nonumber\\
 k&=\frac{\sqrt{\lambda(m_A^2,m_B^2,m_C^2)}}{2m_A}
\end{align}
where $\lambda(x,y,z)=x^2+y^2+z^2-2xy-2xz-2yz$ is the K\"allén function. Assuming that the three intermediate particles are scalars with no width, the
loop integral yields,

\begin{align}
 I = i\,\int \frac{d^4q}{(2\pi)^4} \frac{1}{(P-q)^2-m_1^2+i\varepsilon}\frac{1}{q^2-m_2^2+i\varepsilon}\frac{1}{(P-q-k)^2-m_3^2+i\varepsilon}
\end{align}

Considering that the three particles can be on-shell, we have $(q^2-m^2+i\varepsilon)^{-1}\to (2\omega(q^0-\omega))^{-1}$, with $\omega=\sqrt{q^2+m^2}$. Then, the integral is,

\begin{align}
 I = i\,\int \frac{d^4q}{(2\pi)^4} \frac{1}{8\omega_1\omega_2\omega_3}\frac{1}{[(P^0-q^0)-\omega_1(q)][q^0-\omega_2(q)][(P^0-q^0-k^0)-\omega_3(q)]}
\end{align}
where $\omega_{i=1,2}(q)=\sqrt{q^2+m_i^2}$ and $\omega_3(q)=\sqrt{q^2+k^2+2qkz+m_3^2}$, with $z$ the cosine of the angle of the vectors
$\vec q$ and $\vec k$.

If one integrates carefully in $q^0$ and in $z$~\cite{Bayar:2016ftu}, the amplitude diverges when these conditions are met:

\begin{align}
&P^0-\omega_1-\omega_2=0,\nonumber\\
&P^0-k^0-\omega_2-\sqrt{m_3^2+q^2+k^2\pm 2qk}=0\,,
\end{align}
that's it, all $P_1$, $P_2$ and $P_3$ are on-shell and $P_2$ and $P_3$ are collinear (so $\vec q\cdot\vec k=\pm1$). This leads to the so-called triangle singularity, which appears as a enhancement near the $BC$ threshold.  This effect is thought to be responsible, e.g., for the origin of the isovector $a_1(1420)$ resonance detected by COMPASS~\cite{COMPASS:2020yhb}.

In summary, all these threshold effects reinforce the idea that coupled channels need to be considered for the systems where exotic states have been discovered, which theoretically pushes us to go beyond the naive description of hadrons as simple $q\bar q$ or $qqq$ structures.

\section{Multiquark states}

The description of the hadron spectroscopy based on potential quark models detailed so far assumes that
the only relevant configurations are $q\bar q$ for mesons and $qqq$ for baryons.
This is a reasonable approach because, as it was originally commented by Gell-Mann~\cite{Gell-Mann:1964ewy,Gell-Mann:1962yej},
it is enough to describe the {\bf 1}, {\bf 8} and {\bf 10} representations for the lowest light baryons
and the {\bf 1} and {\bf 8} representations for the lowest light mesons.
However, these are not the only quark combinations compatible with color confinement, as further
$q\bar q$ pairs can be added to the lowest meson and baryon configurations giving rise to
tetraquark ($qq\bar q\bar q$), pentaquark ($qqqq\bar q$) or higher multiquark structures, already suggested by Gell-Mann in his original work~\cite{Gell-Mann:1964ewy,Gell-Mann:1962yej}.

This multiquark states differ from the exotic states described in the previous section as they are thought to be tightly bound, in opposition to molecular states or dynamically generated structures, however they can mix with the conventional meson or baryon spectrum as the molecular states.

\subsection{Tetraquarks}

The tetraquark states were the first exotic states to be analyzed in the pioneer work by R.~L.~Jaffe~\cite{Jaffe:1976ig}. These states are colorless compact states made of two quarks and two antiquarks.
The color wave function of a tetraquark can be obtained using two different, but connected, schemes: a ``molecular-like'' scheme, $[(q\bar q)(q\bar q)]$, and a diquark-antidiquark scheme, $[(qq)(\bar q\bar q)]$. The latter one is more convenient to impose the Pauli principle between identical particles, whereas the first one is more useful to analyze asymptotic meson-meson channels.

In the molecular-like scheme, the color SU(3) fundamental representation is the product of two meson (colorless) representations, $ {\bf 3\otimes\bar 3=1\oplus 8}$, so for a four-quark state one has $\bf (1\oplus8)(1\oplus 8)$,
\begin{align}
 \bf 1\otimes 1 &= \bf 1,\nonumber\\
 \bf 1\otimes 8 &= \bf 8,\nonumber\\
 \bf 8\otimes 1 &= \bf 8,\nonumber\\
 \bf 8\otimes 8 &= \bf 1\oplus 8\oplus 8'\oplus 10\oplus\overline{10}\oplus 27.
\end{align}

As physical states must be color singlets, the only allowed combinations are the $\bf [1\otimes 1]$ and $\bf [8\otimes 8]$ configurations giving the singlet representation $\bf 1$. The $\bf [1\otimes 1]$ is, indeed, the color configuration of a meson-meson molecule, that's it, two interacting color singlets, whereas the $[\bf 8\otimes 8]$ represents a hidden-color configuration, or the interaction of two colored meson-like states.

In the diquark-antidiquark picture, the color SU(3) fundamental representation of two quarks is given by $\bf 3\otimes 3=6\oplus \bar 3$, while two antiquarks give $\bf \bar 3\otimes\bar 3=\bar 6\oplus 3$. Then, the four-quark state has the following color combinations:
\begin{align}
 \bf 6\otimes \bar 6 &= \bf 1\oplus 8\oplus 27,\nonumber\\
 \bf 6\otimes 3 &= \bf 8\oplus 10,\nonumber\\
 \bf \bar 3\otimes \bar 6 &= \bf 8'\oplus \overline{10},\nonumber\\
 \bf 3\otimes \bar 3 &= \bf 1\oplus 8
\end{align}
Again, the only colorless combinations are given by $[\bf 6\otimes \bar 6]$ and $[\bf 3\otimes \bar 3]$. The $\bf [6\otimes\bar 6]$ is symmetric under the exchange of both quarks and antiquarks, while the $\bf [3\otimes\bar 3]$ is antisymmetric.

The two schemes are, nevertheless, related via a basis transformation. Labeling the quarks as $q_1q_2\bar q_3\bar q_4$ and combining $(q_1\bar q_3)(q_2\bar q_4)$,

\begin{align}
 [{\bf 1}_{13}\otimes {\bf 1}_{24}] &=\sqrt{\frac{1}{3}}[\bar {\bf 3}_{12}\otimes{\bf 3}_{34}]+\sqrt{\frac{2}{3}}[{\bf 6}_{12}\otimes\bar {\bf 6}_{34}],\nonumber\\
 [{\bf 8}_{13}\otimes {\bf 8}_{24}] &=-\sqrt{\frac{2}{3}}[\bar{\bf 3}_{12}\otimes{\bf 3}_{34}]+\sqrt{\frac{1}{3}}[{\bf 6}_{12}\otimes\bar {\bf 6}_{34}],
\end{align}
and, as we can couple as well $(q_1\bar q_4)(q_2\bar q_3)$,
\begin{align}
 [{\bf 1}_{14}\otimes {\bf 1}_{23}] &=-\sqrt{\frac{1}{3}}[\bar{\bf 3}_{12}\otimes{\bf 3}_{34}]+\sqrt{\frac{2}{3}}[{\bf 6}_{12}\otimes\bar {\bf 6}_{34}],\nonumber\\
 [{\bf 8}_{14}\otimes {\bf 8}_{23}] &=\sqrt{\frac{2}{3}}[\bar{\bf 3}_{12}\otimes{\bf 3}_{34}]+\sqrt{\frac{1}{3}}[{\bf 6}_{12}\otimes\bar {\bf 6}_{34}],
\end{align}

\begin{figure}[t!]
\begin{center}
 \includegraphics[width=.85\textwidth]{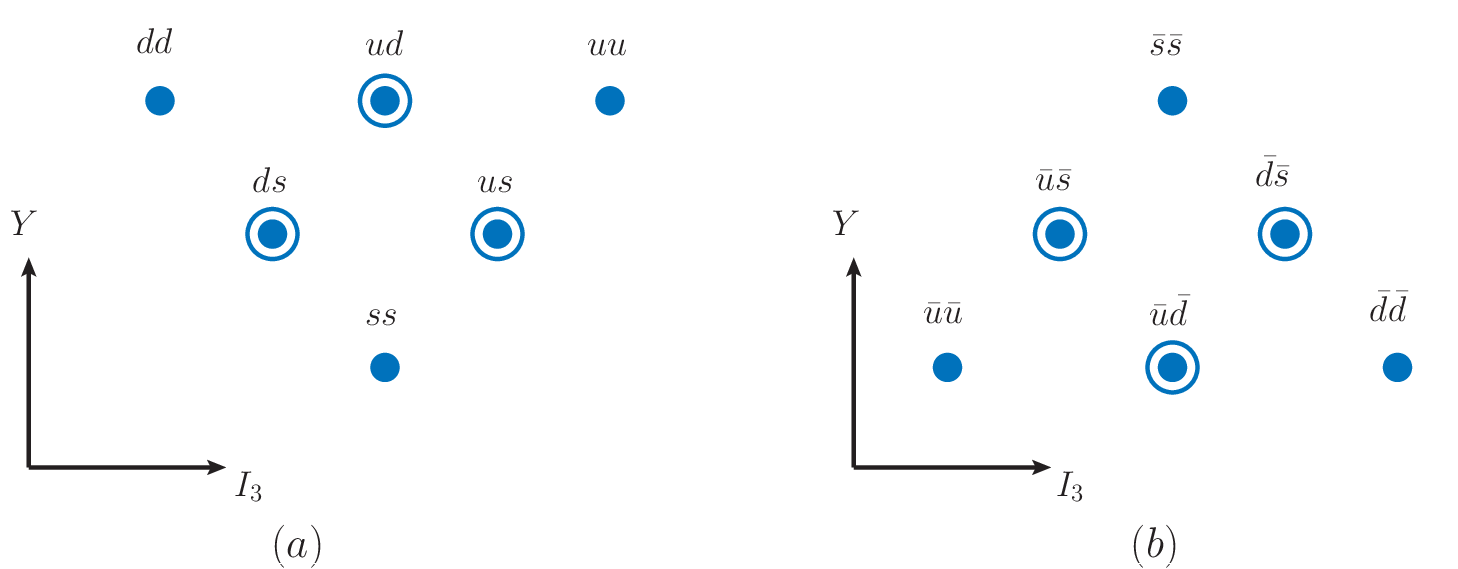}
 \end{center}
\caption{\label{fig:su3f} SU(3)$_f$ representations for diquarks ($qq$) and antidiquarks ($\bar q\bar q$): Panel \emph{(a)} shows the diquark $\bf 6$ (solid circles) and $\bf \bar 3$ (open circles) representations, while panel \emph{(b)} is the antidiquark $\bf \bar 6$ (solid circles) and $\bf 3$ (open circles) representations.}
\end{figure}

If we deal only with light quarks, that's it $q=\{u,d,s\}$, considering SU$_f$(3) the possible flavor configurations are the same described above for flavor, but now the representations allowed are not only the full singlet representations. In the diquark-antidiquark scheme we can build the $\bf 3(\bar 3)$ and $\bf 6(\bar 6)$ representations, shown in Figs.~\ref{fig:su3f}. Then, as the constituent quark mass $m_n\sim 330$ MeV we would expect that the ground state tetraquarks to have a mass of around $1.3$ GeV.

The most interesting scenario where tetraquarks can play a role are the scalar meson spectrum, $J^{PC}=0^{++}$. The $0^{++}$ mesons need to be in a $P$-wave and that extra unit of angular momentum pushes the lightest scalars to be around $1$ GeV, in the same energy region as the lightest tetraquarks in $S$-wave, which can naturally couple to $0^{++}$ quantum numbers. In other words, the energy cost of exciting the angular momentum up to $L=1$ is similar as creating an extra $q\bar q$ light pair, so both $q\bar q$ and $qq\bar q\bar q$ can coexist and mix.

\begin{figure}[t!]
\begin{center}
 \includegraphics[width=.85\textwidth]{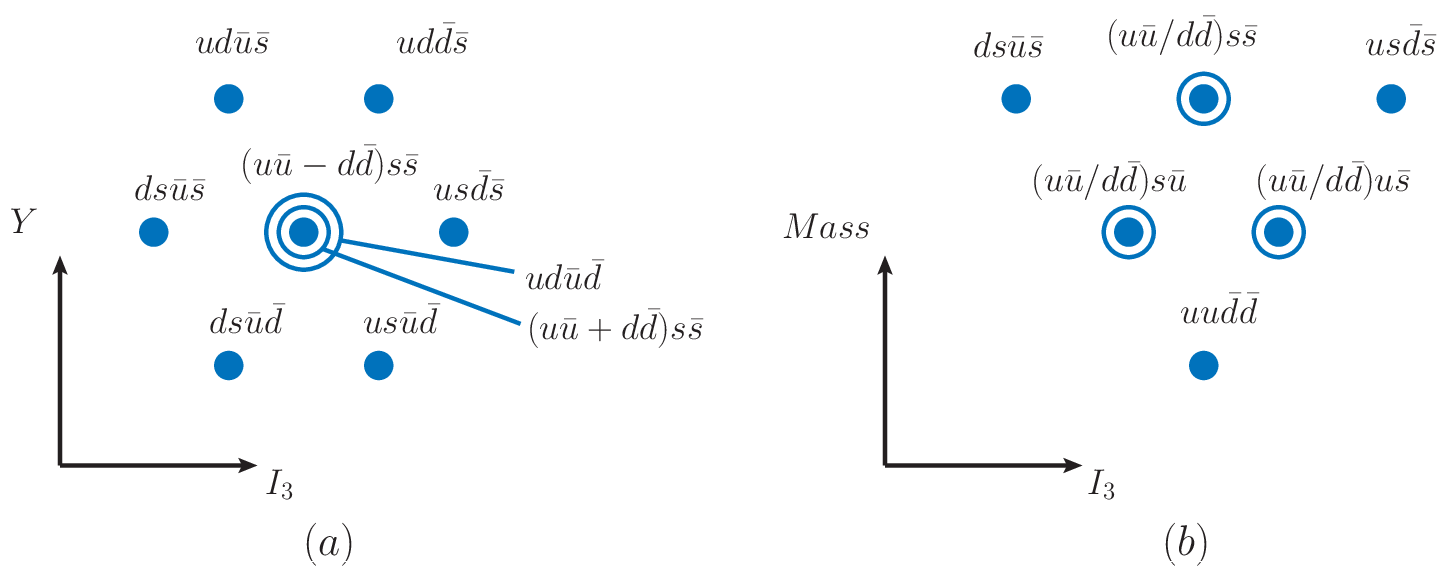}
 \end{center}
\caption{\label{fig:su3fnonet} \emph{(a)}: SU(3)$_f$ Tetraquark flavor nonet representation. \emph{(b)}: Mass inverted spectrum of the tetraquark flavor nonet.}
\end{figure}

Furthermore, one needs to consider the color hyperfine splitting which arise from the one gluon exchange (OGE),

\begin{align}
V_{\rm hyp}  &= -a\,(\vec\lambda^c_i\cdot\vec \lambda^c_j)(\vec\sigma_i\cdot\vec\sigma_j)
\end{align}
with $a\sim 30$ MeV,
which favor states in which quarks and antiquarks are antisymmetric in flavor and spin. Then, the
lowest configuration for $qq\bar q\bar q$ is the $\bf [\bar 3\otimes 3]_f$, the flavor nonet with spin $0$ shown in Fig.~\ref{fig:su3fnonet}. Indeed, in this case, $S=0$ for the two quarks and the two antiquarks and $\langle \vec \sigma_i\cdot\vec\sigma_j\rangle = -3$ for $ij$ both quarks or antiquarks and $\langle \vec \sigma_i\cdot\vec\sigma_j\rangle = 0$ for any quark-antiquark couple, and $\langle \vec\lambda_i\cdot\vec\lambda_j\rangle = -\frac{8}{3}$ for $ij$ both quarks or antiquarks, giving, for $(qq)(\bar q\bar q)$

\begin{align}
 \langle V_{\rm hyp}\rangle &= \langle V_{\rm hyp}\rangle_{qq}+\langle V_{\rm hyp}\rangle_{\bar q\bar q}=-16a \sim 0.5\,GeV
\end{align}
so the tetraquark scalar nonet would lie at $4m_n-16a\sim 900$ MeV. Actually, in this energy region there are four scalar mesons, $f_0(500)$, $K_0^*(700)$, $a_0(980)$ and $f_0(980)$, which has been traditionally related as members of this nonet. In this tetraquark picture, their quark content would be,

\begin{align}
 f_{\rm 0}(500) &= u\bar ud\bar d,\nonumber,\\
 K_{\rm 0}^*(700) &= (u\bar sd\bar d,d\bar su\bar u,\bar d su\bar u,\bar u sd\bar d),\nonumber\\
 f_{\rm 0}(980) &= s\bar s(n\bar n)_{I=0} ,\nonumber\\
 a_{\rm 0}(980) &= s\bar s(n\bar n)_{I=1}
\end{align}
with $n=\{u,d\}$, $(n\bar n)_{I=0}=\frac{1}{\sqrt{2}}(u\bar u+d\bar d)$ and  $(n\bar n)_{I=1}=(u\bar d,\frac{1}{\sqrt{2}}(d\bar d-u\bar u),-d\bar u)$.

The most surprising property of this nonet compared to the meson nonet is its inverted spectrum (see Fig.~\ref{fig:su3fnonet}), whose resemblance with the experimental situation of the light scalar states is evident. Unfortunately, the energy region where these tetraquarks are predicted is overpopulated with many meson resonances. As their quantum numbers are the same as the $0^{++}$ mesons, in a relative $^3P_0$ partial wave, these $2q2\bar q$ states are expected to strongly mix with the conventional $q\bar q$ strongly, complicating the proper identification of their exotic nature.

\begin{figure}[t!]
\begin{center}
 \includegraphics[width=.85\textwidth]{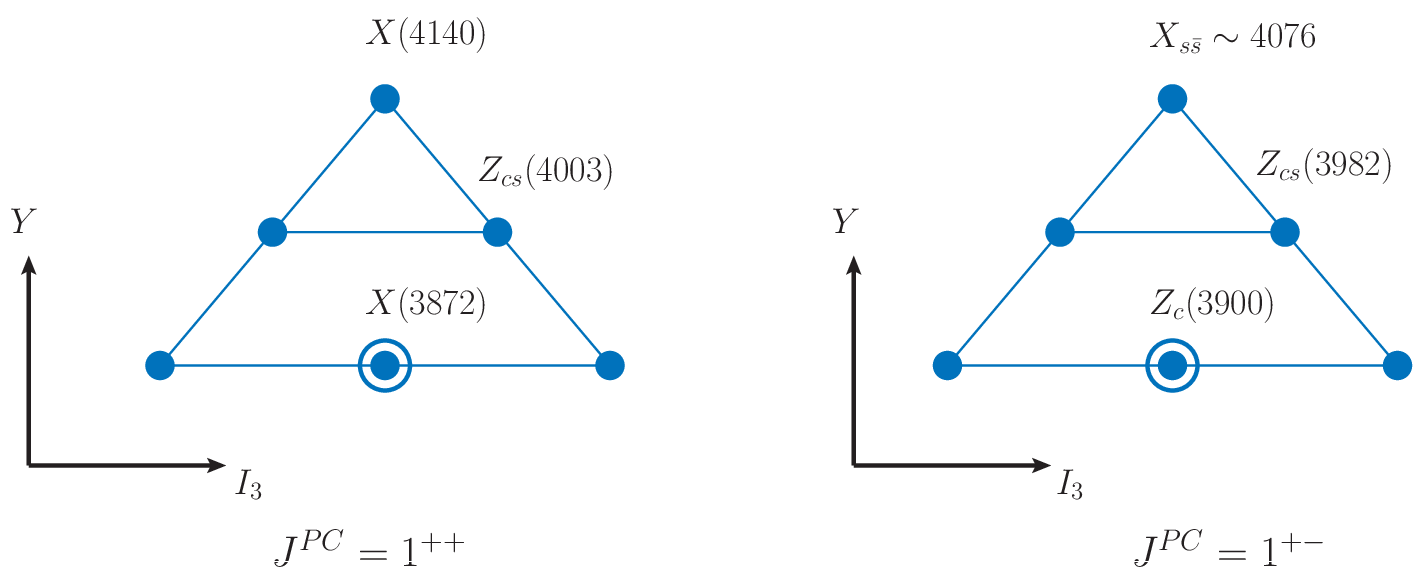}
 \end{center}
\caption{\label{fig:multipl} Possible hidden charm tetraquark nonets from Ref.~\cite{Maiani:2021tri}.}
\end{figure}

The situation is similar in the heavy spectrum, where no undoubted candidate has been detected yet. Most of the signals found with exotic quantum numbers or non-conventional properties lie close to meson-meson or meson-baryon thresholds, which suggests a molecular nature or alternative coupled-channels effects, as described in the previous section. However, there are studies~\cite{Maiani:2021tri} which claim that some of the $Q\bar Qq\bar q$ doubly-heavy exotics can form two SU$_f$(3) multiplets, with mass difference determined by the $s$ quark mass difference $m_s-m_u\sim 120-150$ MeV. Their proposed assignment is shown in Fig.~\ref{fig:multipl}. In our humble opinion, although there is a clear mass splitting in the $c\bar c q\bar q$, this is not an indication of a compact tetraquark nature.
If those resonances are created due to their closeby meson-meson thresholds, their masses are expected to follow the mass splitting of these thresholds. In this case, e.g., we would have,

\begin{align}
 m_{X(4140)}-m_{Z_{cs}(4003)} &\approx m_{D_s^*\bar D_s^*}-m_{D\bar D_s^*} \approx m_{s}-m_u,\nonumber\\
 m_{Z_{cs}(4003)}-m_{X(3872)} &\approx m_{D\bar D_s^*}-m_{D\bar D^*} \approx m_s-m_u
\end{align}

Although the compact tetraquark picture for those detected exotic candidates has not been ruled out, the discussion now is focused on developing observables that allow us to clearly discriminate between hadronic molecules and compact multiquarks. For example, the analysis of the production and decays in prompt and non-prompt hadron collisions or femtoscopic correlations can shed light to the inner structure of this resonances.

\subsection{Pentaquarks}

As for the tetraquarks, the first mention to pentaquarks (states with four quarks plus an antiquark) rolls back to the original works of Gell-Mann~\cite{Gell-Mann:1964ewy,Gell-Mann:1962yej}. The first dynamical studies were scarce at that time, with the pioneer work of D.~Strottman~\cite{Strottman:1979qu} in the MIT bag model. However, it was with the discovery of the controversial $\Theta(1540)^+$ signal when the pentaquarks came into the spotlight.

Let's briefly summarize its history (more details can be found, e.g., in Refs~\cite{Praszalowicz:2024mji,Kim:2024tae}). In 2003, the DIANA~\cite{DIANA:2003uet} and LEPS~\cite{LEPS:2003wug} Collaborations announced the detection of a narrow baryon with positive strangeness (ie, explicitly exotic) with a mass of around $1.54$ GeV in the $K^0p$ and $K^+n$ invariant mass spectrum with more than $4\sigma$ of C.L.. The minimum quark content of the $\Theta(1540)^+$ state was $uudd\bar s$, and its mass was in agreement with previous predictions of the lightest pentaquarks of the $\bf \overline{10}$ flavor multiplet of positive parity performed in the 90's~\cite{Diakonov:1997mm}, theoretical studies which actually motivated the experimental search on the first place.

The next years after the announcement of LEPS and DIANA, other nine experiments claimed the observation of narrow $K^+n$ and $K^0p$ structures between $1.52$ and $1.56$ GeV, which motivated the inclusion of the $\Theta^+$ in the Particle Data Group with a 3-star status in 2004~\cite{ParticleDataGroup:2004fcd}. This status was low due to concerns about the results in the community. Nonetheless, further ten experiments searched for the $\Theta^+$ with no success. In particular, Belle~\cite{Belle:2005thz} and CLAS~\cite{CLAS:2005koo} Collaborations mimicked the DIANA and SAPHIR~\cite{SAPHIR:2003lnh} experiments, which gave positive results, but didn't find any structure.
The repetition of the strongest observation of the $\Theta^+$ with negative outcome had a large impact, questioning since then the existence of this light pentaquark. Thus, the significance of the $\Theta^+$ was reduced to two stars in the 2005 Review of the PDG, it was removed from the summary tables in 2007 and it disappeared from the PDG lists in 2008. There are still some attempts to detect this state, as the experiments cannot rule out its existence but set an upper limit for its production. The interest has risen again since the discovery of heavy pentaquark candidates in the charm and bottom sectors.

Similarly to the tetraquark case, we can derive the general color structure of the pentaquarks. The possible color representations of a $4q$ state can be obtained from the product of two quark representations $\bf 3\otimes 3=6\oplus\bar 3$, so
\begin{align}
 \bf 3\otimes 3\otimes 3\otimes 3 &= \bf (\bar 3\oplus 6)(\bar 3\oplus 6) = 3\oplus\bar 6\oplus 15\oplus 15'
\end{align}

Combining the previous representations with the additional antiquark $\bar q$ representation ($\bar 3$), we have,
\begin{align}
\bf  3\otimes\bar 3 &=\bf 1\oplus 8,\nonumber \\
 \bf \bar 6\otimes\bar 3 &=\bf 8\oplus \overline{10},\nonumber \\
 \bf 15\otimes\bar 3 &=\bf 8\oplus 10\oplus 27,\nonumber \\
 \bf 15'\otimes\bar 3 &=\bf 10\oplus 35,
\end{align}

The color representation must be colorless, so the only allowed SU$_c$(3) representation is $\bf 3\otimes\bar 3$.

\begin{figure}[t!]
\begin{center}
 \includegraphics[width=.5\textwidth]{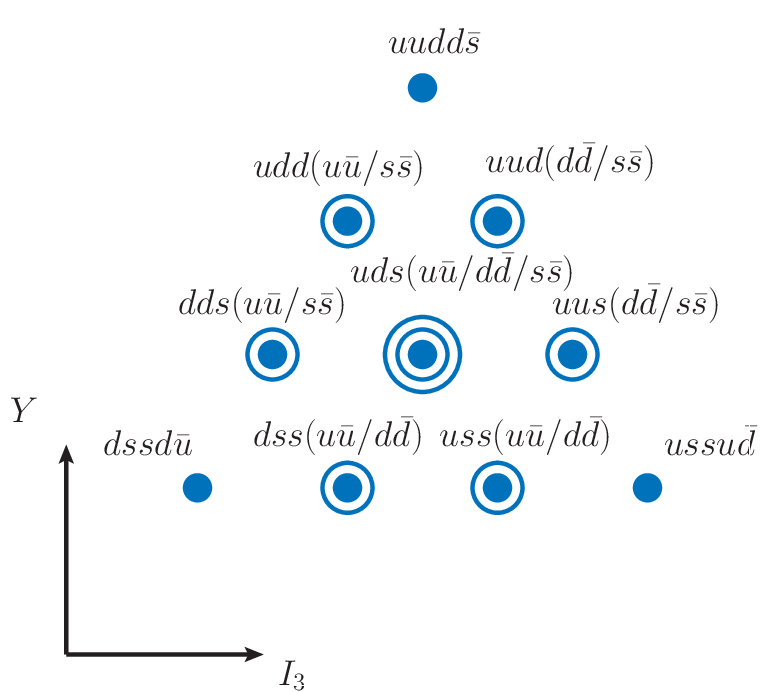}
 \end{center}
\caption{\label{fig:pentaq} Pentaquark $\bf \overline{10}$ (solid circles) and $\bf 8$ (open circles) SU(3) flavor representation.}
\end{figure}

For the flavor structure, if one only considers $\{u,d,s\}$ quarks, we have the same SU$_f$(3) representations.
As already mentioned, the hypothetical $\Theta^+$ has a minimal quark content of $uudd\bar s$. The lowest representation including positive-parity strangeness is $\bf \overline{10}$, which appears in the direct product of the flavor $\bf \bar 6$ of $4q$ and those of the $\bar q$: $\bf \bar 6\otimes\bar 3 =8\oplus \overline{10}$ (See Fig.~\ref{fig:pentaq}).

Some of the states in this representation can, indeed, mix with conventional baryons. For example, we have neutron-like and proton-like pentaquarks in both the antidecuplet and octet representations, with quark content,
\begin{align}
 |p\rangle_{\bf \overline{10}} &=\sqrt{\frac{1}{3}}|uudd\bar d\rangle+\sqrt{\frac{2}{3}}|uuds\bar s\rangle  ,\nonumber\\
 |p\rangle_{\bf 8} &= - \sqrt{\frac{2}{3}}|uudd\bar d\rangle+\sqrt{\frac{1}{3}}|uuds\bar s\rangle\,,
\end{align}
whereas the $\Theta^+$ is located at the top of the $\bf \overline{10}$ triangle.

The $\Theta^+$ signal from LEPS and DIANA turned out to be an illusion. But the hope of discovering a true pentaquark remained in the hadronic physics community for years. The first hint of an exotic baryon with a minimum quark content of 5 quarks was spotted in the $\Lambda_b^0\to J/\psi K^-p$ decay by the LHCb Collaboration in 2015~\cite{LHCb:2015yax}. The so-called $P_c(4380)$ and $P_c(4450)$ have $uudc\bar c$ minimum quark content, so they were firm candidates for pentaquark states. These states where reanalyzed in 2019 with a larger data sample, and the $P_c(4450)$ was found to be composed of two different resonances, the $P_c(4440)$ and the $P_c(4457)$. In addition, a new pentaquark $P_c(4312)$ was observed in the same process~\cite{LHCb:2019kea}. Additionally, the LHCb Collaboration has recently announced the discovery of the $P_c(4337)$ in the $J/\psi p$ invariant mass distribution~\cite{LHCb:2021chn}.

Soon after, in 2020, the LHCb Collaboration identified a hidden-charm pentaquark with strangeness, dubbed $P_{cs}(4459)$, in the $J/\psi \Lambda$ invariant mass distribution in the decay of the $\Xi_b$~\cite{LHCb:2020jpq}, with minimum quark content $udsc\bar c$, whose signal is also compatible with the mixture of two resonances close to the $\bar D^{*\,0}\Xi_c^0$ threshold. Another hidden-charm strange pentaquark, dubbed $P_{cs}(4338)$, followed, discovered in the $J/\psi\Lambda$ invariant mass spectrum~\cite{LHCb:2022ogu}.

Similarly as for the tetraquarks, most of these resonance structures lie close to meson-baryon thresholds. For example, the $P_c$ structures are close to the $\bar D\Sigma_c^*$  ($\sim 4.38$ GeV) or the $\bar D\Lambda_c^*$-$\bar D^*\Sigma_c$ ($\sim 4.46$ GeV) thresholds, while the $P_{cs}$ pentaquarks are near the $\bar D\Xi_c$ ($\sim 4.34$ GeV) and $\bar D\Xi_c'$-$\bar D\Xi_c^*$ ($\sim 4.4$-$4.5$ GeV) thresholds. Thus, their molecular nature cannot be ruled out~\cite{Ortega:2016syt,Ortega:2022uyu}. Hence, the inner structure of these multiquark states remains unknown.
Whether they are meson-baryon molecules, threshold cusps, triangle singularities or real compact states, the exploration of their nature will give us plenty of information about the strong interaction and the possible combinations that quarks can form beyond the conventional meson and baryon structures.

\section{Baryon Baryon interaction: Dibaryons}

We end this chapter with a section concerning the two baryon interaction and dibaryon resonances. Besides the applications discussed so far, the constituent quark model was used to study the nucleon-nucleon interaction. Constituent quark models were able to generate the $NN$ hard core due to antisymmetry~\cite{FAESSLER1983145,FAESSLER1983555} and when chiral models were introduced they also
include in a natural way the well known OPE tail of the interaction. This is the reason why the first application
of the ST model introduced in Sec.~\ref{sec7} was the $NN$ interaction~\cite{Fernandez_1993}. 

In the constituent
quark model the hadron-hadron interaction is a residual interaction due to the quark interactions between their
constituents. To obtain it, the Resonating Group Method~\cite{PhysRev.52.1083} (RGM) was traditionally used.
Antisymmetry effects introduced high non-localities in the potentials which are more complicate to treat in
coordinate space and for this reason the momentum space version was first introduced for the $NN$ system in Ref.~\cite{PhysRevC.62.034002}. Here we will briefly sketch it as an example of how Eq.(\ref{Hmult}) is obtained
for hadron-hadron interactions.

The two baryon wave function can be written as
\begin{eqnarray}
    \psi_{B_1B_2} &=& {\mathcal A}\big[ \phi_{B_1}(\vec p_{\xi_1})\phi_{B_2}(\vec p_{\xi_2}) \chi(\vec P) \chi_{B_1B_2}^{ST} \xi_c[2^3]\big]
\end{eqnarray}
where ${\mathcal A}$ is the antisymmetrizer operator, $\phi_{B_i}$ are the spatial internal wave functions of baryon $i$
in terms of its Jacobi coordinates $\vec p_{\xi_i}$, $\chi(\vec P)$ is the relative wave functions of the two baryons,
$\chi_{B_1B_2}^{ST}$ is the spin-isospin wave function of the two baryons coupled to total spin $S$ and total
isospin $T$ and $\xi_c[2^3]$ is the product of the two color singlet wave functions. Antisymmetry effects are introduce
through the operator ${\mathcal A}$ that for a 6 identical quarks with total antisymmetric wave functions for quarks 1 to 3 and 4 to 6 can be written as
\begin{eqnarray}
    {\mathcal A} &=& \frac 1 2 (1-P)(1-9P_{36})
\end{eqnarray}
with $P$ the operator that exchanges the three quarks of one baryon by the three quarks of the other baryon and
$P_{36}$ the operator that exchanges quarks 3 and 6. The operator $\frac 1 2 (1-P)$ gives the correct symmetry at baryon level and the the operator
$1-9P_{39}$ gives the direct term (for the term $1$) and exchange terms (for the term in $P_{36}$).

One starts from the Schr\"odinger equation of the six quark system and uses the variational principle
\begin{eqnarray}
    (H-E_T) |\psi \rangle = 0 \quad \Rightarrow  \langle \delta \psi | (H-E_T) |\psi \rangle = 0
\end{eqnarray}
In order to obtain an Schr\"odinger like equation for the relative motion, one does variations on it and uses the trick
\begin{eqnarray}
    \psi_{B_1B_2} &=& \int {\mathcal A}\big[ \phi_{B_1}(\vec p_{\xi_1})\phi_{B_2}(\vec p_{\xi_2}) \delta^3(\vec P-\vec P_i) \chi_{B_1B_2}^{ST} \xi_c[2^3]\big]
    \chi(\vec P_i) d^3 P_i
\end{eqnarray}
and obtains the equation
\begin{eqnarray}
    \left( \frac{P'^2}{2\mu} -E \right) \chi(\vec P') + \int \left[\,^{RGM}V_D(\vec P',\vec P_i)+\,^{RGM}K(\vec P',\vec P_i)\right] \chi(\vec P_i) d^3 P_i = 0
\end{eqnarray}
where $\mu$ is the reduced mass of the system, $E$ the relative energy of the baryons and $\,^{RGM}V_D(\vec P',\vec P_i)$ and $\,^{RGM}K(\vec P',\vec P_i)$ are the direct and exchange terms respectively. The direct
term is given by
\begin{eqnarray}
    \,^{RGM}V_D(\vec P',\vec P_i) &=& \sum_{i \in B_1, j\in B_2} \int \phi^*_{B_1}(\vec p'_{\xi_1})\phi^*_{B_2}(\vec p'_{\xi_2}) 
    V_{ij}(\vec P',\vec P_i) \phi_{B_1}(\vec p_{\xi_1})\phi_{B_2}(\vec p_{\xi_2})
    d\vec p'_{\xi_1}d\vec p_{\xi_2} d\vec p'_{\xi_1}d\vec p_{\xi_2}
\end{eqnarray}
Here all the internal degrees of freedom are integrated out and so the result depends on the initial and final relative momenta. $V_{ij}$ are matrix elements of the interaction operator of quarks
$i$ and $j$ between spin-isospin-color wave functions of the baryons. The operator $V_{ij}$ depends on the relative momentum (initial and final) of quarks $i$ and $j$ and is written in terms of the
Jacobi momenta and the relative momenta between baryons to integrate on the Jacobi momenta. Similarly the exchange operator is
\begin{eqnarray}
    \,^{RGM}K(\vec P',\vec P_i) &=& \,^{RGM}H_E(\vec P',\vec P_i)-E_T\,^{RGM}N(\vec P',\vec P_i)
\end{eqnarray}
being $\,^{RGM}N(\vec P',\vec P_i)$ the normalization kernel and
\begin{eqnarray}
    \,^{RGM}H_E(\vec P',\vec P_i) &=& \int \phi^*_{B_1}(\vec p'_{\xi_1})\phi^*_{B_2}(\vec p'_{\xi_2}) 
    H P_{36} \left[ \phi_{B_1}(\vec p_{\xi_1})\phi_{B_2}(\vec p_{\xi_2}) \delta^3(\vec P-\vec P_i)\right]
    d^3P d\vec p'_{\xi_1}d\vec p_{\xi_2} d\vec p'_{\xi_1}d\vec p_{\xi_2}
\end{eqnarray}
Details of the calculation can be found in the Appendix of Ref.~\cite{PhysRevC.62.034002}. One can project into partial waves and include more than one two-baryon channel
to end up with a system of coupled integral equations at baryon level. The equations are just Schr\"odinger like equations that can be transformed into Lippman-Schwinger like equations for scattering
that are solved in a standard way. The only difference with a usual non-relativistic two-baryon equation is that the exchange kernel is energy dependent through the term of the normalization kernel. However
in momentum space this is not a problem.
In Ref.~\cite{PhysRevC.62.034002} the scattering phase-shifts up to $J=6$ were shown finding an overall good agreement with empirical phase shifts.

Bound states of two baryons are called dibaryons. The only well stablished dibaryon is the deuteron, a $NN$ bound state with $J^P=1^+$. In Ref.~\cite{PhysRevC.62.034002} the deuteron
properties were evaluated finding a good agreement with experimental data. However other dibaryons can be found and already in 1989, Goldman predicted possible $\Delta\Delta$ dibaryons
in any constituent quark model base on OGE and confinement~\cite{Goldman:1989zj}. In 2011, the WASA-at-COSY Collaboration~\cite{WASA-at-COSY:2011bjg} found a candidate for these dibaryons from
the double pionic fusion reaction $pn\to d \pi^0\pi^0$, called the $d^*(2380)$, although there is still controversy~\cite{6416a7575db420433b7b9874}. The long history of the search for dibaryons in the light
quark sector can be found in Ref.~\cite{Clement:2016vnl}.

Having possible light dibaryon candidates,
one would expect that baryons with heavy quarks would also appear. 
There are already two Lattice QCD calculations with $\Omega_{ccc}\Omega_{ccc}$~\cite{Lyu_2021} and $\Omega_{bbb}\Omega_{bbb}$~\cite{Mathur_2023}
candidates and many constituent quark models predicts such states (see Ref.~\cite{PhysRevD.111.054002} and references there in).

%%%%%%%%%%%%%%%%%%%%%%%%%%%%%%%%%%%%%%%%%
%% Mandatory: A concluding paragraph summing up your main points in the chapter
%% Optional: Also include big questions in the field that are still to be answered. What topics/methods/questions are researchers like to focus on next?
\section{Conclusions}
\label{sec:conclusions}

The constituent quark model was build as a phenomenological model based on the symmetries of the naive quark model. A constituent quark mass was
introduced for light quarks, that now is understood to be an effect of the spontaneous breaking of the Chiral Symmetry of QCD. It has been used in many
problems from hadron spectra to hadron-hadron interactions with quite success. Modern constituent quark models mix one and two-hadron configurations,
and are used to study more complex states as tetraquark or pentaquark states, that are beyond the naive quark model, and it is one of the hot topics
in particle physics nowadays.

\begin{ack}[Acknowledgments]%
This work has been partially funded by 
Ministerio Espa\~nol de Ciencia e Innovaci\'on under grant
nos. PID2022-141910NB-I00 and PID2022-140440NB-C22;
Junta de Castilla y Leon under Grant
No. SA091P24;
and Junta de Andalucía under contract Nos. PAIDI FQM-370 and PCI+D+i under the title: ”Tecnologías avanzadas para la exploración del universo y sus componentes” (Code AST22-0001).
\end{ack}

%%%%%%%%%%%%%%%%%%%%%%%%%%%%%%%%%%%%%%%%%%%%
%% Optional: A list of references to other relevant works/articles/websites which are not cited in the text but that would further enhance a readers understanding of this topic
%\seealso{article title article title}

%%%%%%%%%%%%%%%%%%%%%%%%%%%%%%%%%%%%%%%%%
%% Mandatory: Bibliography using bibtex 
\bibliographystyle{Numbered-Style} %% for Numbered Reference Style
\bibliography{reference}

\end{document}